\documentclass[dvipdfm]{aa}
\usepackage{graphicx}
\usepackage{txfonts}
\usepackage{natbib}
\usepackage{longtable}
\bibpunct{(}{)}{;}{a}{}{,}

\begin{document}

\title{A Sino-German $\lambda$6~cm polarization survey of the Galactic plane} 
\subtitle{VII. Small supernova remnants}

\author{X. H. Sun\inst{1,2}
        \and P. Reich\inst{2}
        \and W. Reich\inst{2} 
        \and L. Xiao\inst{1}
        \and X. Y. Gao\inst{1}
        \and J. L. Han\inst{1}
        }

\institute{National Astronomical Observatories, CAS, Jia-20 Datun Road, 
           Chaoyang District, Beijing 100012, China\\
           \email{[xhsun;hjl]@nao.cas.cn}
            \and Max-Planck-Institut f\"{u}r Radioastronomie, 
                 Auf dem H\"ugel 69, 53121 Bonn, Germany\\
           \email{[preich;wreich]@mpifr-bonn.mpg.de}}

\date{Received / Accepted}

\abstract
{} 
{We study the spectral and polarization properties of supernova remnants (SNRs) 
based on our $\lambda$6~cm survey data.
}
{The observations were taken from the Sino-German $\lambda$6~cm polarization 
survey of the Galactic plane. By using the integrated flux densities at 
$\lambda$6~cm together with measurements at other wavelengths from the 
literature we derive the global spectra of 50 SNRs. In addition, we use the 
observations at $\lambda$6~cm to present the polarization images of 24 SNRs. 
}
{We derived integrated flux densities at $\lambda$6~cm for 51 small SNRs with 
angular sizes less than $1\degr$. Global radio spectral indices were 
obtained in all the cases except for Cas~A. For SNRs G15.1$-$1.6, G16.2$-$2.7, 
G16.4$-$0.5, G17.4$-$2.3, G17.8$-$2.6, G20.4+0.1, G36.6+2.6, G43.9+1.6, 
G53.6$-$2.2, G55.7+3.4, G59.8+1.2, G68.6$-$1.2, and G113.0+0.2, the spectra 
have been significantly improved. From our analysis we argue that the object 
G16.8$-$1.1 is probably an \ion{H}{II} region instead of a SNR. Cas~A shows a 
secular decrease in total intensity, and we measured a 
flux density of 688$\pm$35~Jy at $\lambda$6~cm between 2004 and 2008. Polarized emission from 25 SNRs 
were detected. For G16.2$-$2.7, G69.7+1.0, G84.2$-$0.8 and G85.9$-$0.6, the 
polarized emission is detected for the first time confirming them as SNRs. 
} 
{High frequency observations of SNRs are rare but important to establish 
their spectra and trace them in polarization in particular towards the 
inner Galaxy where Faraday effects are important.}

\keywords{ISM: supernova remnants -- Surveys -- Polarization -- Radio 
continuum: general -- Methods: observational}

\maketitle

\section{Introduction}

Supernovae release enormous energy into the interstellar medium (ISM). The 
ejected material as well as the material swept up by the shock wave form the 
supernova remnants (SNRs). SNRs can be traced up to several 
ten-thousand years 
\citep[e.g.][for a review]{rei02}. To study the evolution of SNRs is a key to 
understand the interaction between the blast wave from the supernova explosion 
and the ISM. Since SNRs are primary radio objects, the morphology, spectrum, 
and magnetic field configuration obtained from radio observations provide vital 
input to figure out the evolution of SNRs. 

The Sino-German $\lambda$6~cm polarization survey of the Galactic plane covers 
the broad band of the Galactic plane of $10\degr\leq l\leq230\degr$ and 
$|b|\leq5\degr$ \citep{shr+07,grh+10,srh+11,xhr+11}. The polarization 
structures such as Faraday screens, voids and canals revealed in the survey 
have advanced the understanding of the ISM. Meanwhile many SNRs were detected 
as strong discrete sources in the survey. A study of large SNRs with angular 
sizes exceeding about $1\degr$ was already presented by \citet{ghr+11}. In this 
paper, we focus on SNRs with a smaller angular size. Most of these small SNRs 
cannot be resolved in the survey, so that we could not investigate their 
morphology in detail. We obtained their integrated flux densities and 
established their spectra together with data at other radio wavelengths. Although 
spectra for most of the SNRs have already been presented in the literature, we are able to make 
significant improvements particularly for some weak and slightly extended 
SNRs by using the new measured flux densities from the $\lambda$6~cm 
and the Effelsberg $\lambda$11~cm and $\lambda$21~cm surveys 
\citep{rfrr90, frrr90a, rrf90, rrf97}. We also 
obtained polarization images for about half of the SNRs, some of which were 
used to estimate rotation measures (RMs) in context with polarization data 
obtained at other wavelengths such as $\lambda$11~cm and 
$\lambda$2.8~cm. 

The paper is organized as follows: We briefly describe the $\lambda$6~cm survey 
in Sect.~2. The results are reported in Sect.~3, where the integrated flux 
densities and spectra of 50 SNRs are presented, the polarization images of 
24 SNRs are displayed, and a possible mis-identification of G16.8$-$1.1 is 
discussed. Due to the secular decrease in intensity of Cas~A, this 
remnant is studied separately. The summary is given in Sect.~4.

\section{The $\lambda$6~cm survey}

The Sino-German $\lambda$6~cm polarization survey of the Galactic plane has 
been conducted by using the Urumqi 25-m telescope located about 70~km 
south of Urumqi city with the geographic longitude of $87\degr$E and latitude 
of $+43\degr$. The survey has an angular 
resolution of $9\farcm5$ and a system temperature of about 22~K towards the 
zenith. The central frequency was set to either 4.8~GHz or 4.963~GHz with 
corresponding bandwidths of 600~MHz and 295~MHz. The system gain is 
$T_{\rm B}[{\rm K}]/S[{\rm Jy}]=0.164$. Detailed information about the 
receiving system was already presented by \citet{shr+07}.

The Galactic plane was mapped in raster scans in both longitude and latitude 
directions. The separation of sub-scans was $3\arcmin$, and the scan velocity 
was $4\degr/{\rm min}$. Observations were made at night time with clear sky. 
The primary calibrator was 3C~286 with an assumed flux density 
at $\lambda$6~cm $S_{\rm 6\,cm}=7.5$~Jy consistent with that by 
\citet{bgpw77}, a 
polarization percentage of 11.3\%, and a polarization angle 
${\rm PA}=33\degr$. The sources 3C~48 
and 3C~138 were used as secondary calibrators, and 3C~295 and 3C~147 as 
unpolarized calibrators. 

The raw data from the receiving system contain maps of $I$, $U$, and $Q$ stored 
in \textsc{NOD2}-format \citep{has74}. Data processing follows the standard 
procedures developed for continuum observations with the Effelsberg 100-m 
telescope as detailed by \citet{shr+07} and \citet{grh+10}. In the final maps, 
the typical rms noise level is about 1~mK~$T_{\rm B}$ for total 
intensity $I$, and 0.5~mK~$T_{\rm B}$ for Stokes $U$ and $Q$, and polarized 
intensity (PI). 

\section{Results}

According to Green's catalogue\footnote{The Catalog of Galactic Supernova 
Remnants is available at the web site http://www.mrao.cam.ac.uk/surveys/snrs} 
\citep{gre09}, there are 99 small SNRs with angular sizes less than $1\degr$ in 
the survey region. Most of them are smaller than about $30\arcmin$ in size.  
A large fraction (80\%) of these SNRs is located in the region of $10\degr<l<60\degr$, 
where the diffuse emission is the strongest \citep{srh+11}. Because of strong 
confusion along the Galactic plane, it is difficult to determine flux 
densities of some of these SNRs. Patchy structures dominate the polarization 
images \citep{srh+11,xhr+11}, which makes it even more challenging to extract 
polarization information intrinsic to these SNRs. As it is shown 
in Sect.~3.1, we managed to 
obtain integrated flux densities for 51 SNRs and polarization images of 24 SNRs.

The SNR G111.7$-$2.1 (Cas~A) is presently the brightest source beyond the solar system 
in the sky. Cas~A is known to decrease in intensity by a measurable amount 
every year. Therefore we discuss this object separately. 

\subsection{Integrated flux densities}

To determine integrated flux densities of the SNRs, we first removed the 
large-scale diffuse emission using the ``background filtering" technique 
developed by \citet{sr79}. The filter beam size was set to about two or three 
times of the source size, to ensure that no emission from the SNR is 
eliminated. We then made integrations within a polygon region encompassing the 
SNR and subtracted the background estimated by averaging the intensities 
surrounding the edge areas of the SNR. If the SNR had a size comparable to the 
beam, we made a two-dimensional elliptical Gaussian fit to assess the flux 
density. For large circular objects, we performed a ring integration of the 
emission to calculate the total flux density. 
  
We were able to measure the $\lambda$6~cm integrated flux densities of 51 SNRs.
Among the remaining SNRs, some objects, such as G11.0$-$0.0, could not be 
separated from the strong emission along the plane. Emission from other 
objects, such as G83.0$-$0.3, was mixed up with the strong thermal emission 
from the Cygnus region, and some SNRs, such as G54.4$-$0.3 (HC 40), are located 
in a complicated environment and their boundaries could not be well defined. 
For all these objects we could not determine their flux densities with
sufficient precision.  

Flux densities at $\lambda$6~cm for 50 SNRs, excluding Cas~A, are listed in 
the fourth column of Table~\ref{snr_s}. For comparison, previous measurements 
at $\lambda$6~cm and the corresponding references are listed in the second and 
third column, respectively. In case that there are several flux density measurements at $\lambda$6~cm published, we selected the one with the 
highest quality. If the qualities are comparable we used the median. The early measured flux densities have been corrected to conform to the 
scale by \citet{bgpw77}. Some of the corrections were provided by 
\citet{kas89b}. 
The new measurements generally 
are in good agreement, within uncertainties, with previous published 
values.
For 17 SNRs: G15.1$-$1.6, G15.4$+$0.1, G16.2$-$2.7, 
G16.4$-$0.5, G17.4$-$2.3, G17.8$-$2.6, G20.4+0.1, G36.6+2.6, G40.5$-$0.5, 
G43.9+1.6, G53.6$-$2.2 (3C 400.2), G55.7+3.4, G59.8+1.2, G68.6$-$1.2, 
G96.0+2.0, G109.1$-$1.0 (CTB 109), and G113.0+0.2, no $\lambda$6~cm flux 
densities have been obtained up to date.

\subsection{SNR spectra}

We calculated the spectral indices of these SNRs by fitting flux 
densities from the literature together with the new measurements at 
$\lambda$6~cm versus frequencies in logarithmic scale with weighted least 
squares. Some published flux densities could not be corrected to the scale by 
\citet{bgpw77} because no calibration information could be found, which has only 
slight influence to the spectra.   
The newly obtained spectral indices for which we used additional published data 
with references in the eighth column, are listed in the seventh column in 
Table~\ref{snr_s}. We did not include data below 100~MHz to investigate 
the low-frequency spectrum turnover \citep[e.g.][]{kas89a}, as we are only 
interested in the high-frequency spectrum, which the new $\lambda$6~cm 
measurements can help to establish. 
For comparison, previous indices and corresponding 
references are listed in the fifth and sixth column in Table~\ref{snr_s}. If a 
spectral break in a SNR spectrum was firmly established, the spectral indices 
below and above the turnover frequency are both provided. New integrated flux 
densities at $\lambda$11~cm and $\lambda$21~cm were also derived in case either 
no flux density measurements were found or the data available in the literature 
were not sufficient to constrain the spectra. Unless otherwise noted, the flux 
densities at these two bands were measured from the $\lambda$11~cm and 
$\lambda$21~cm Effelsberg surveys\footnote{The data are available in the 
``Survey Sampler" maintained by MPIfR at the web site
http://www.mpifr.de/old\_mpifr/survey.html} \citep{rfrr90, frrr90a, rrf90, rrf97}. The angular resolution and sensitivity are $4\farcm3$ and 20~mK for the 
$\lambda$11~cm survey, and $9\farcm4$ and 40~mK for the $\lambda$21~cm survey. 
Both surveys have total intensity scales consistent with that by 
\citet{bgpw77}.
The 
reference entry (Column 8 in Table~\ref{snr_s}) remains empty if no earlier 
measurements were available. Plots of all SNR spectra are shown in 
Fig.~\ref{spectra}. 

\begin{figure*}[!htbp]
\centering
\resizebox{0.325\textwidth}{!}{\includegraphics[angle=-90]{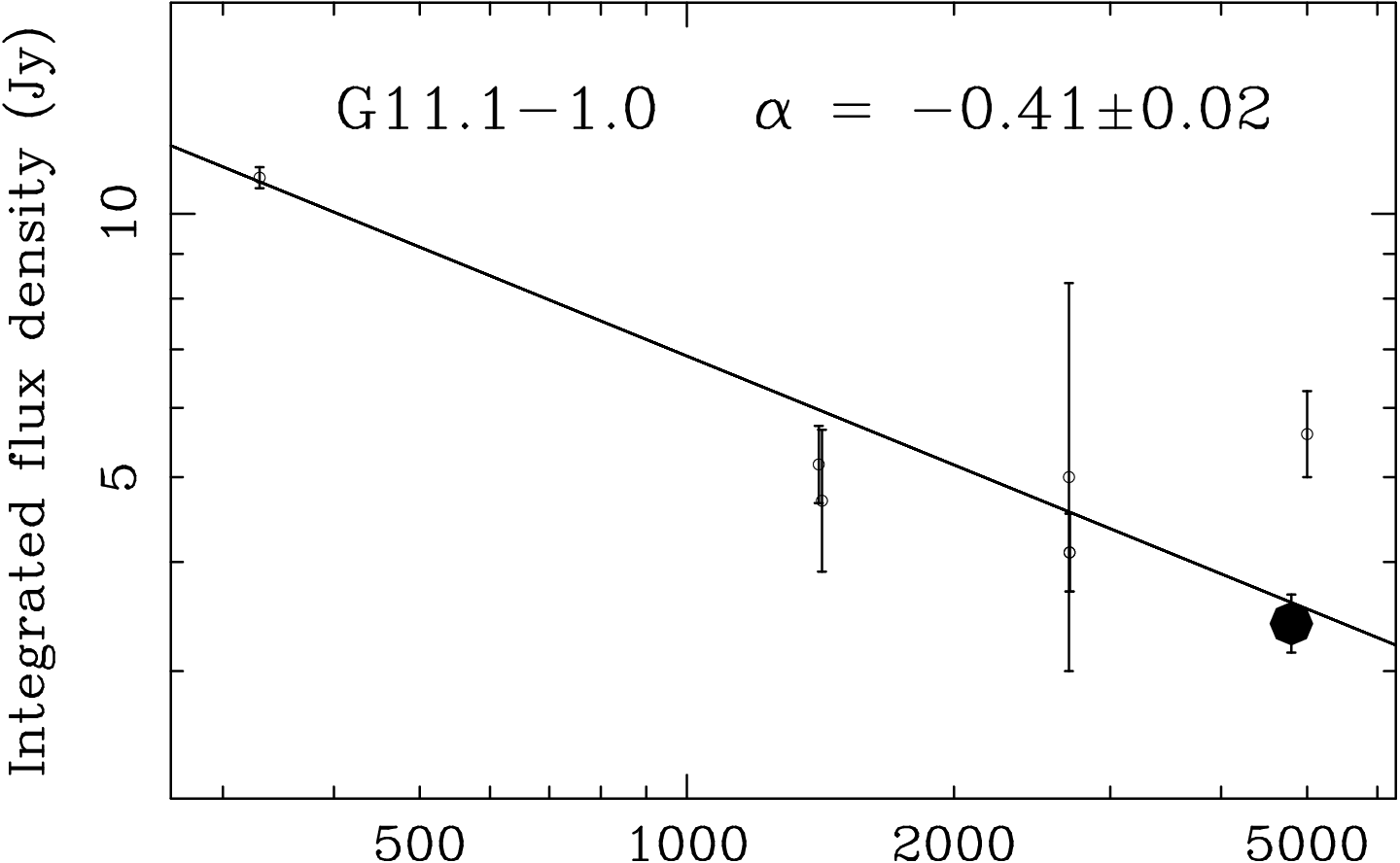}}
\resizebox{0.31\textwidth}{!}{\includegraphics[angle=-90]{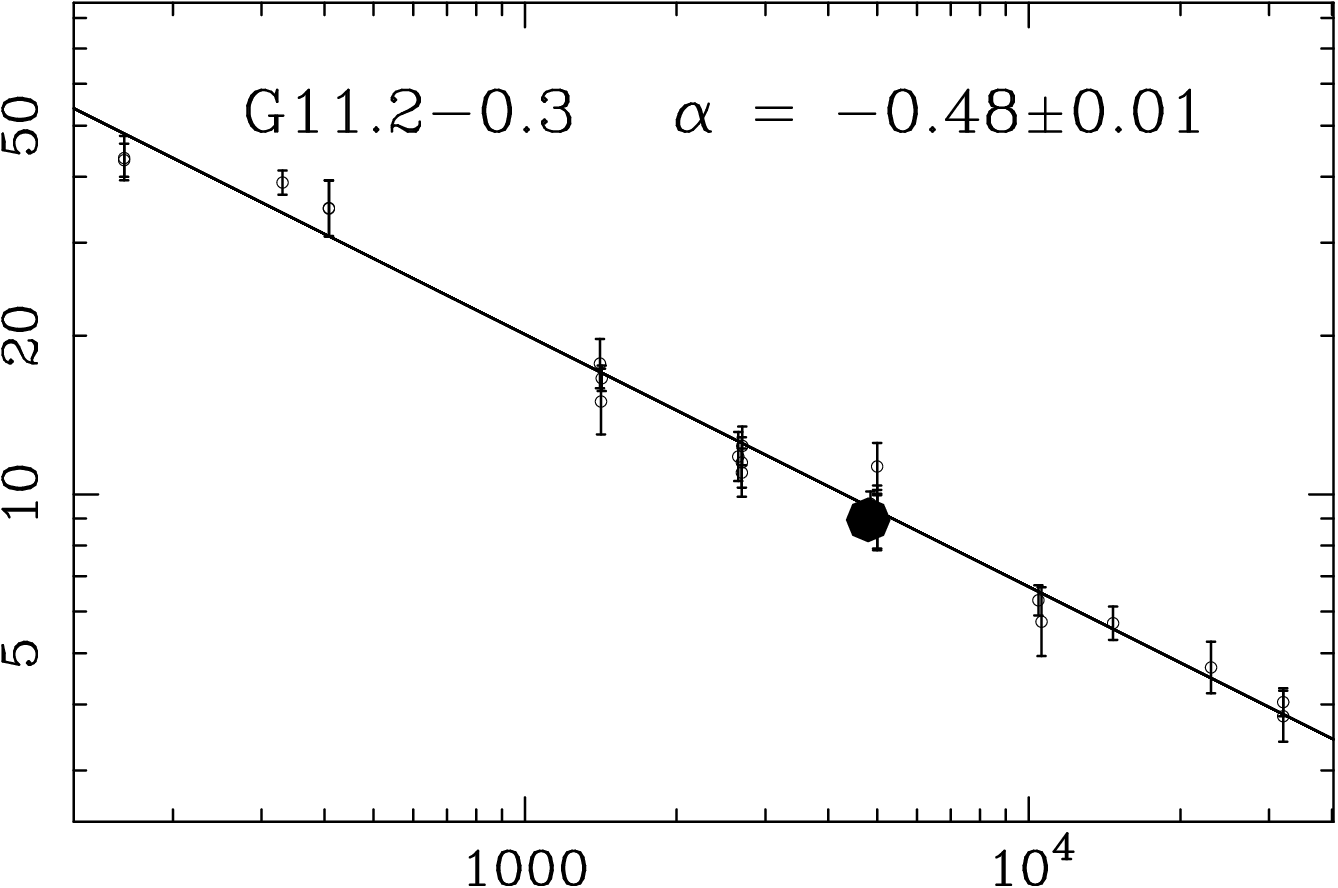}}
\resizebox{0.31\textwidth}{!}{\includegraphics[angle=-90]{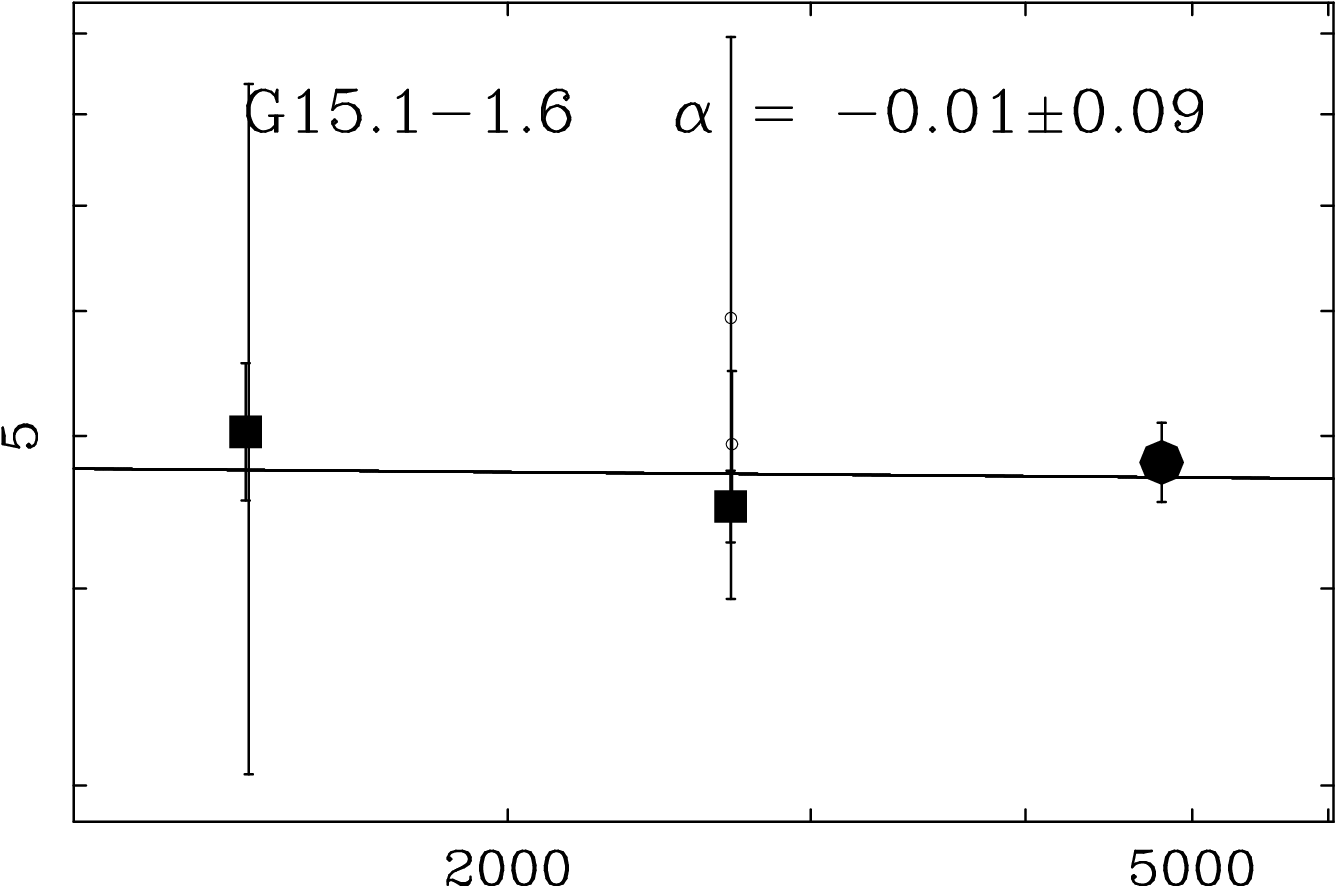}}
\\[1mm]
\resizebox{0.325\textwidth}{!}{\includegraphics[angle=-90]{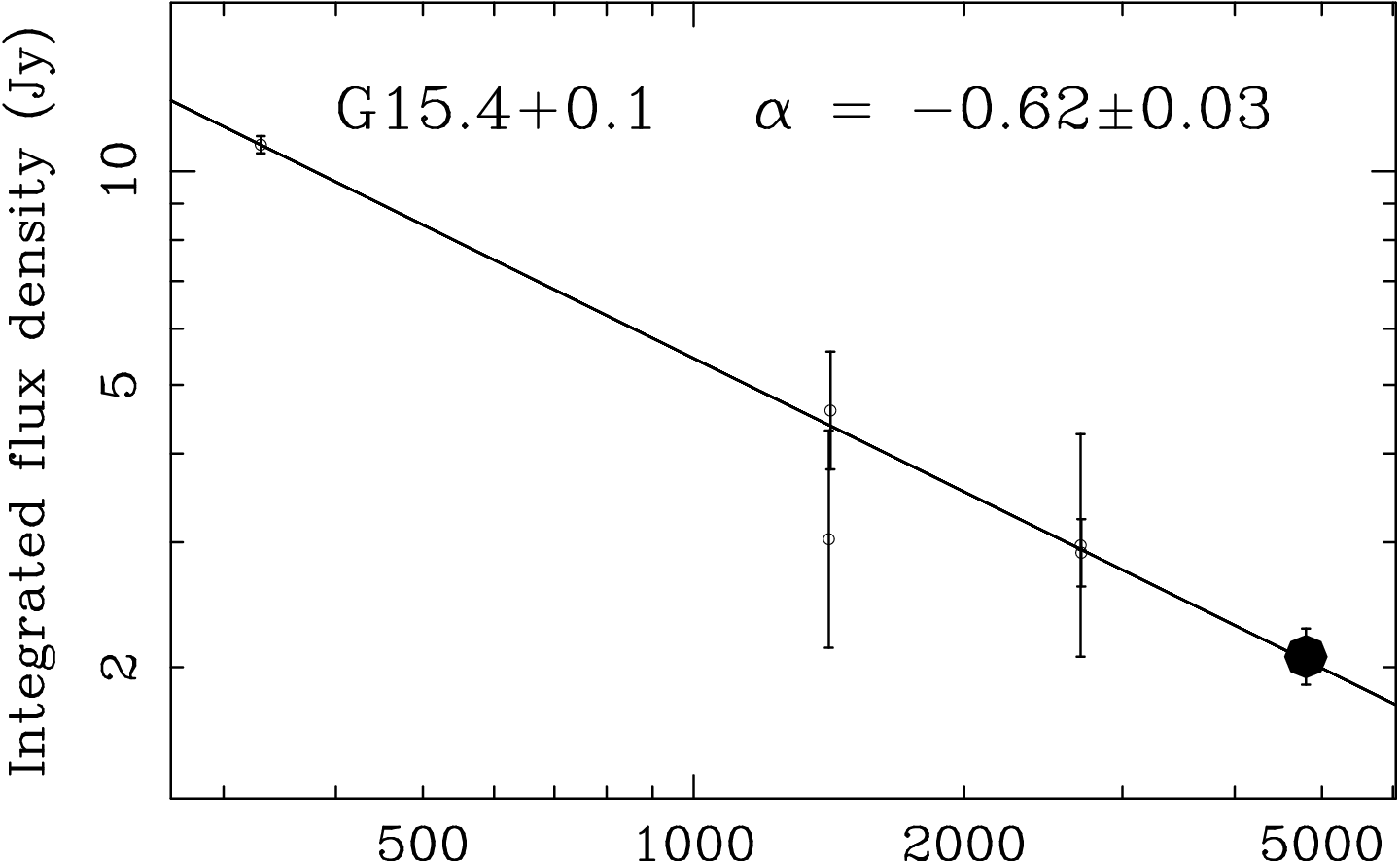}}
\resizebox{0.31\textwidth}{!}{\includegraphics[angle=-90]{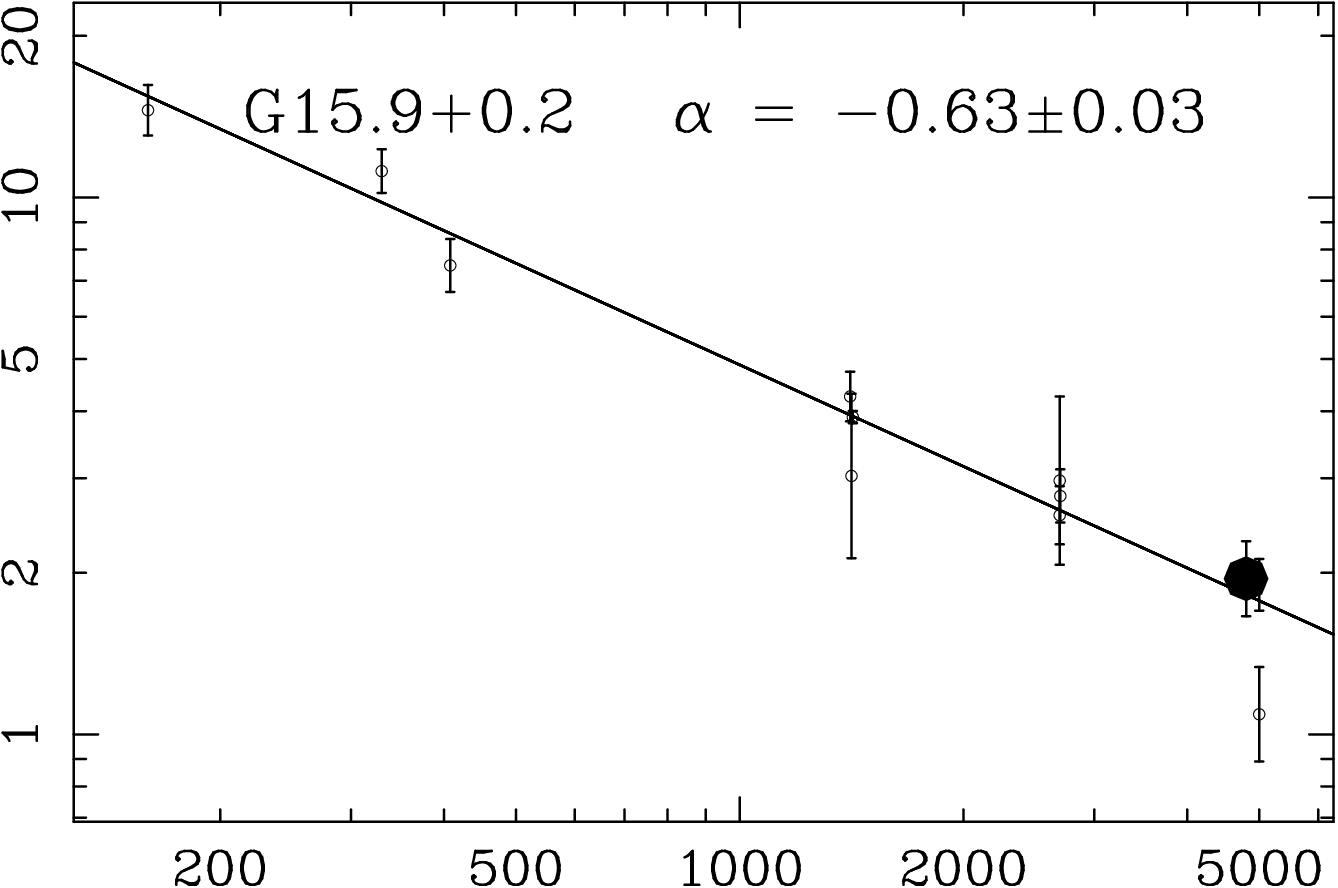}}
\resizebox{0.31\textwidth}{!}{\includegraphics[angle=-90]{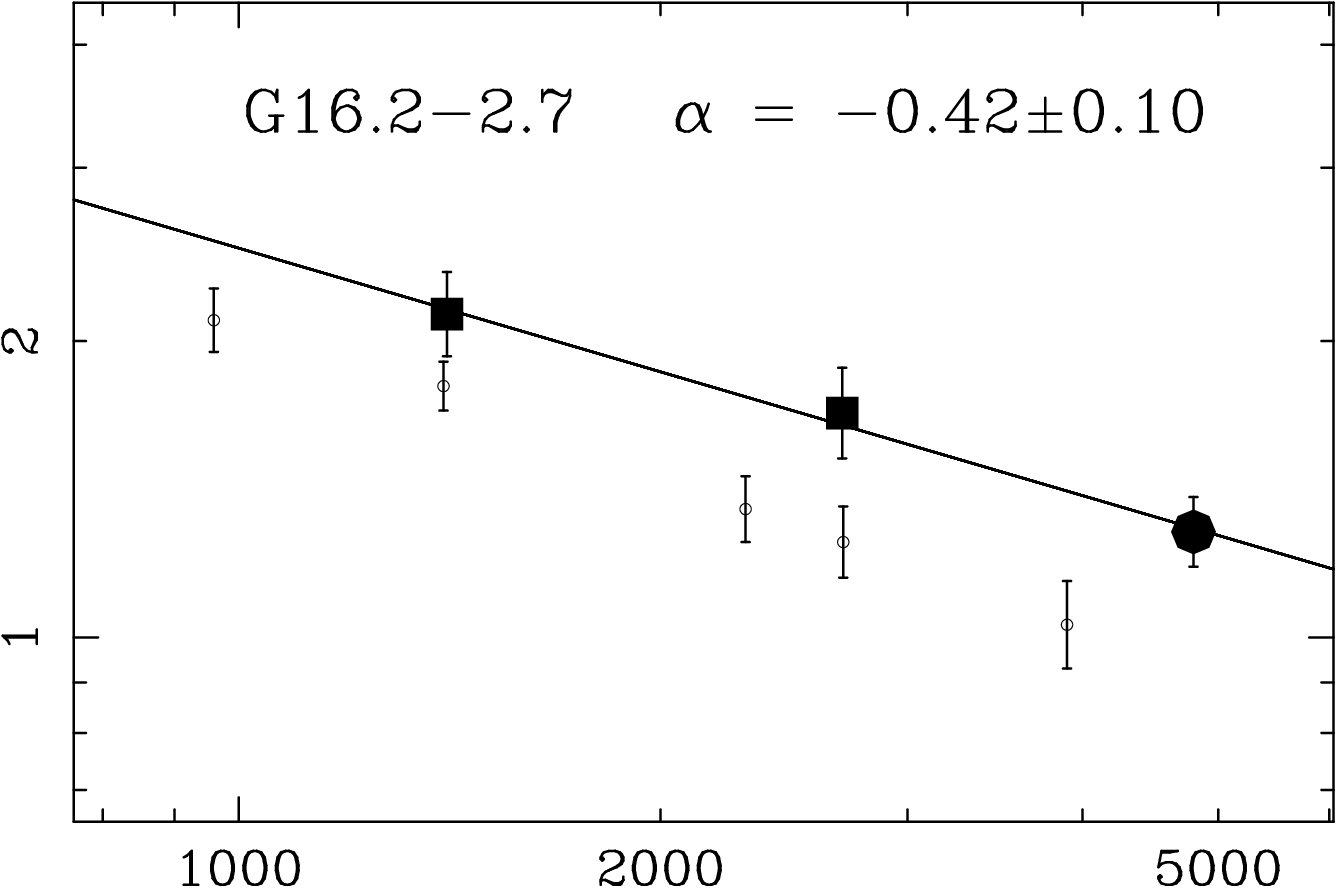}}
\\[1mm]
\resizebox{0.325\textwidth}{!}{\includegraphics[angle=-90]{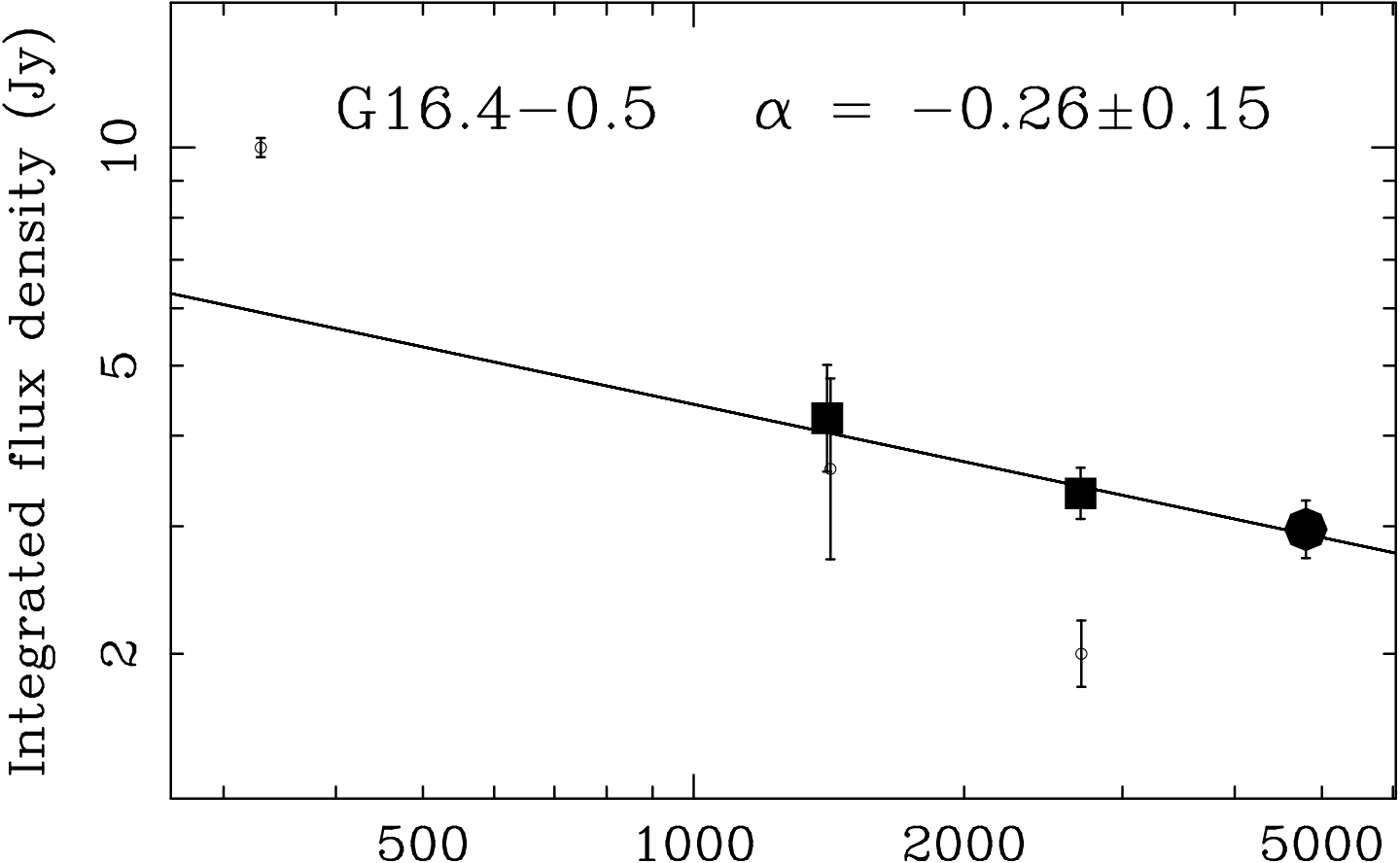}}
\resizebox{0.31\textwidth}{!}{\includegraphics[angle=-90]{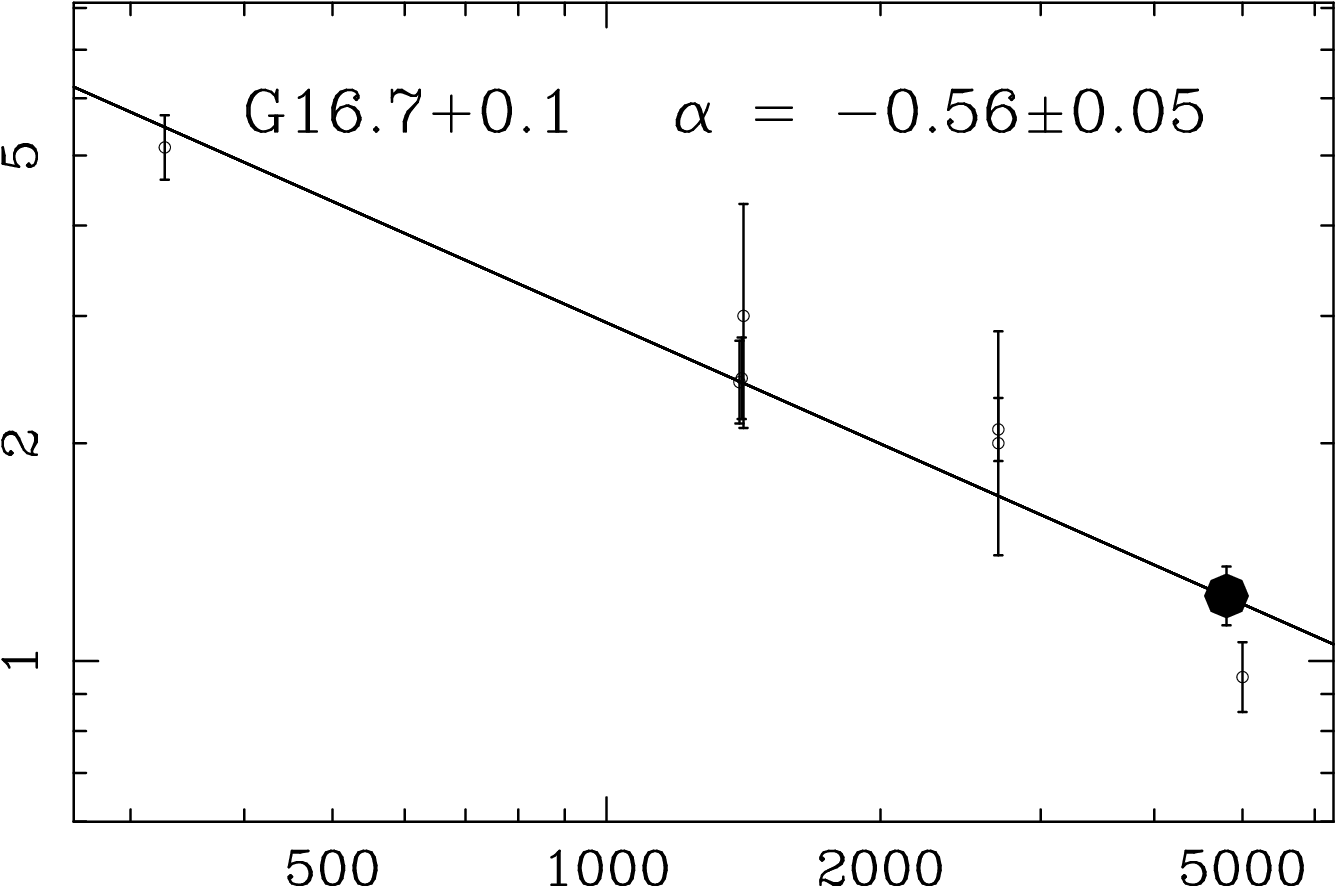}}
\resizebox{0.31\textwidth}{!}{\includegraphics[angle=-90]{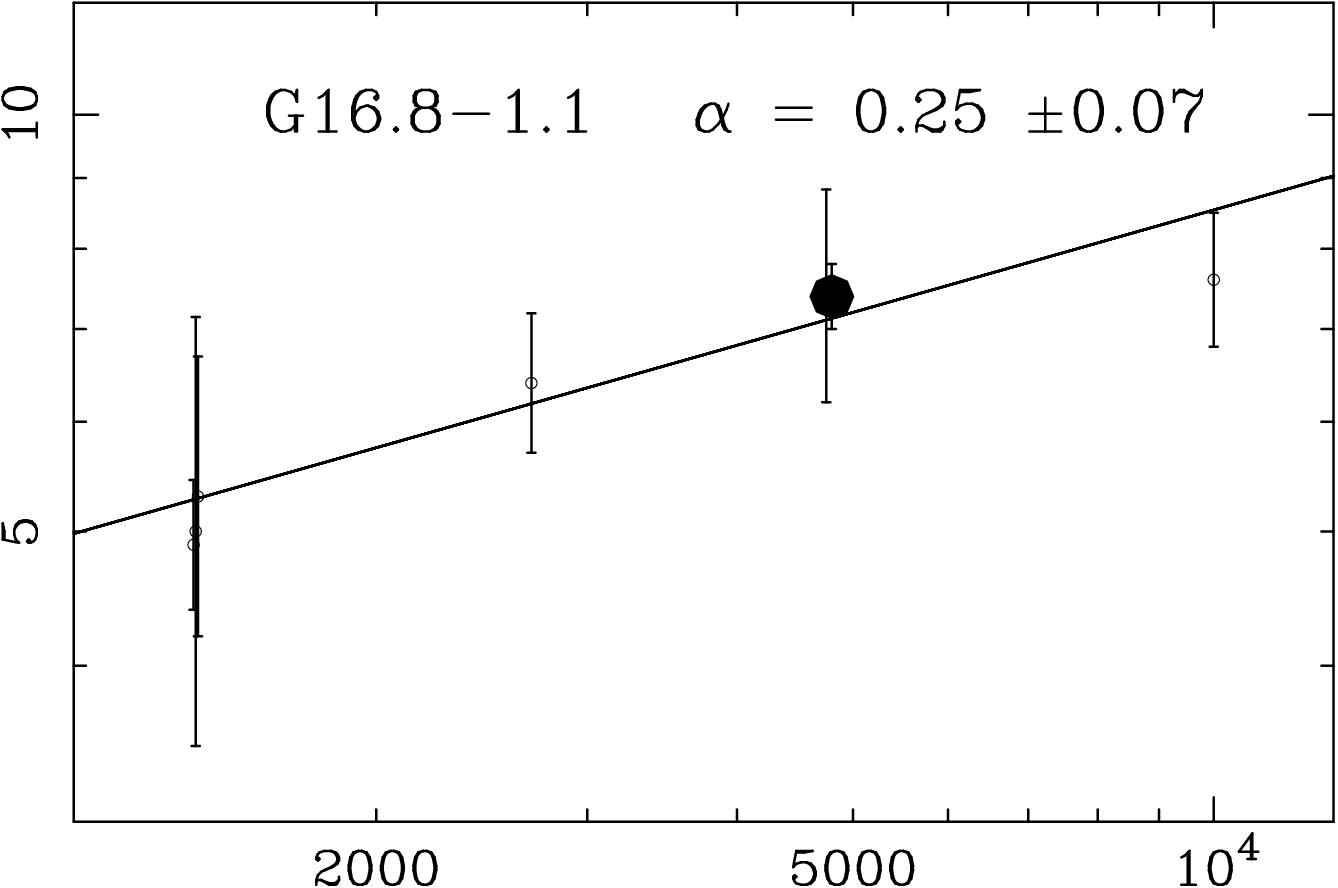}}
\\[1mm]
\resizebox{0.325\textwidth}{!}{\includegraphics[angle=-90]{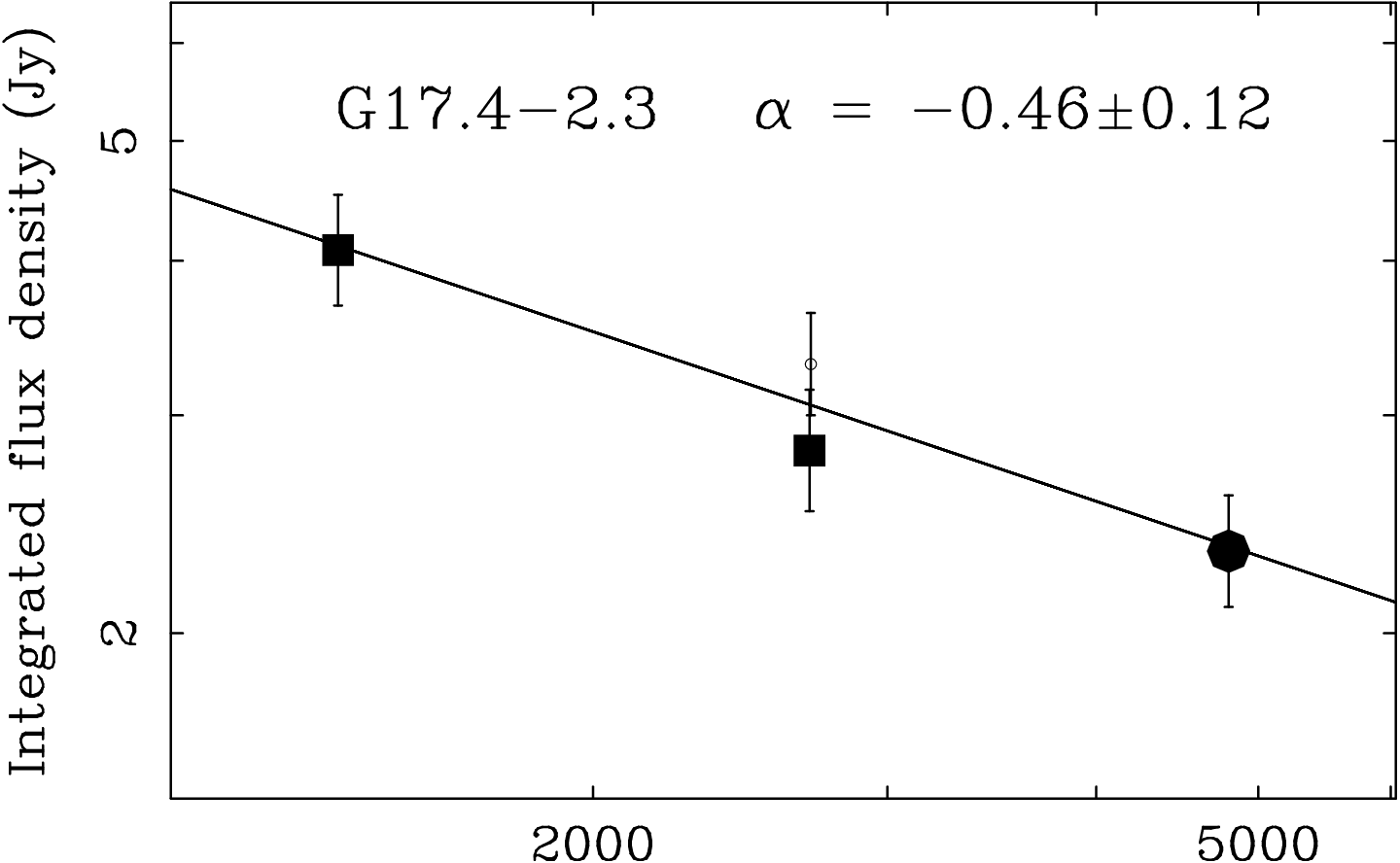}}
\resizebox{0.31\textwidth}{!}{\includegraphics[angle=-90]{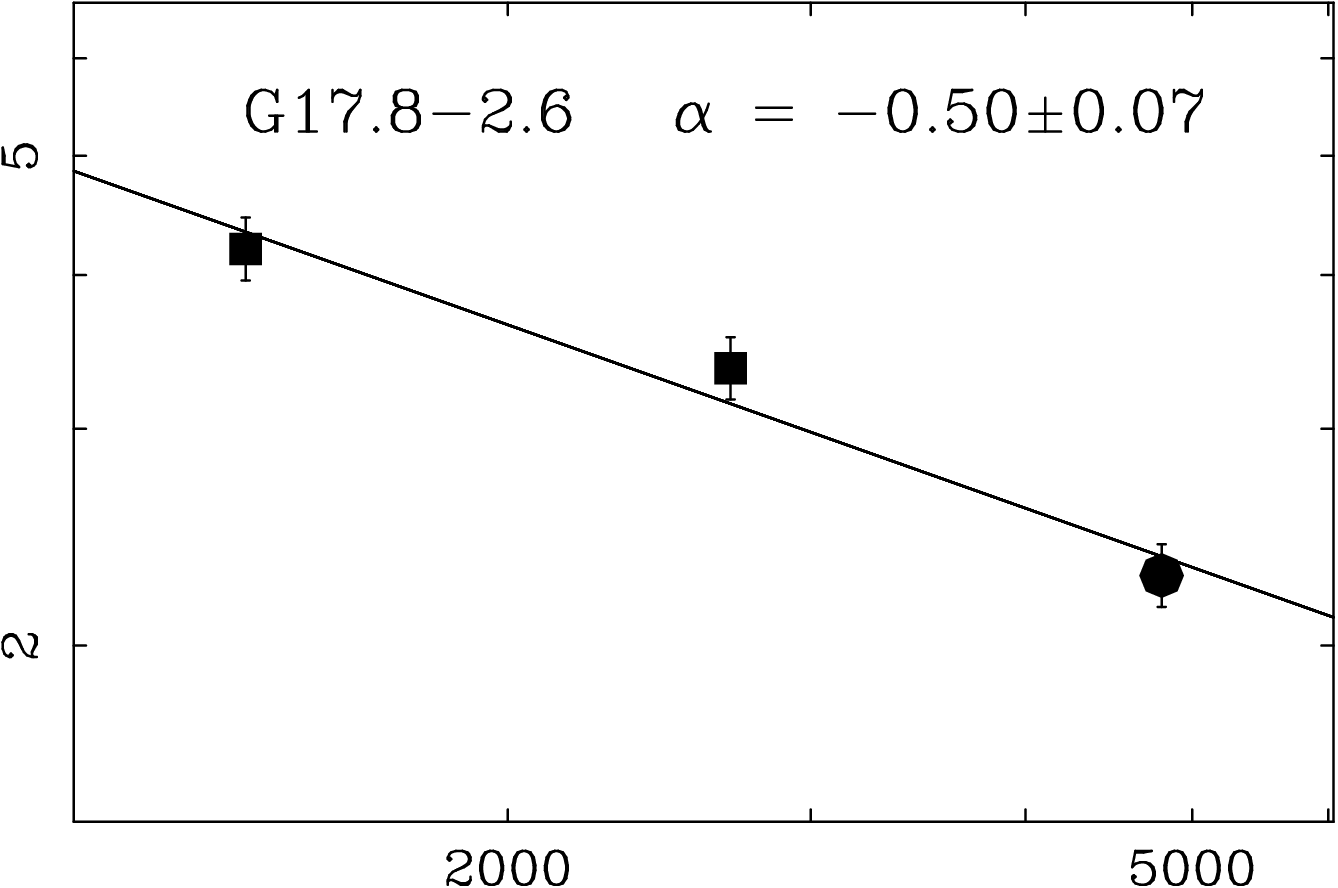}}
\resizebox{0.31\textwidth}{!}{\includegraphics[angle=-90]{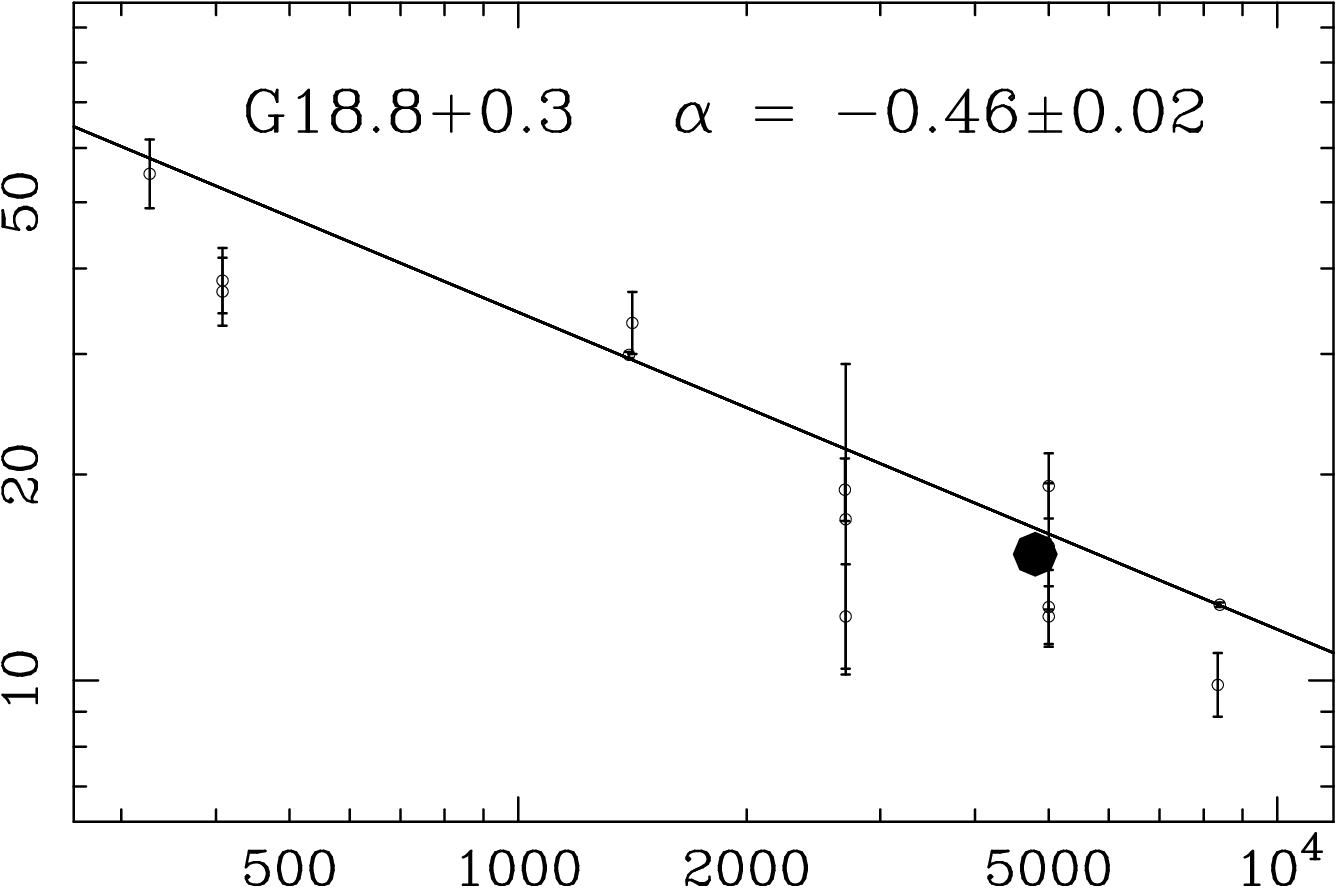}}
\\[1mm]
\resizebox{0.325\textwidth}{!}{\includegraphics[angle=-90]{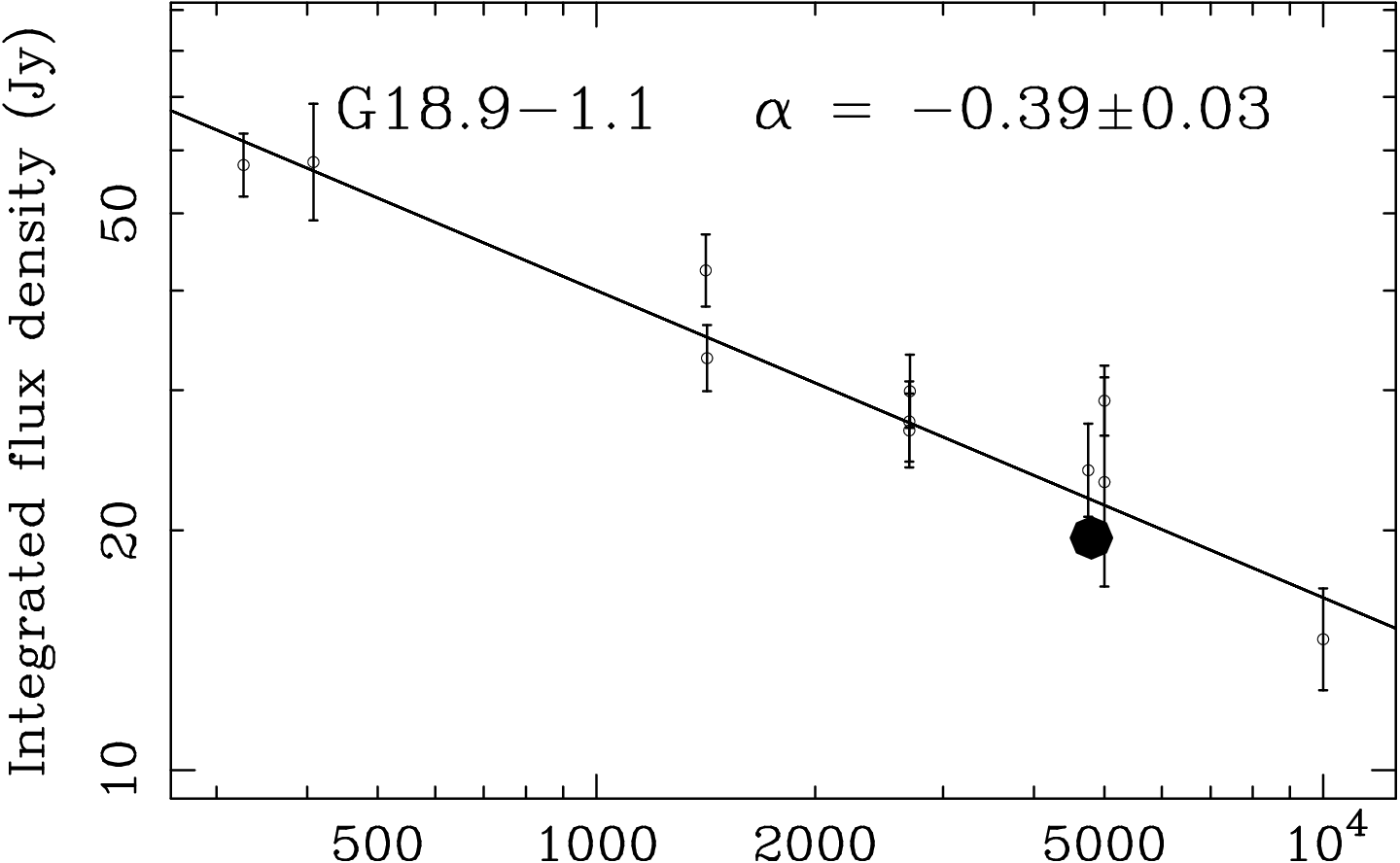}}
\resizebox{0.31\textwidth}{!}{\includegraphics[angle=-90]{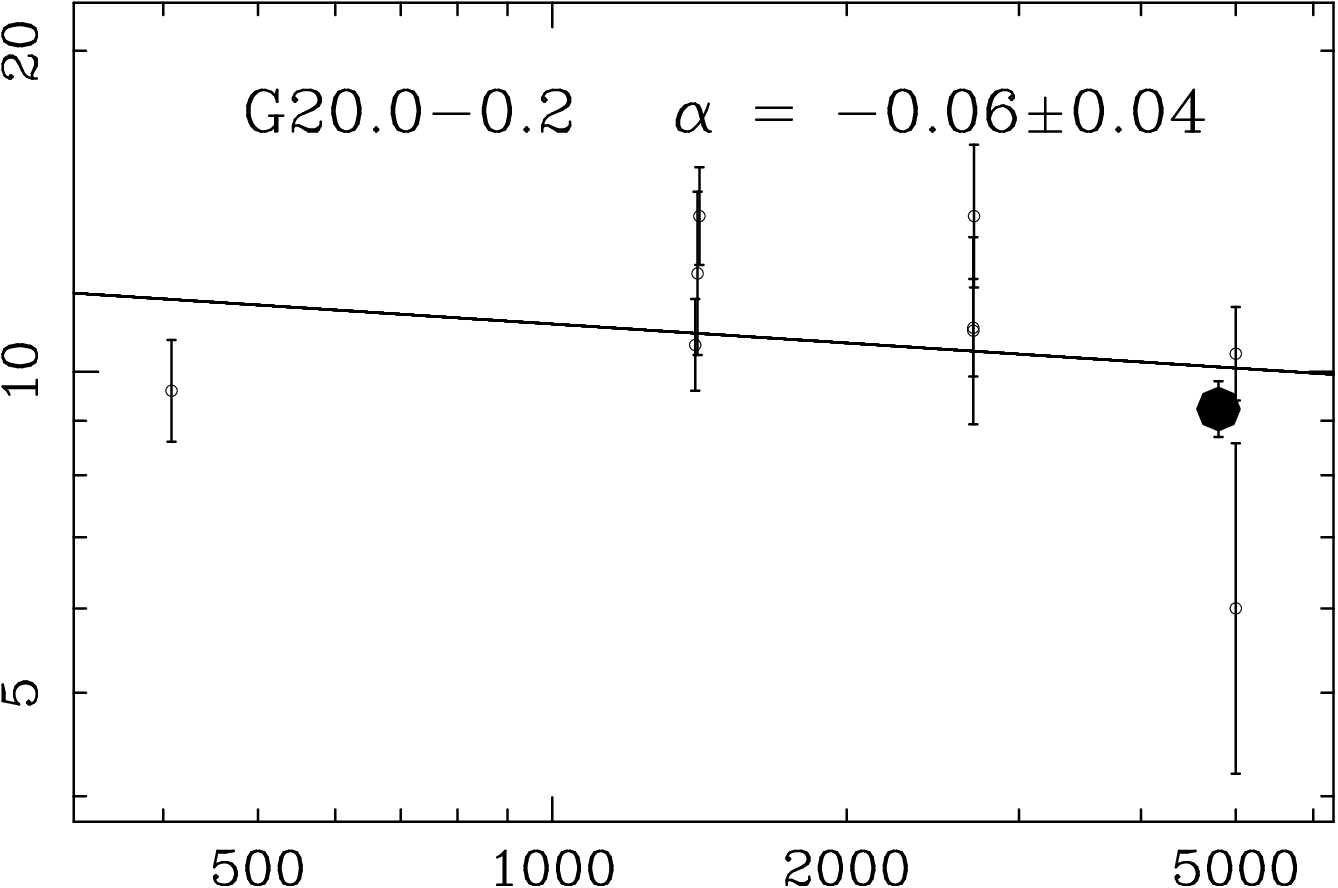}}
\resizebox{0.31\textwidth}{!}{\includegraphics[angle=-90]{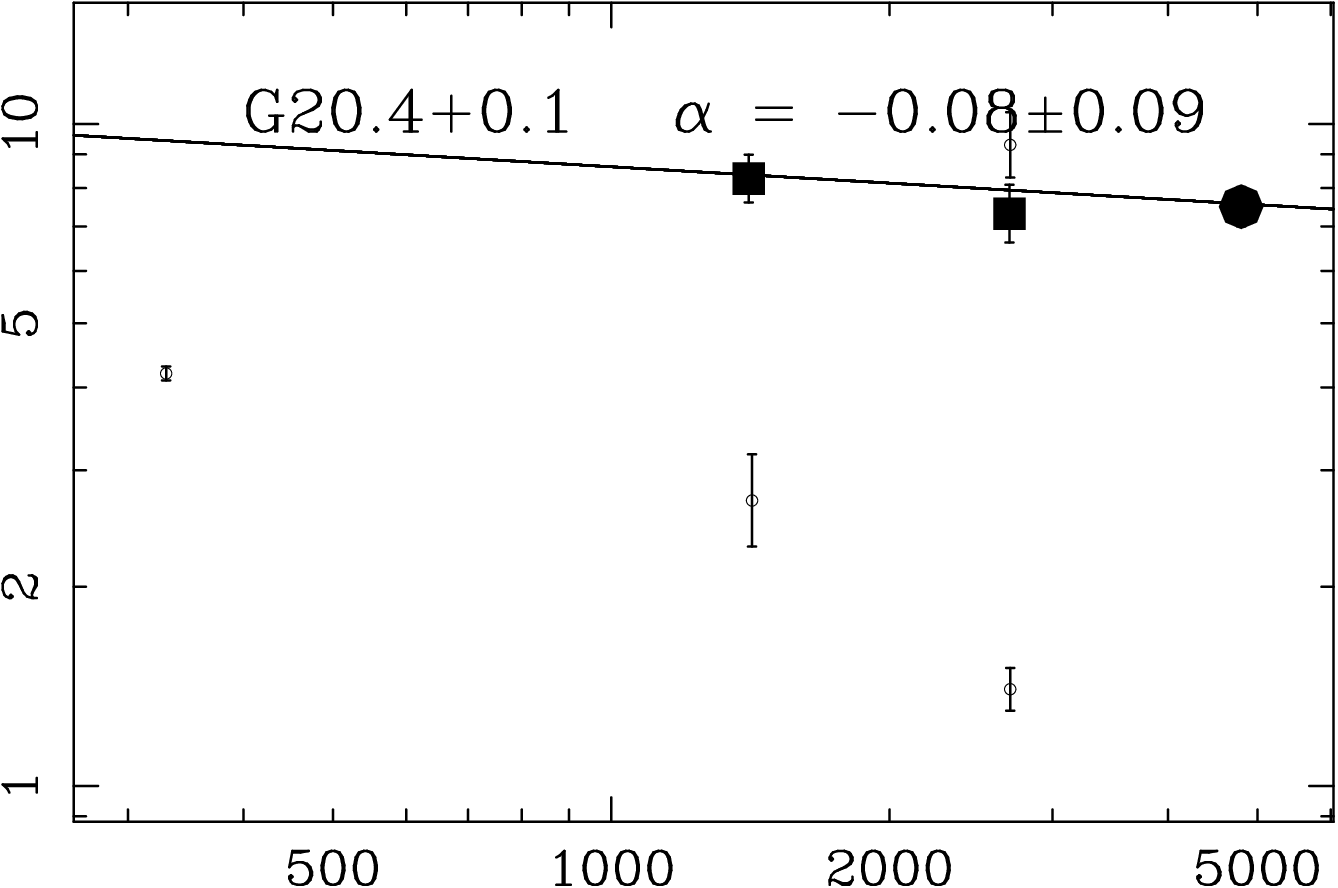}}
\\[1mm]
\resizebox{0.325\textwidth}{!}{\includegraphics[angle=-90]{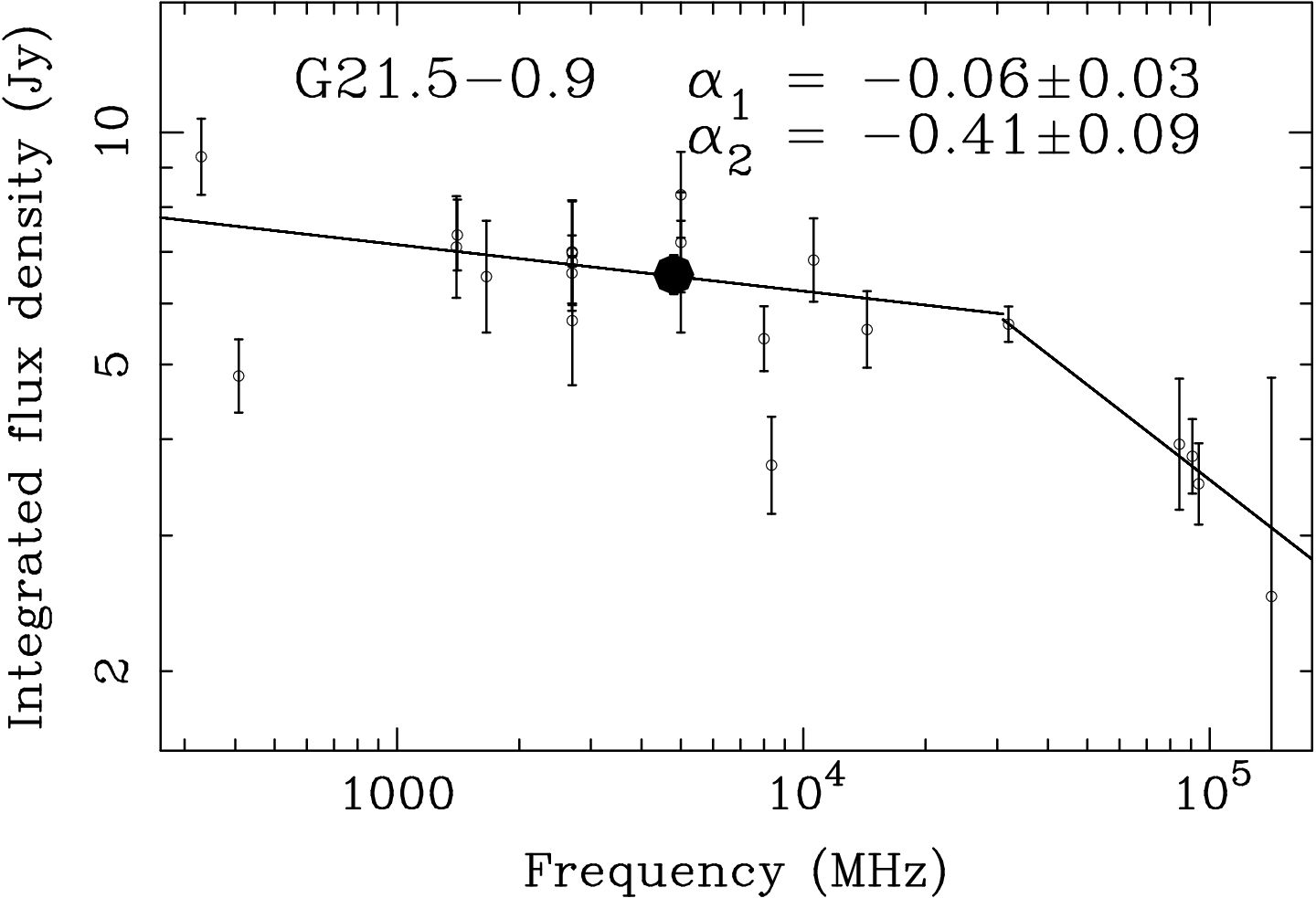}}
\resizebox{0.31\textwidth}{!}{\includegraphics[angle=-90]{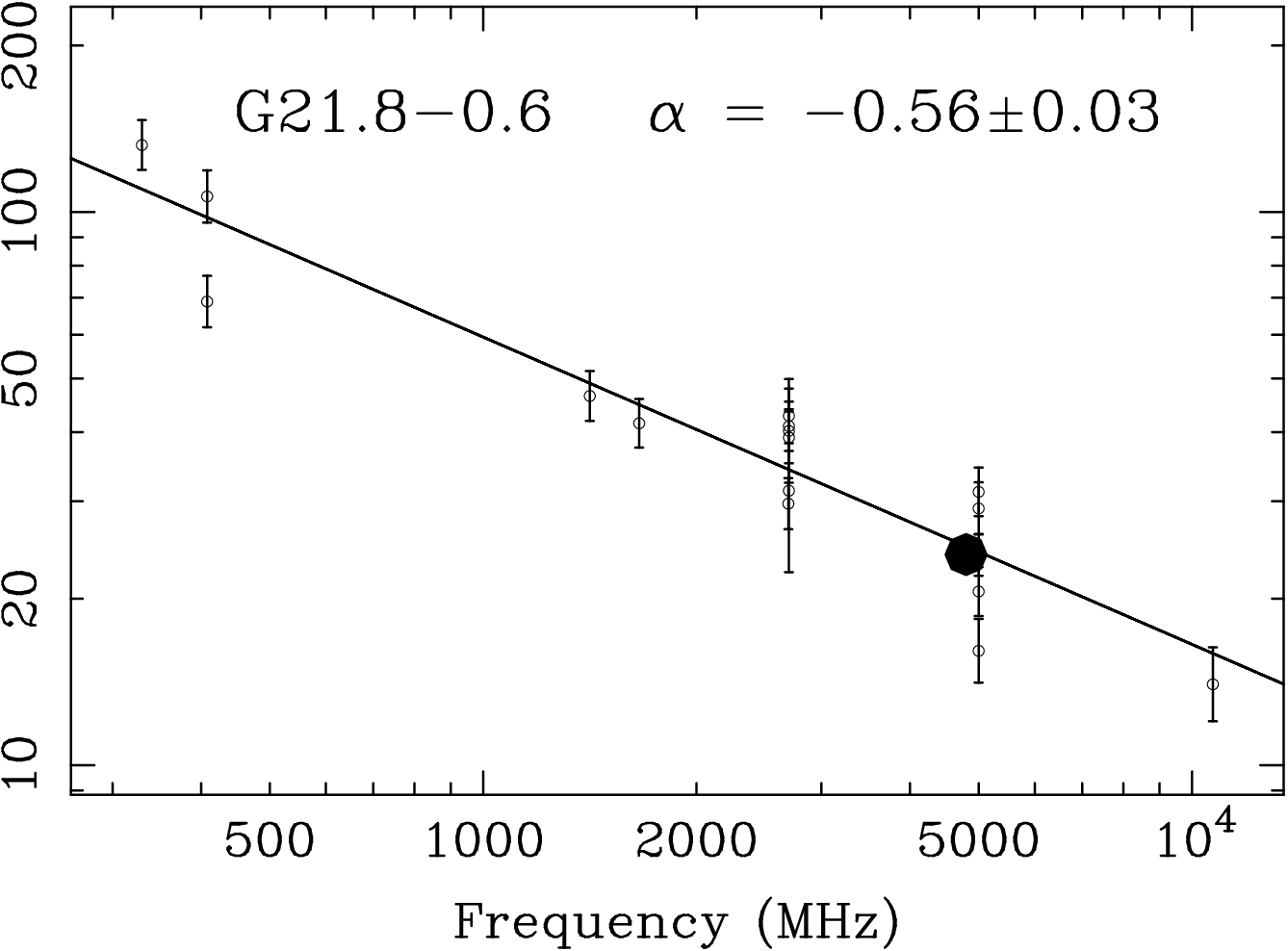}}
\resizebox{0.31\textwidth}{!}{\includegraphics[angle=-90]{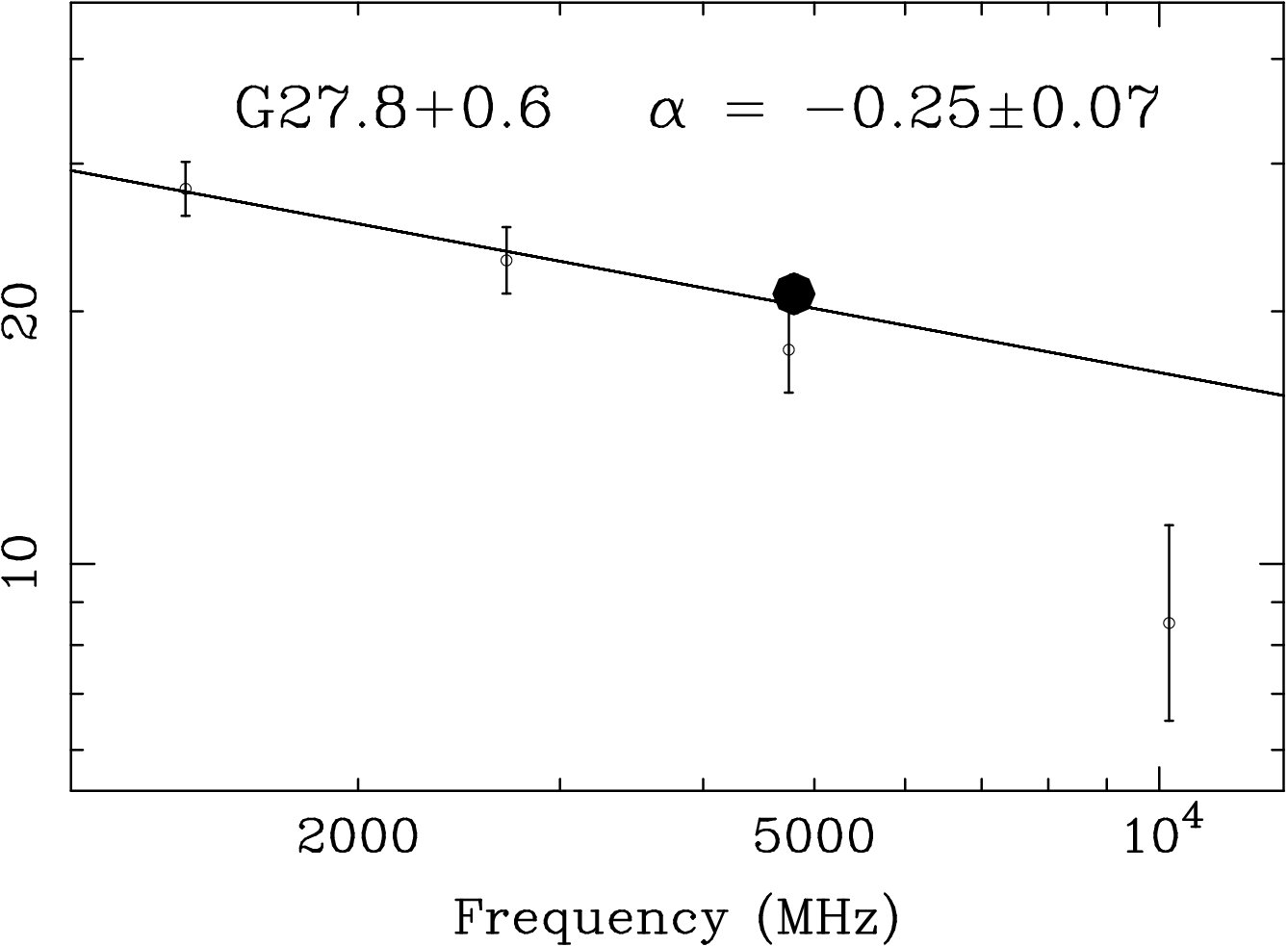}}
\caption{Spectra for 50 SNRs. The present $\lambda$6~cm flux densities are 
indicated by black dots, while the flux densities we derived from the  
$\lambda$11~cm and $\lambda$21~cm Effelsberg surveys are marked by dark 
squares. Other measurements were taken from the references listed in 
Table~\ref{snr_s}.}
\label{spectra}
\end{figure*}\addtocounter{figure}{-1}
\begin{figure*}[!htbp]
\centering
\resizebox{0.325\textwidth}{!}{\includegraphics[angle=-90]{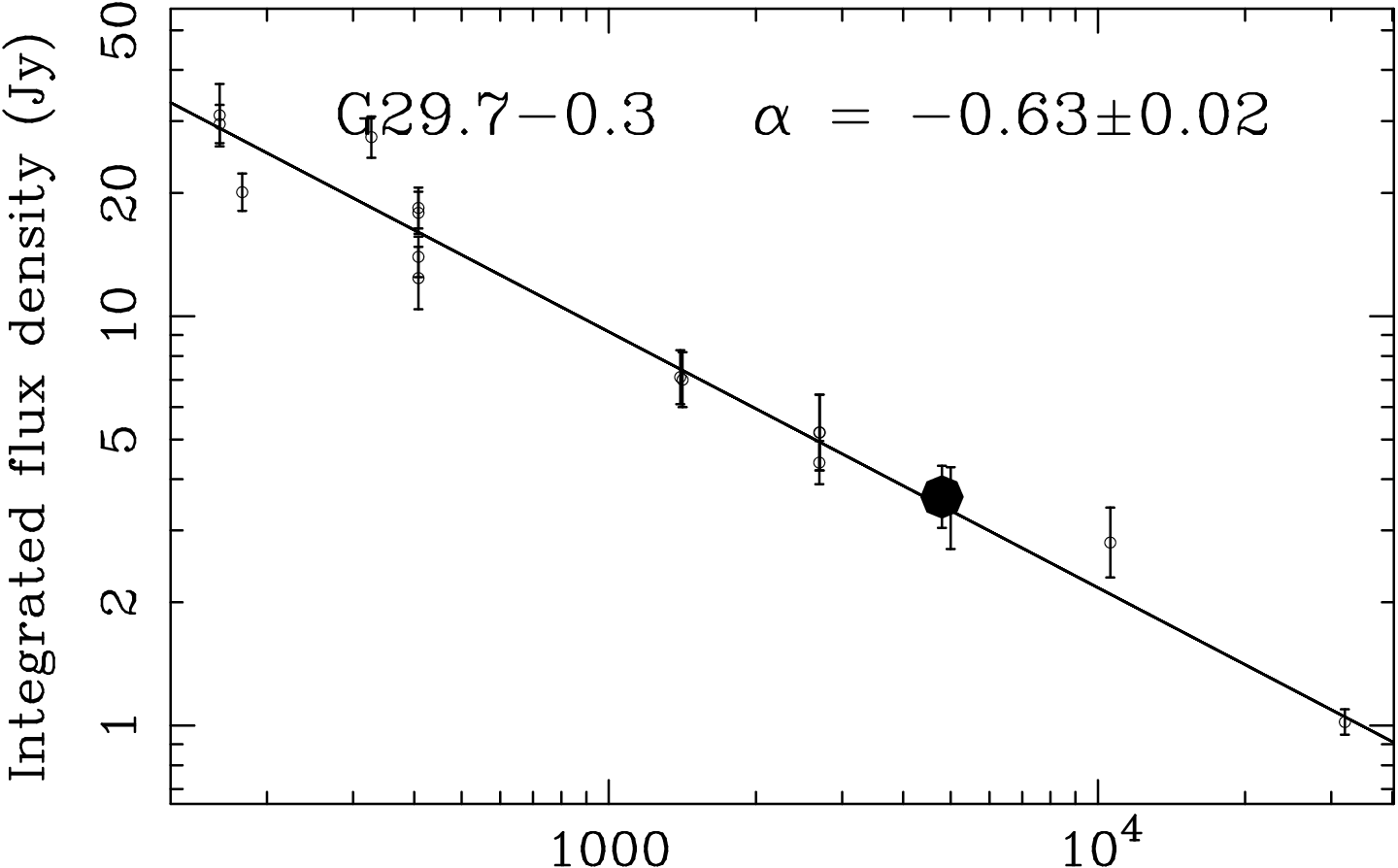}}
\resizebox{0.31\textwidth}{!}{\includegraphics[angle=-90]{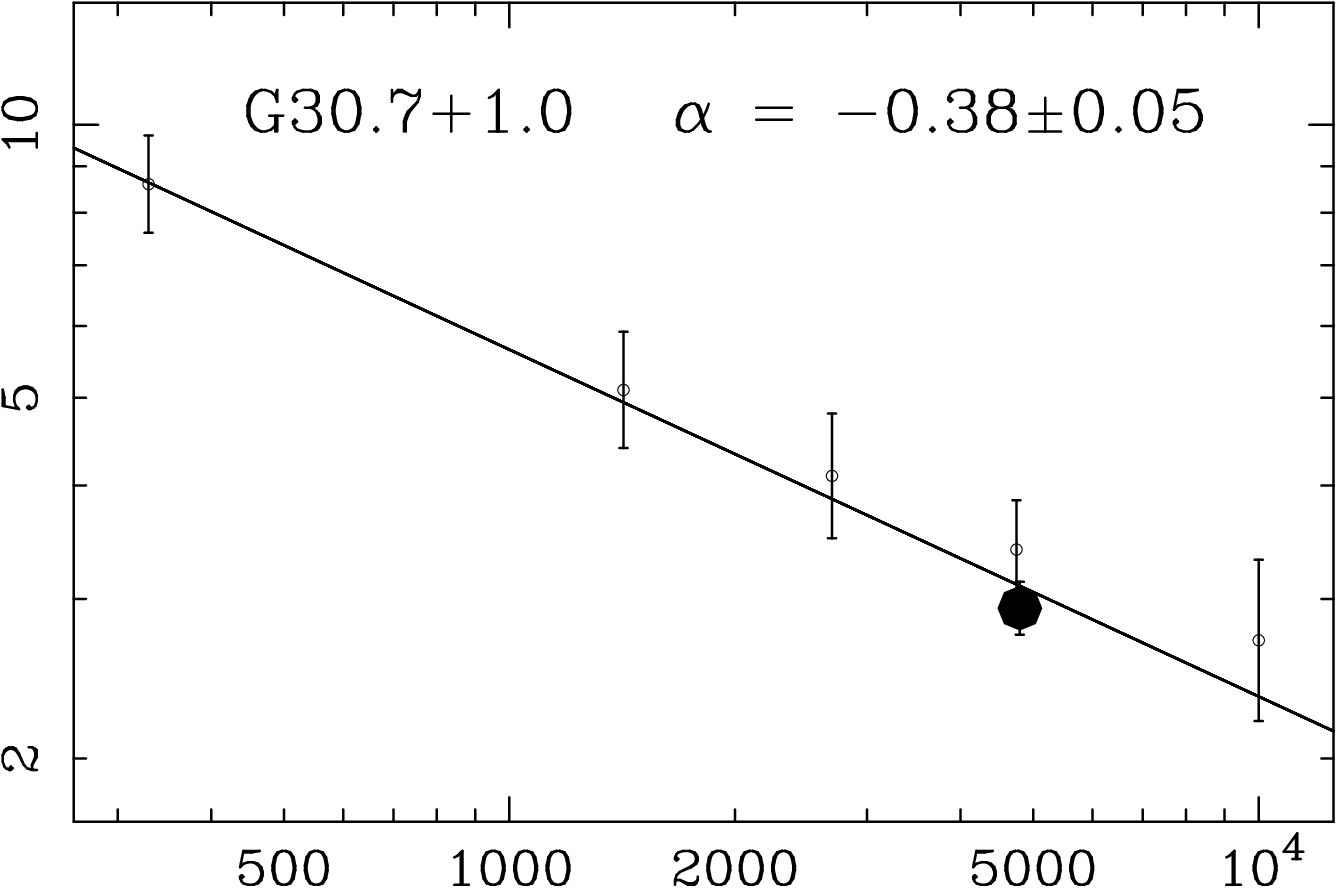}}
\resizebox{0.31\textwidth}{!}{\includegraphics[angle=-90]{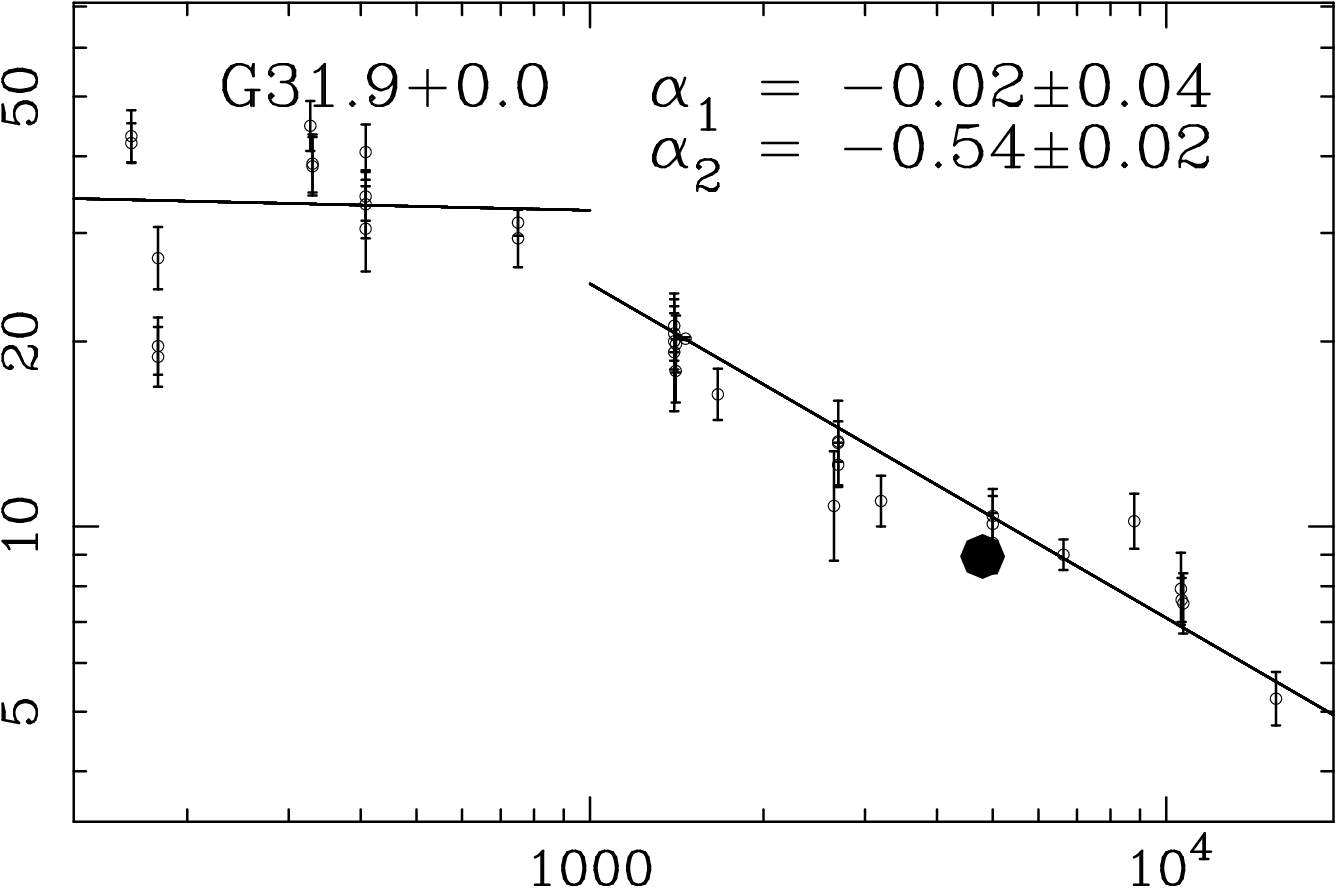}}
\\[1mm]
\resizebox{0.325\textwidth}{!}{\includegraphics[angle=-90]{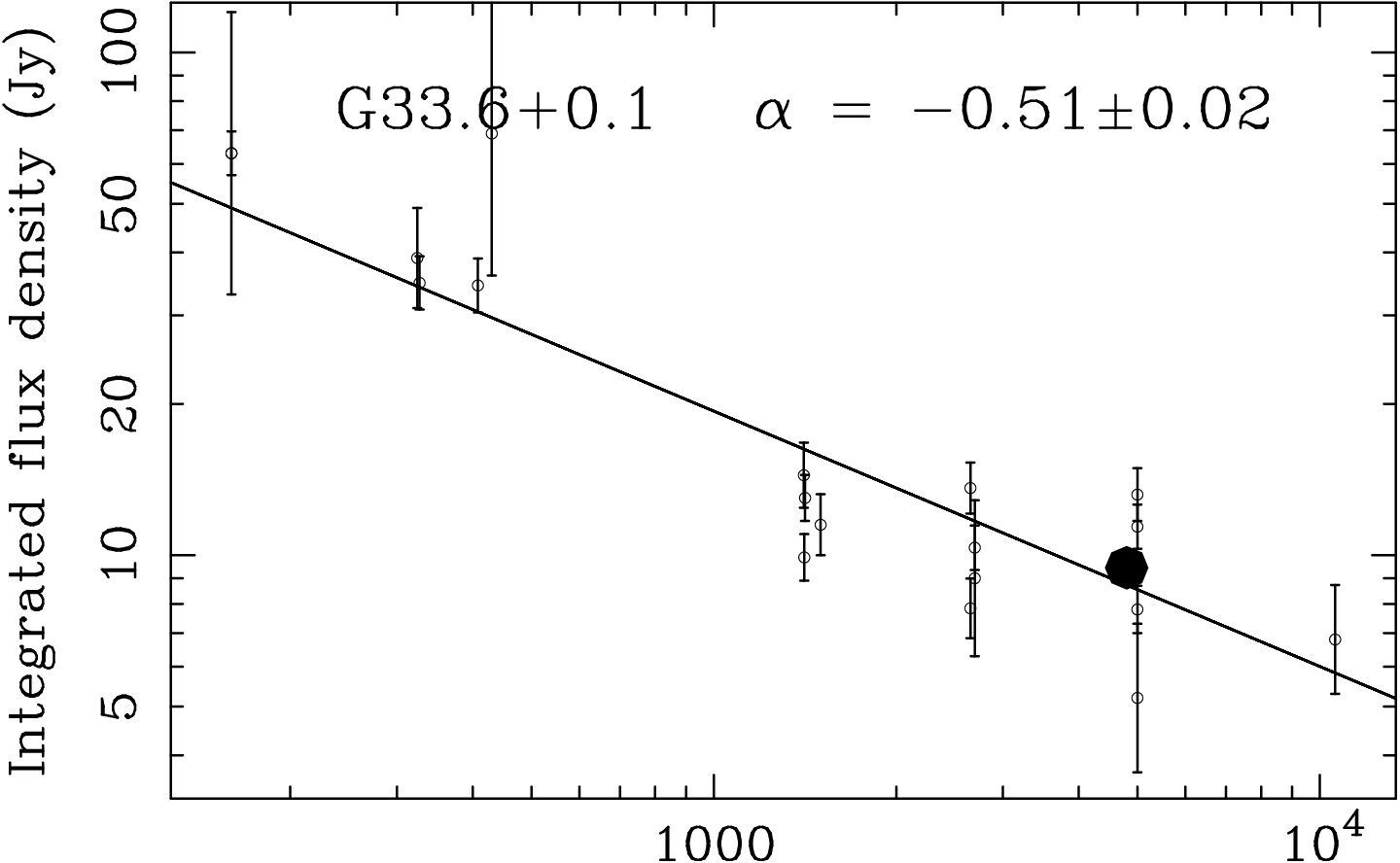}}
\resizebox{0.31\textwidth}{!}{\includegraphics[angle=-90]{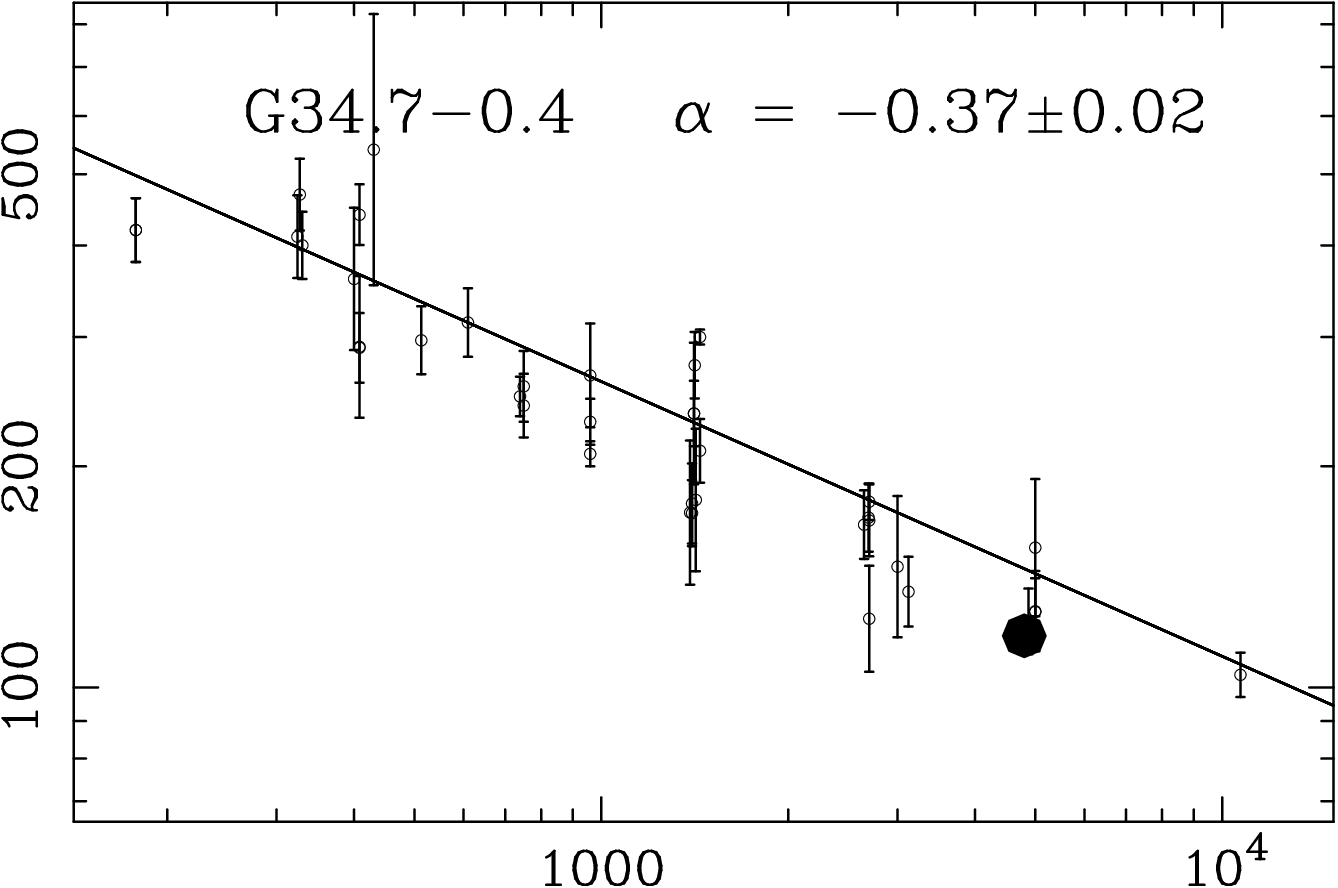}}
\resizebox{0.31\textwidth}{!}{\includegraphics[angle=-90]{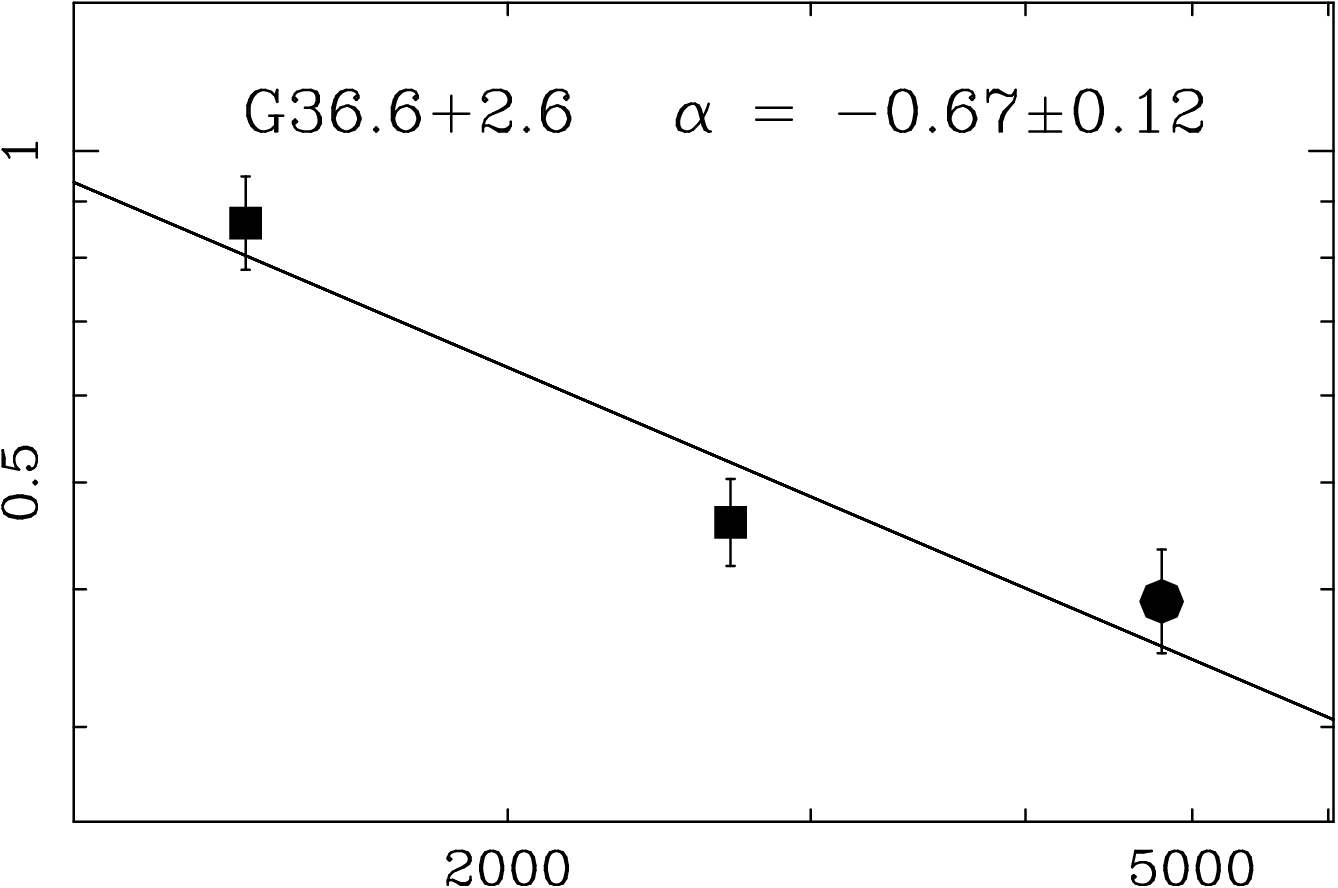}}
\\[1mm]
\resizebox{0.325\textwidth}{!}{\includegraphics[angle=-90]{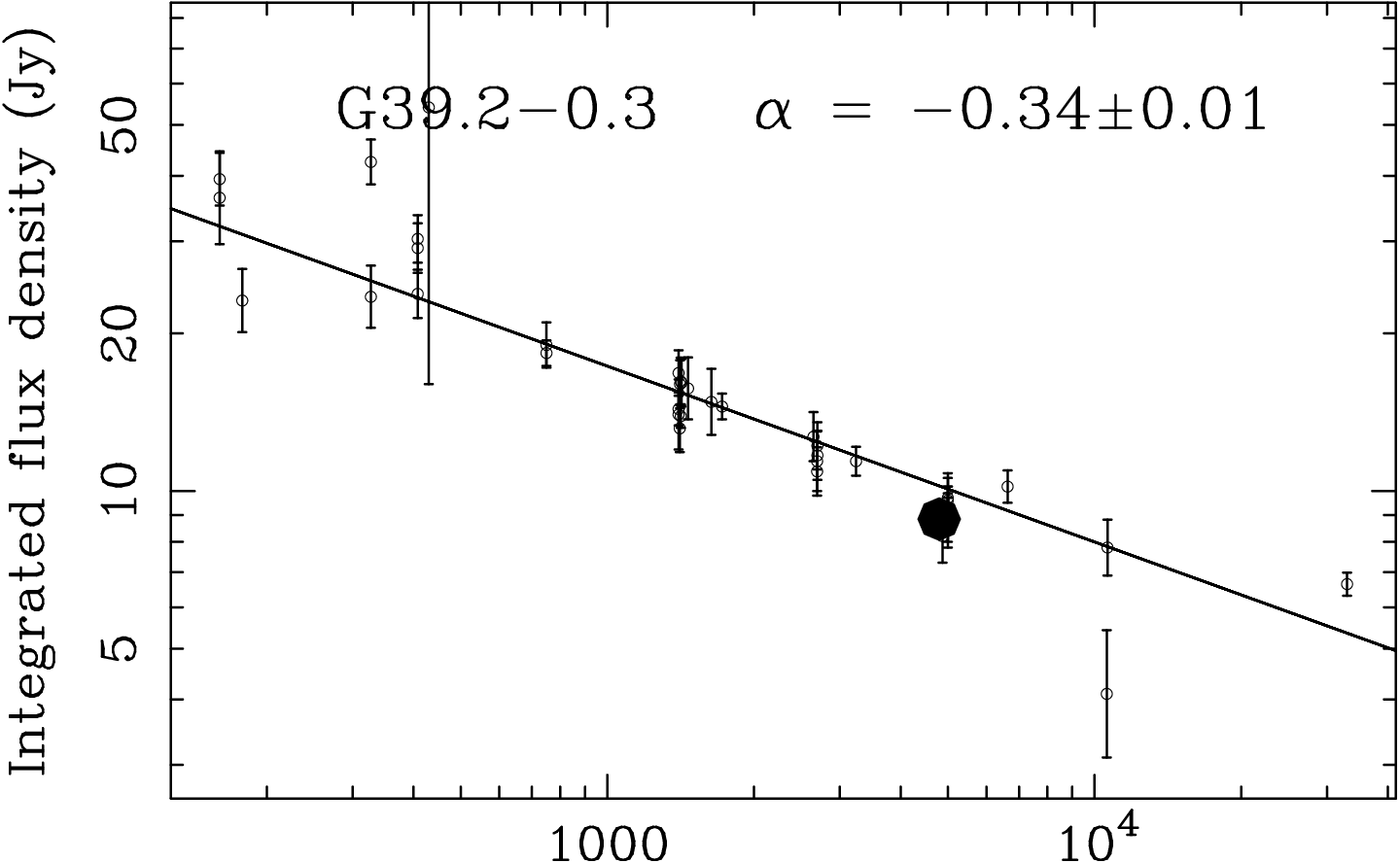}}
\resizebox{0.31\textwidth}{!}{\includegraphics[angle=-90]{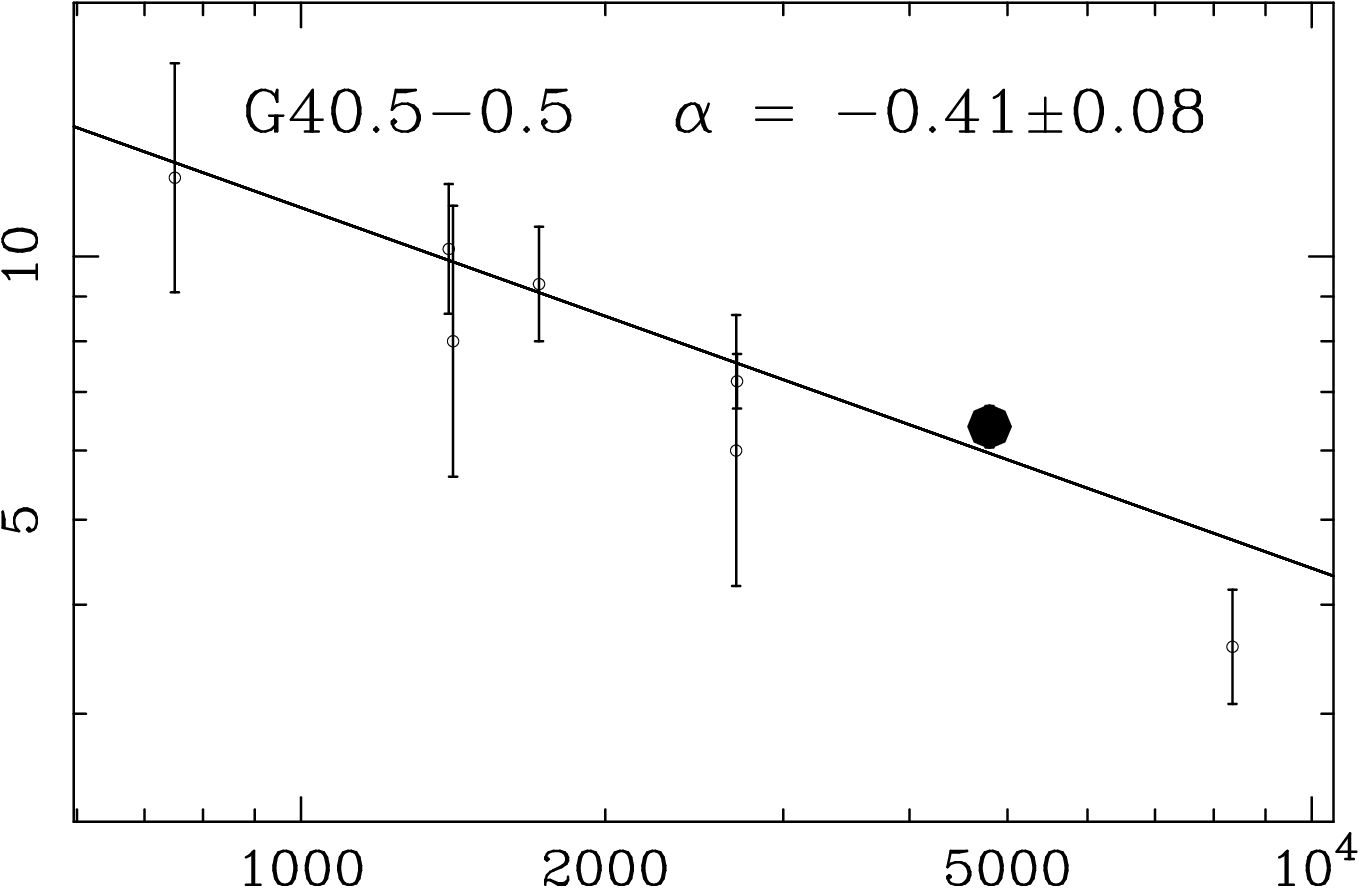}}
\resizebox{0.31\textwidth}{!}{\includegraphics[angle=-90]{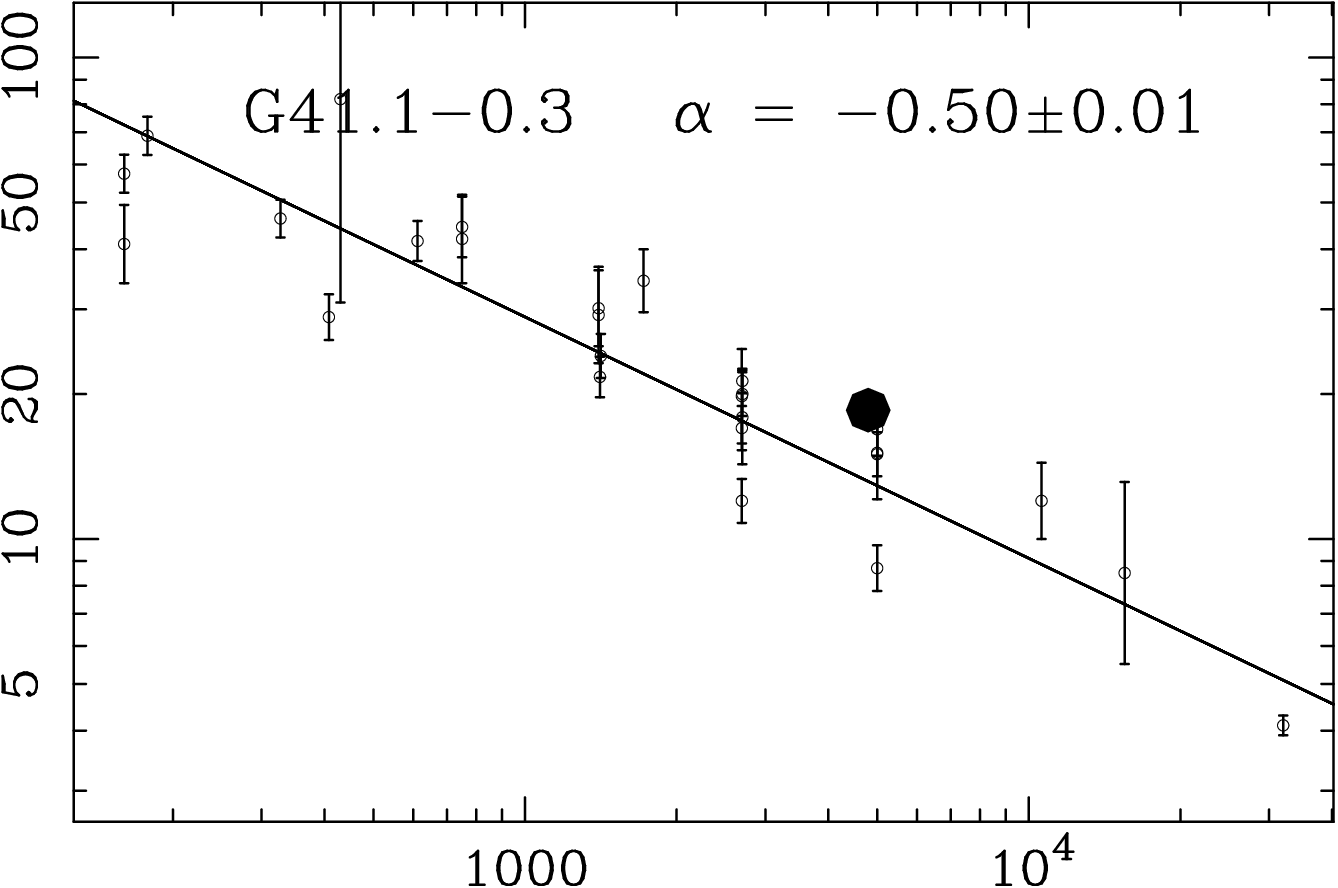}}
\\[1mm]
\resizebox{0.325\textwidth}{!}{\includegraphics[angle=-90]{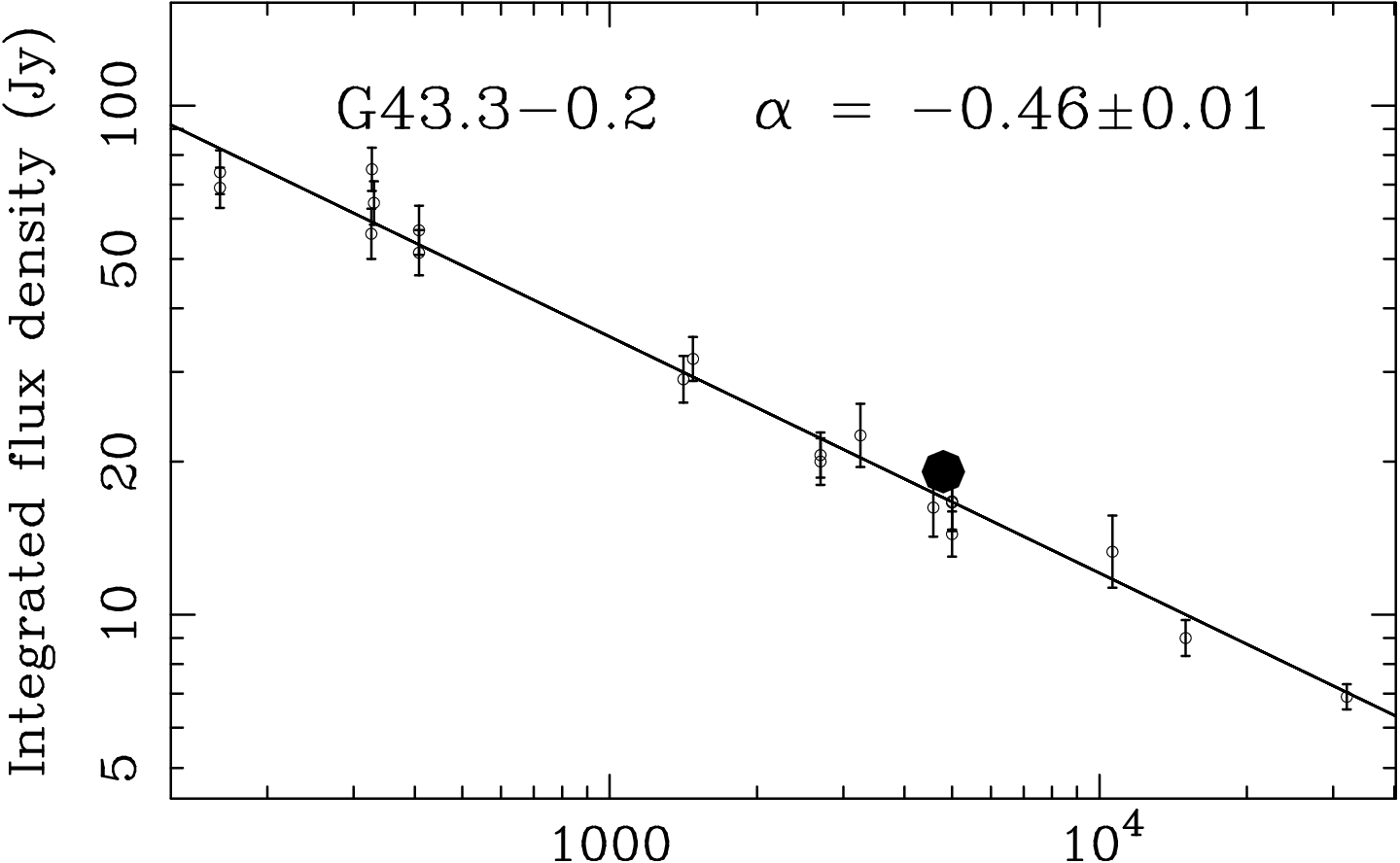}}
\resizebox{0.31\textwidth}{!}{\includegraphics[angle=-90]{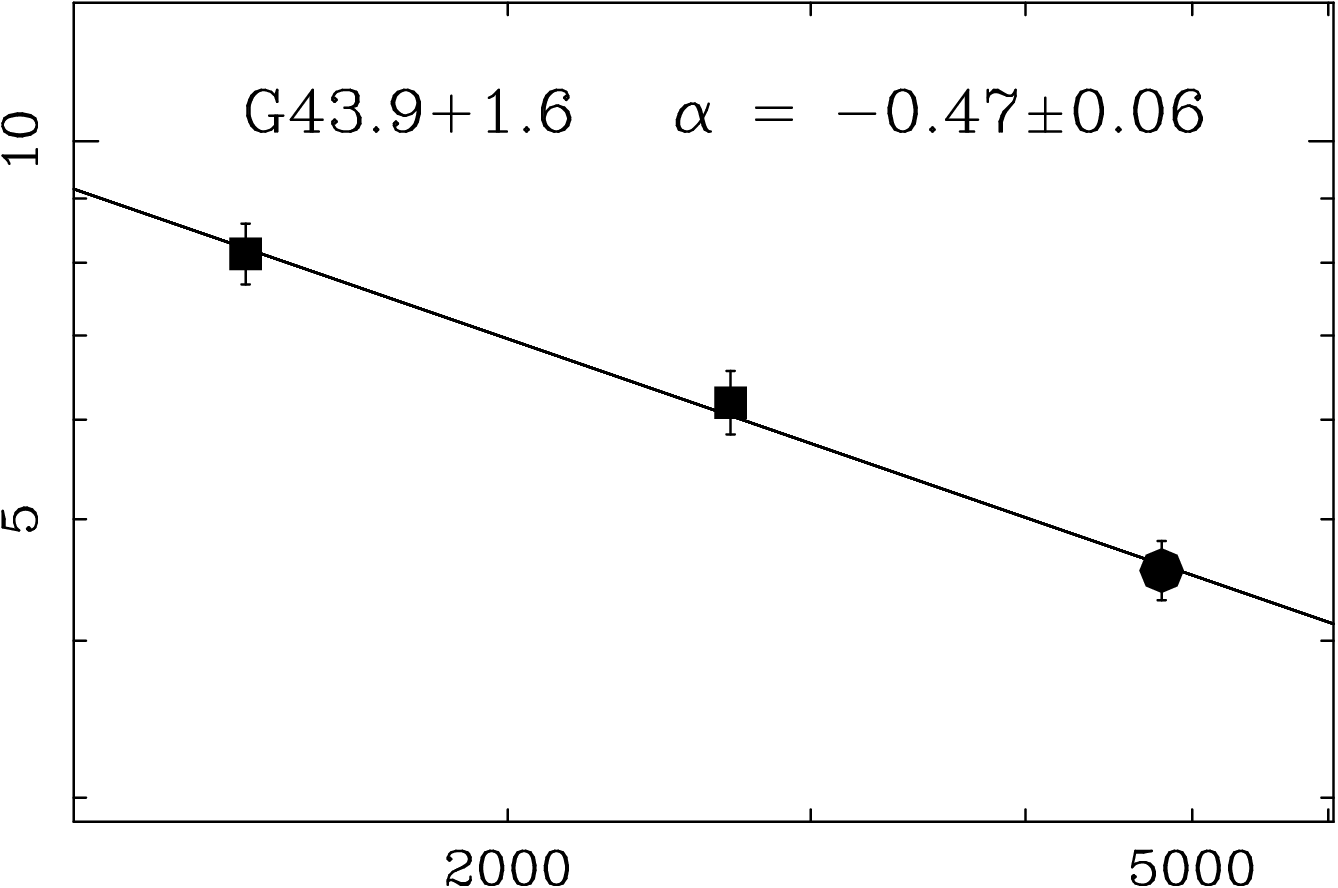}}
\resizebox{0.31\textwidth}{!}{\includegraphics[angle=-90]{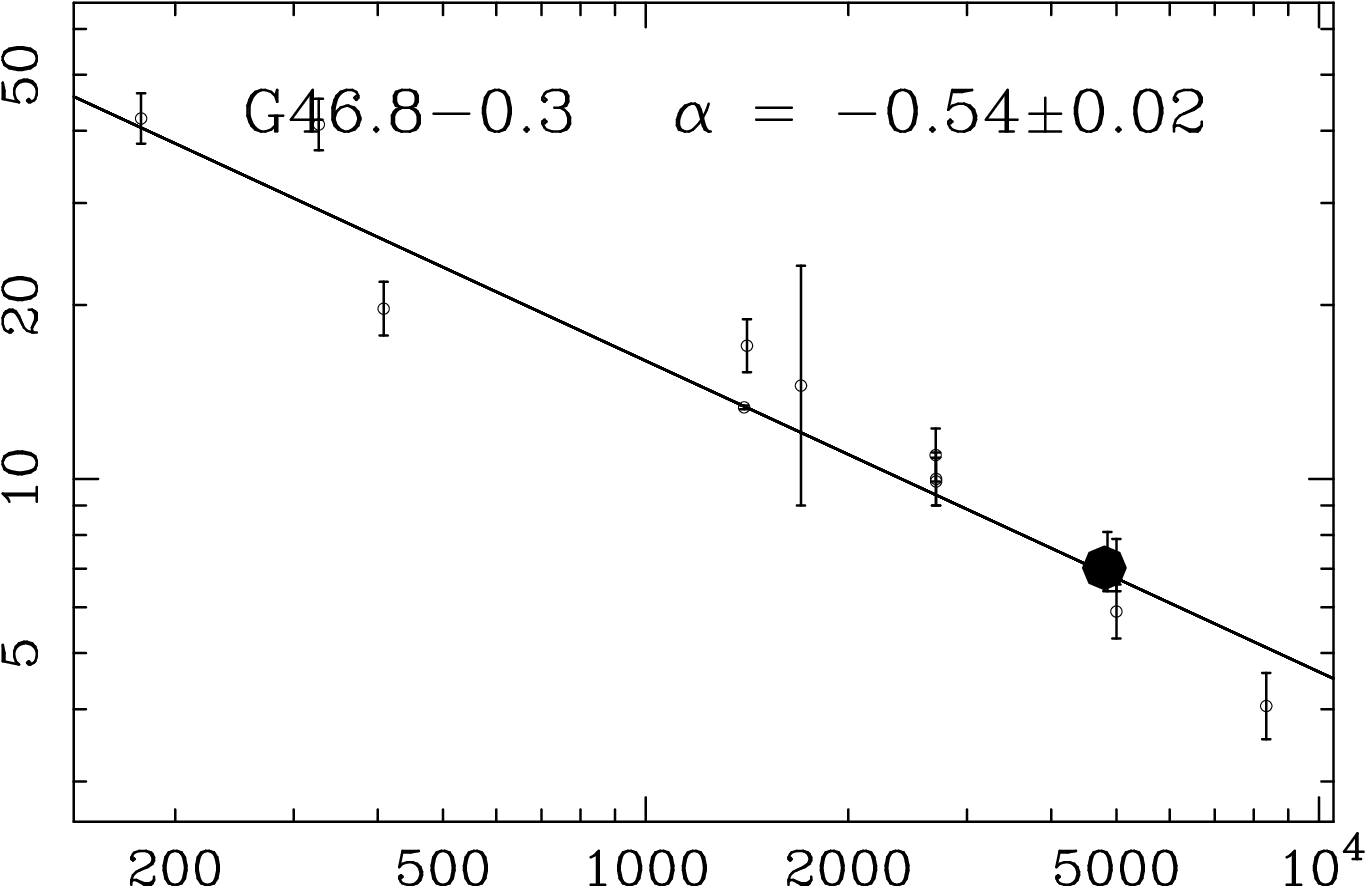}}
\\[1mm]
\resizebox{0.325\textwidth}{!}{\includegraphics[angle=-90]{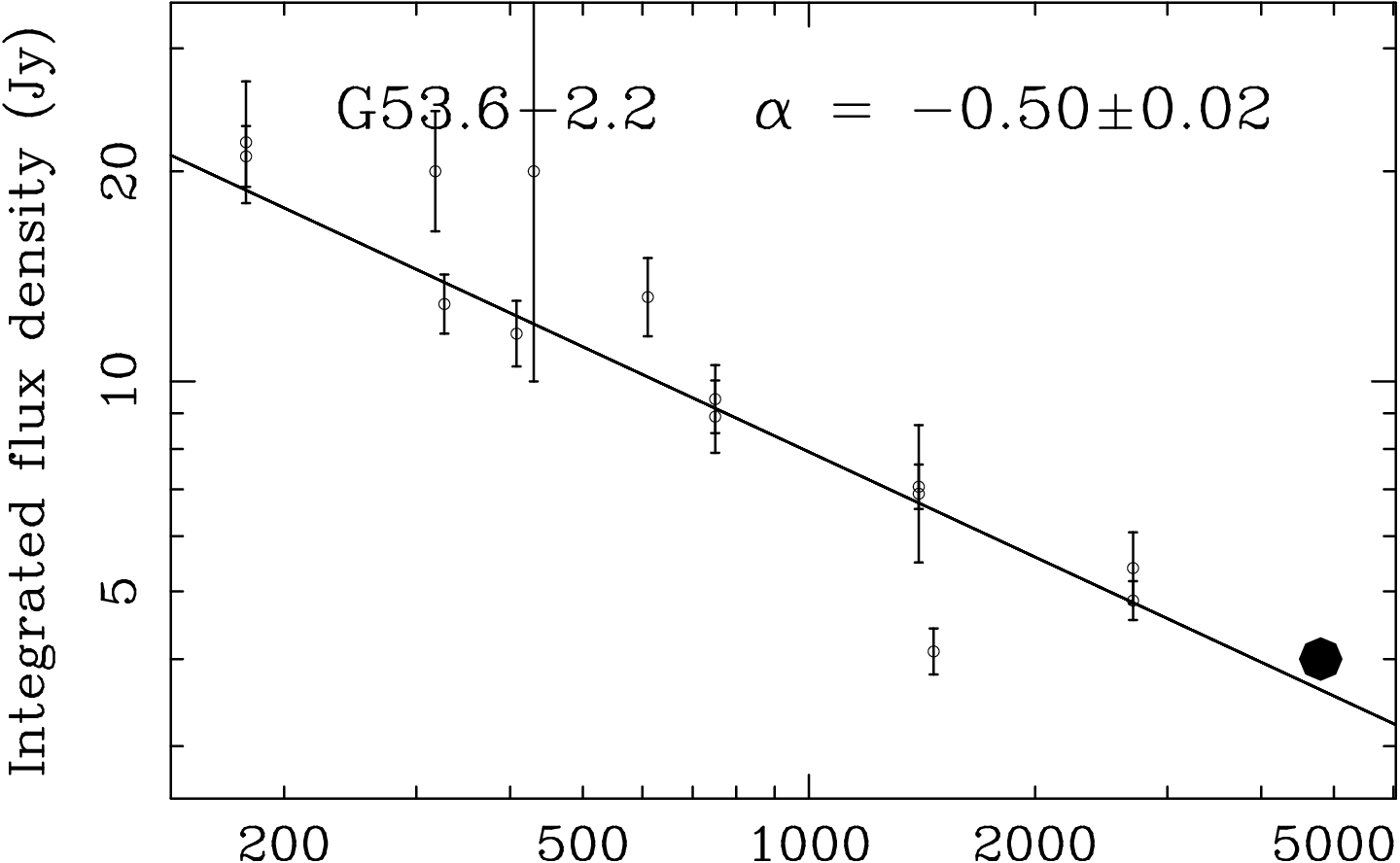}}
\resizebox{0.31\textwidth}{!}{\includegraphics[angle=-90]{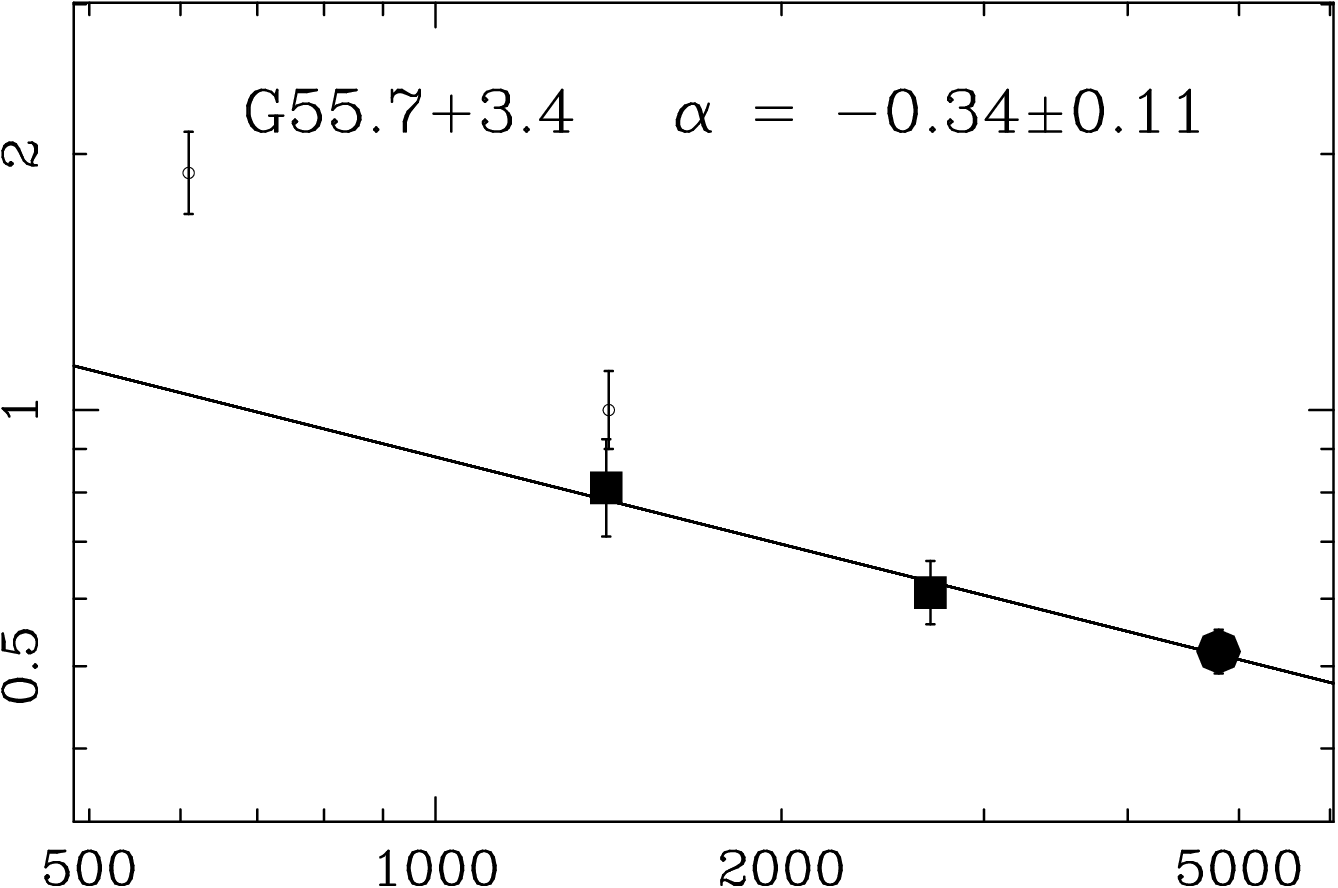}}
\resizebox{0.31\textwidth}{!}{\includegraphics[angle=-90]{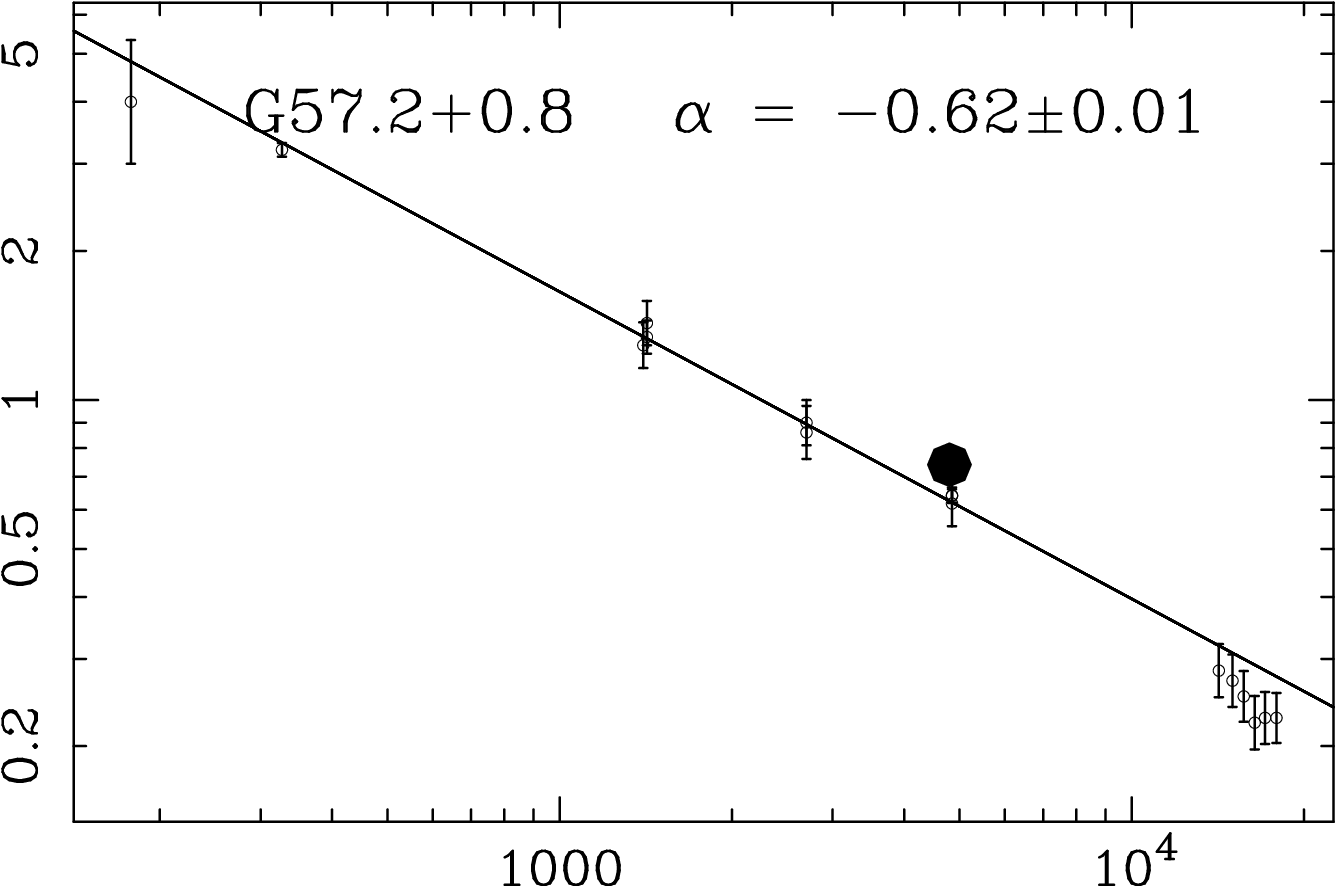}}
\\[1mm]
\resizebox{0.325\textwidth}{!}{\includegraphics[angle=-90]{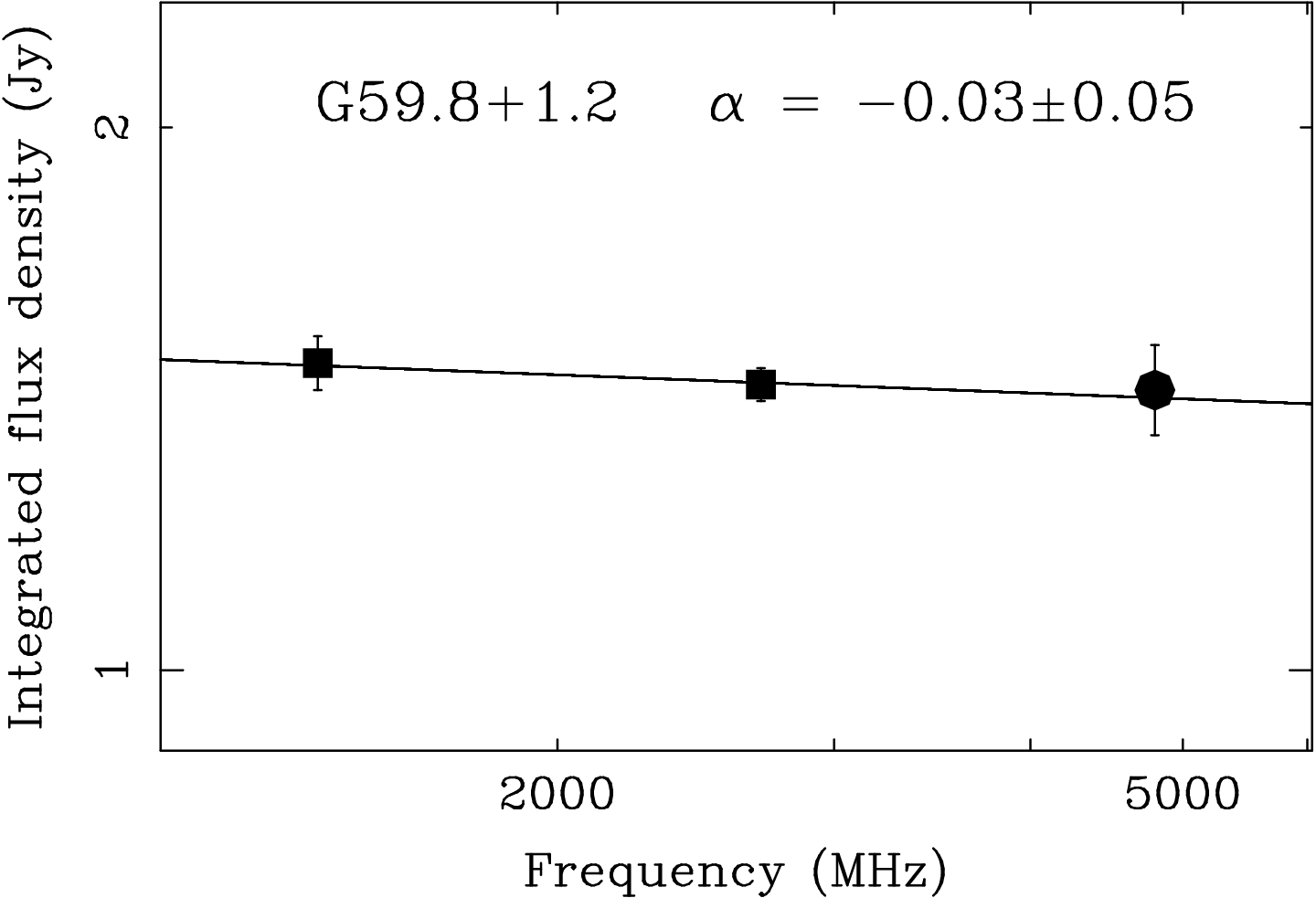}}
\resizebox{0.31\textwidth}{!}{\includegraphics[angle=-90]{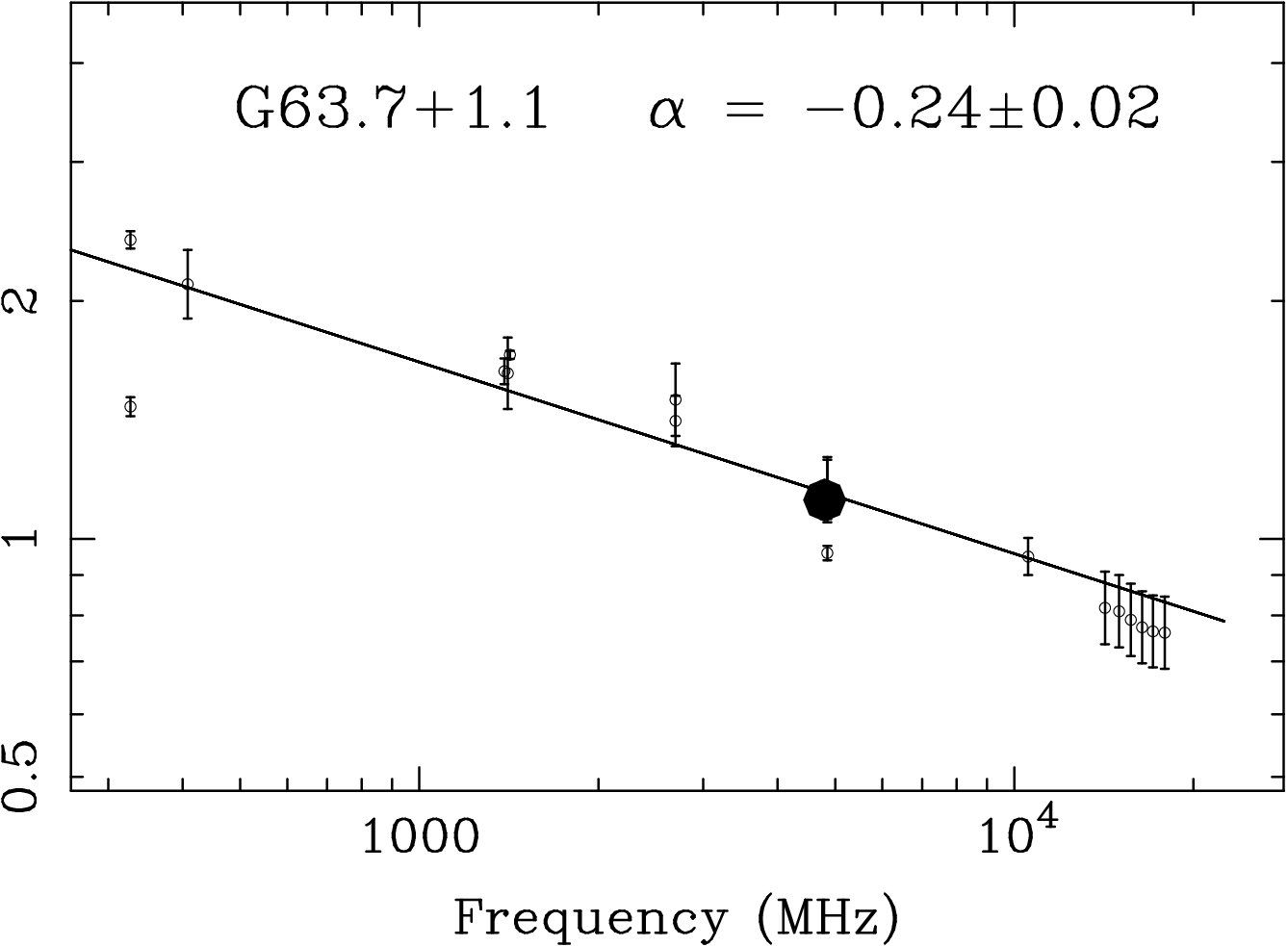}}
\resizebox{0.31\textwidth}{!}{\includegraphics[angle=-90]{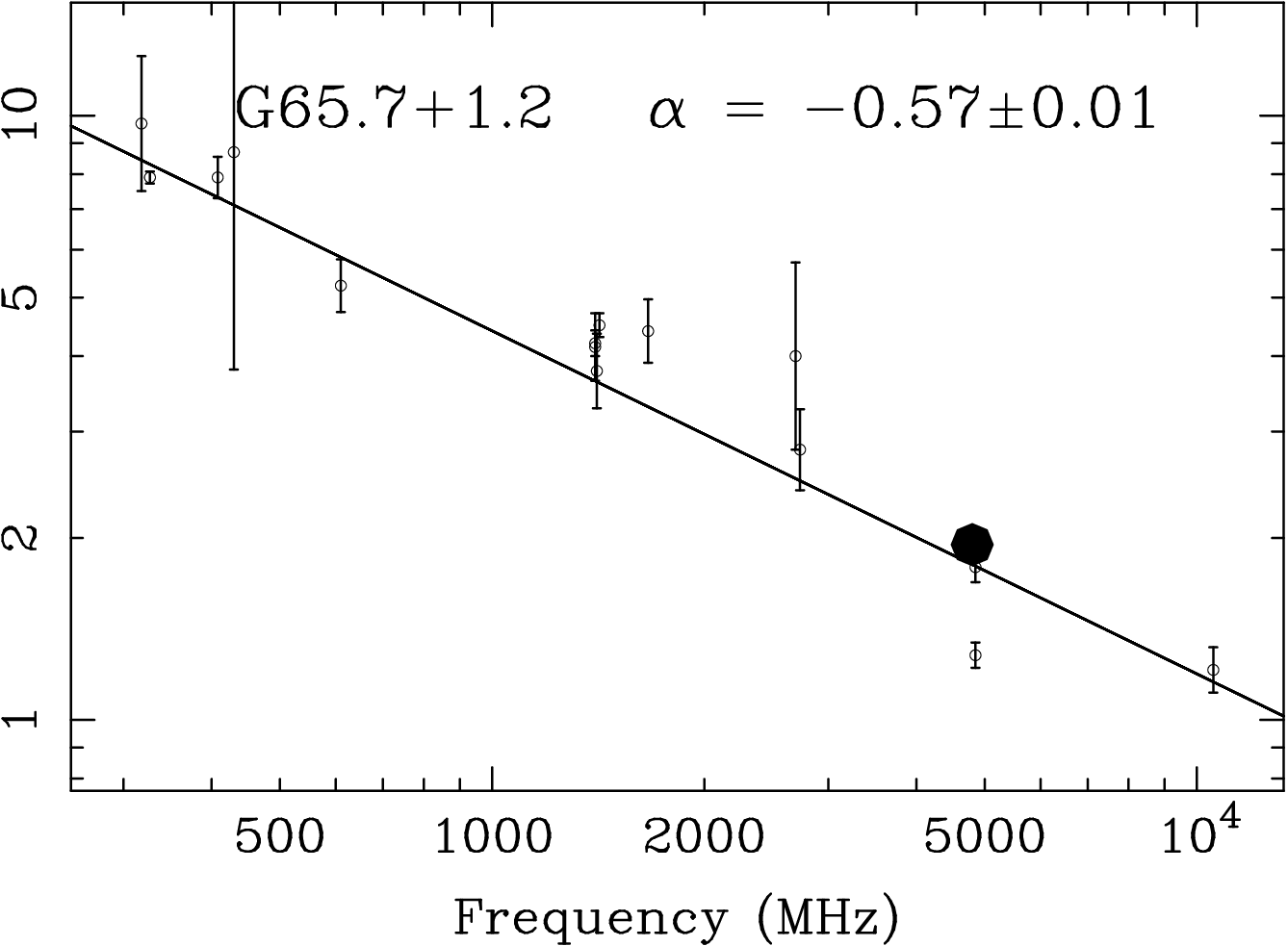}}
\caption{--continued.}
\end{figure*}\addtocounter{figure}{-1}
\begin{figure*}[!htbp]
\centering
\resizebox{0.325\textwidth}{!}{\includegraphics[angle=-90]{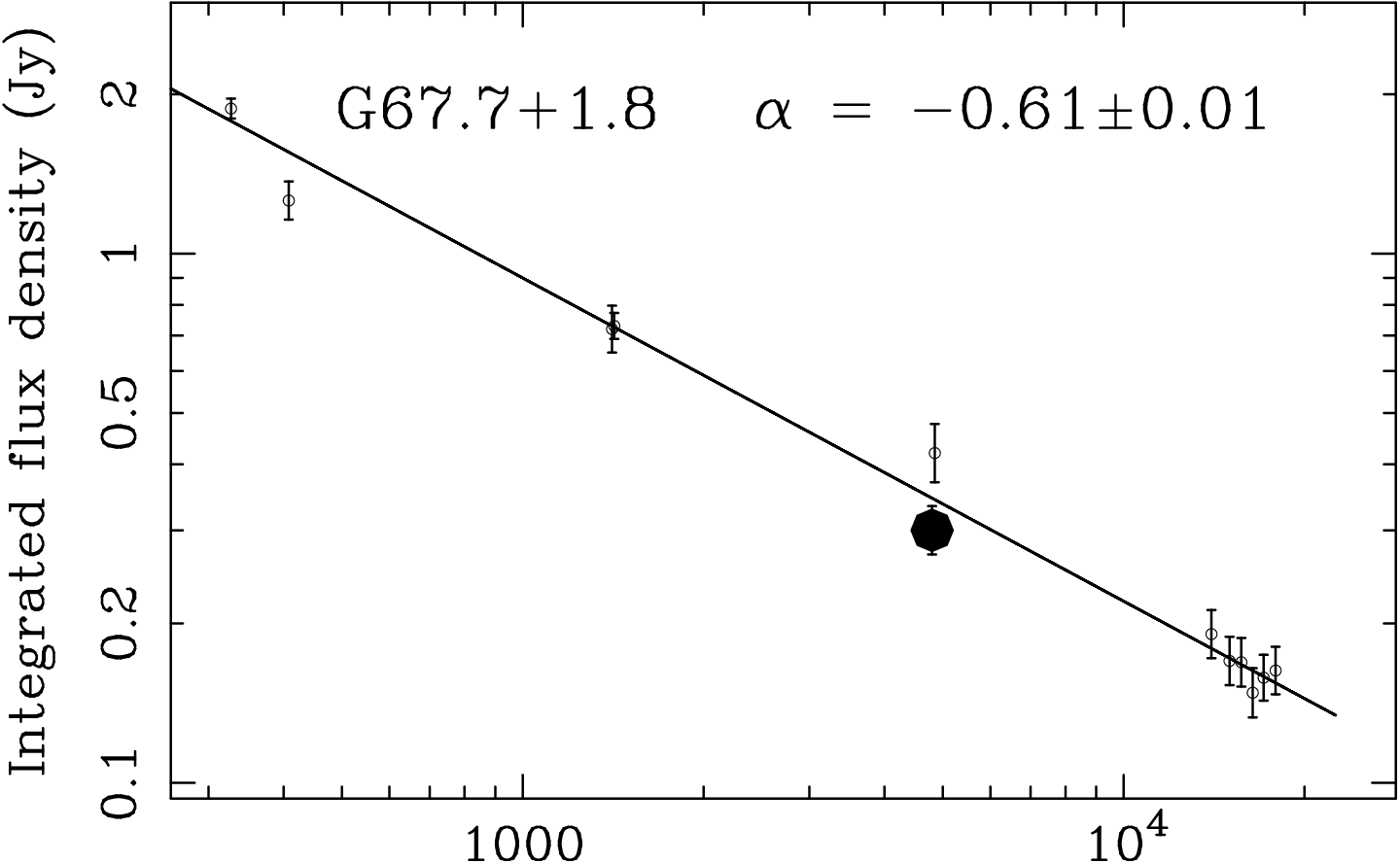}}
\resizebox{0.31\textwidth}{!}{\includegraphics[angle=-90]{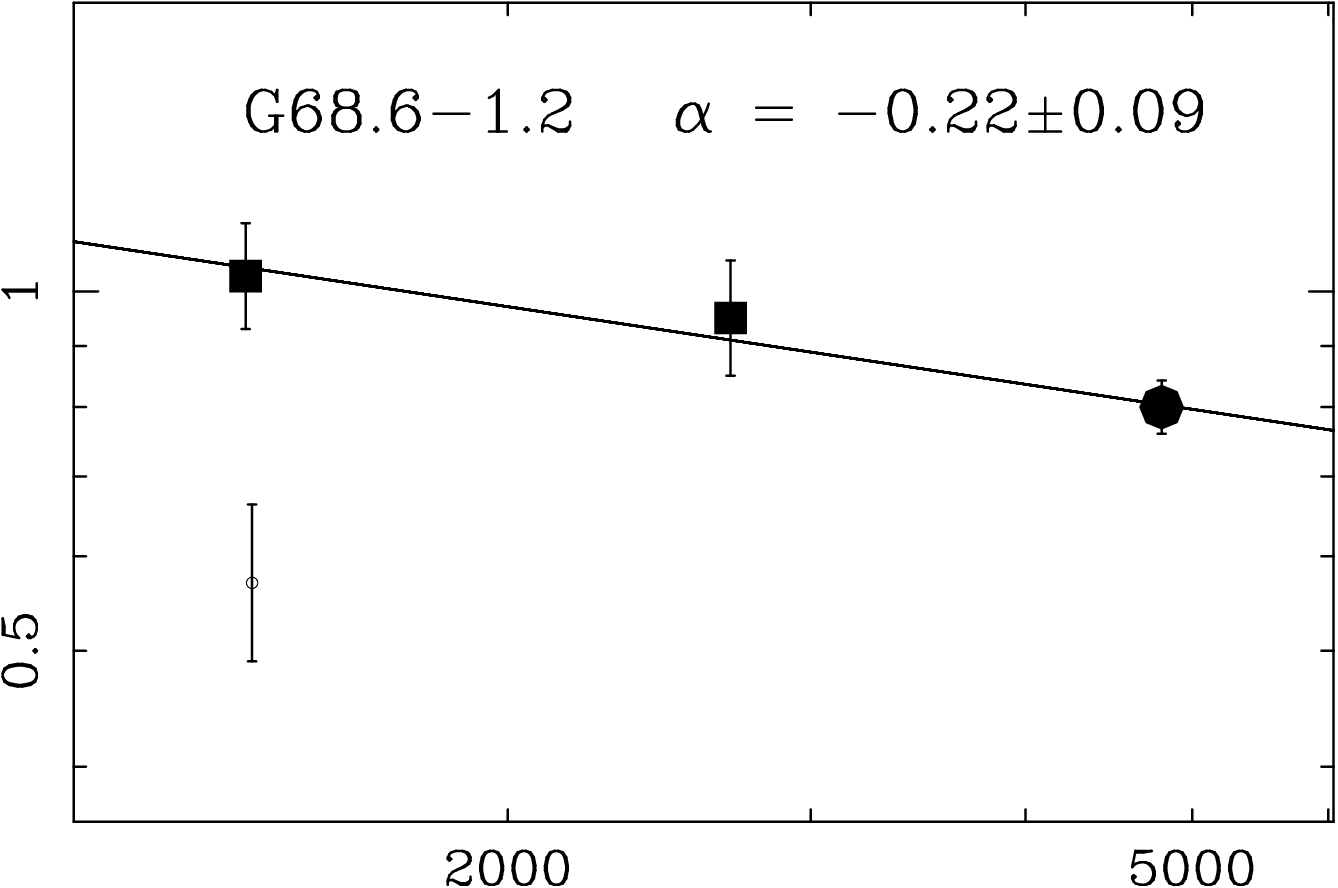}}
\resizebox{0.31\textwidth}{!}{\includegraphics[angle=-90]{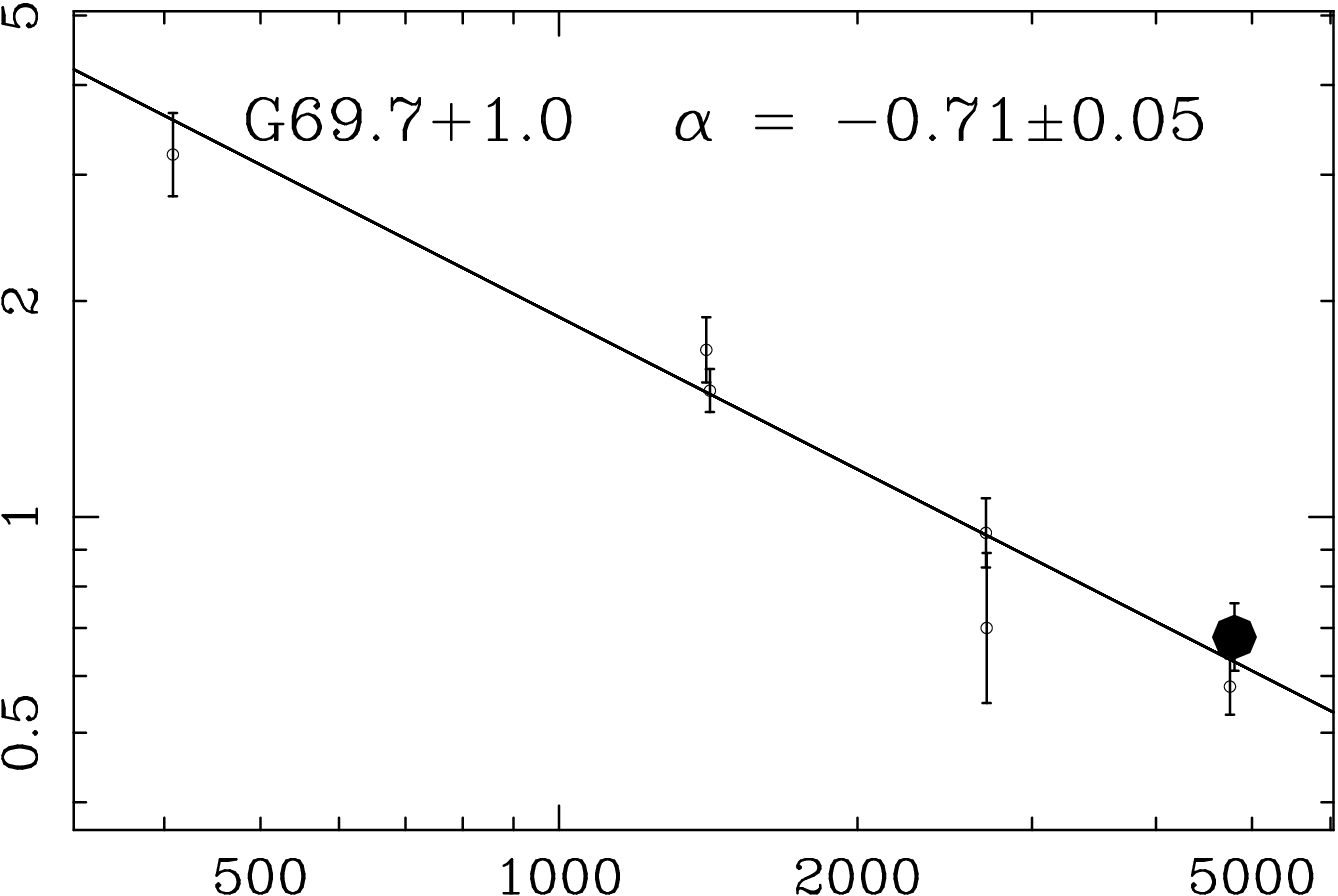}}
\\[1mm]
\resizebox{0.325\textwidth}{!}{\includegraphics[angle=-90]{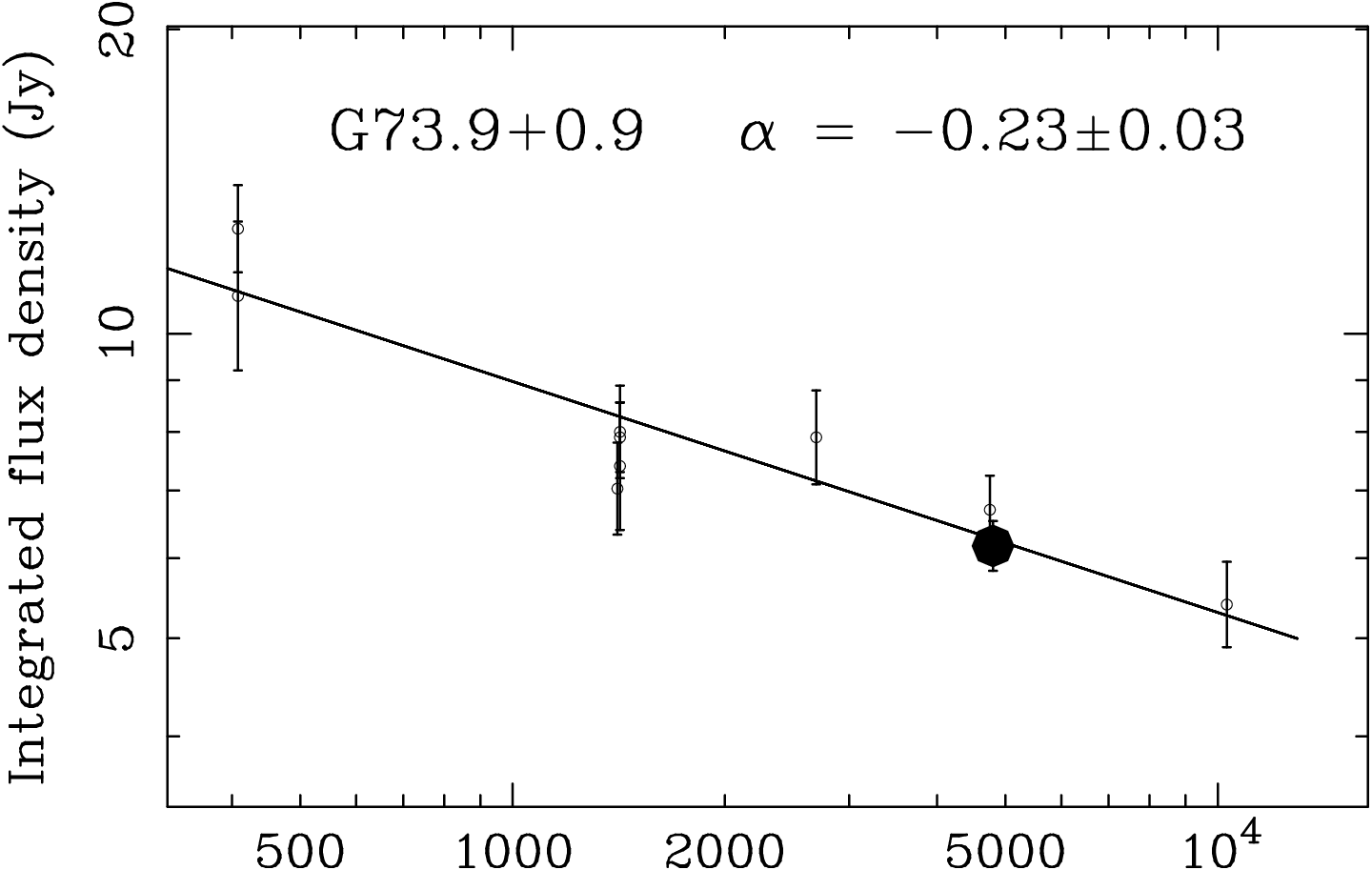}}
\resizebox{0.31\textwidth}{!}{\includegraphics[angle=-90]{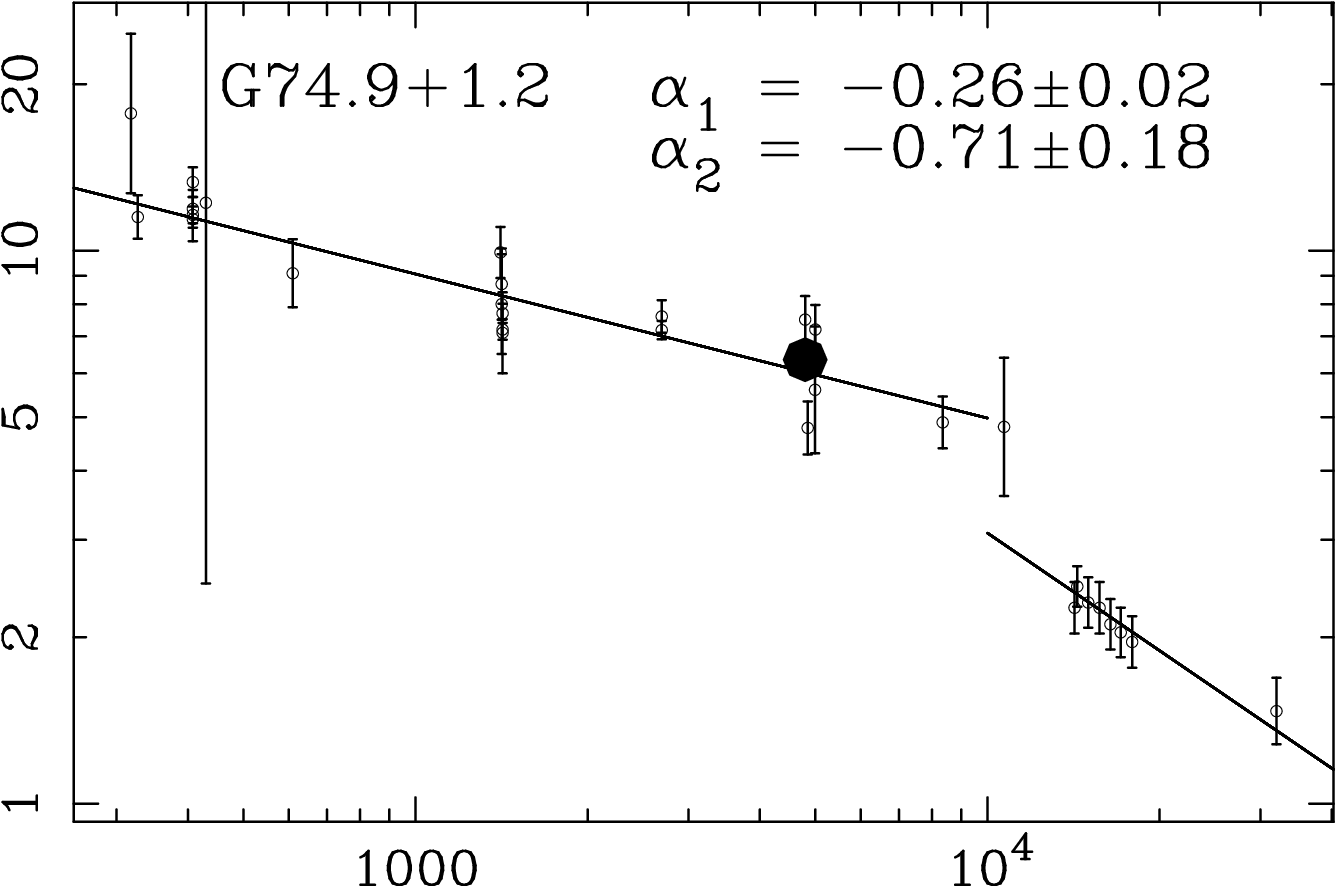}}
\resizebox{0.31\textwidth}{!}{\includegraphics[angle=-90]{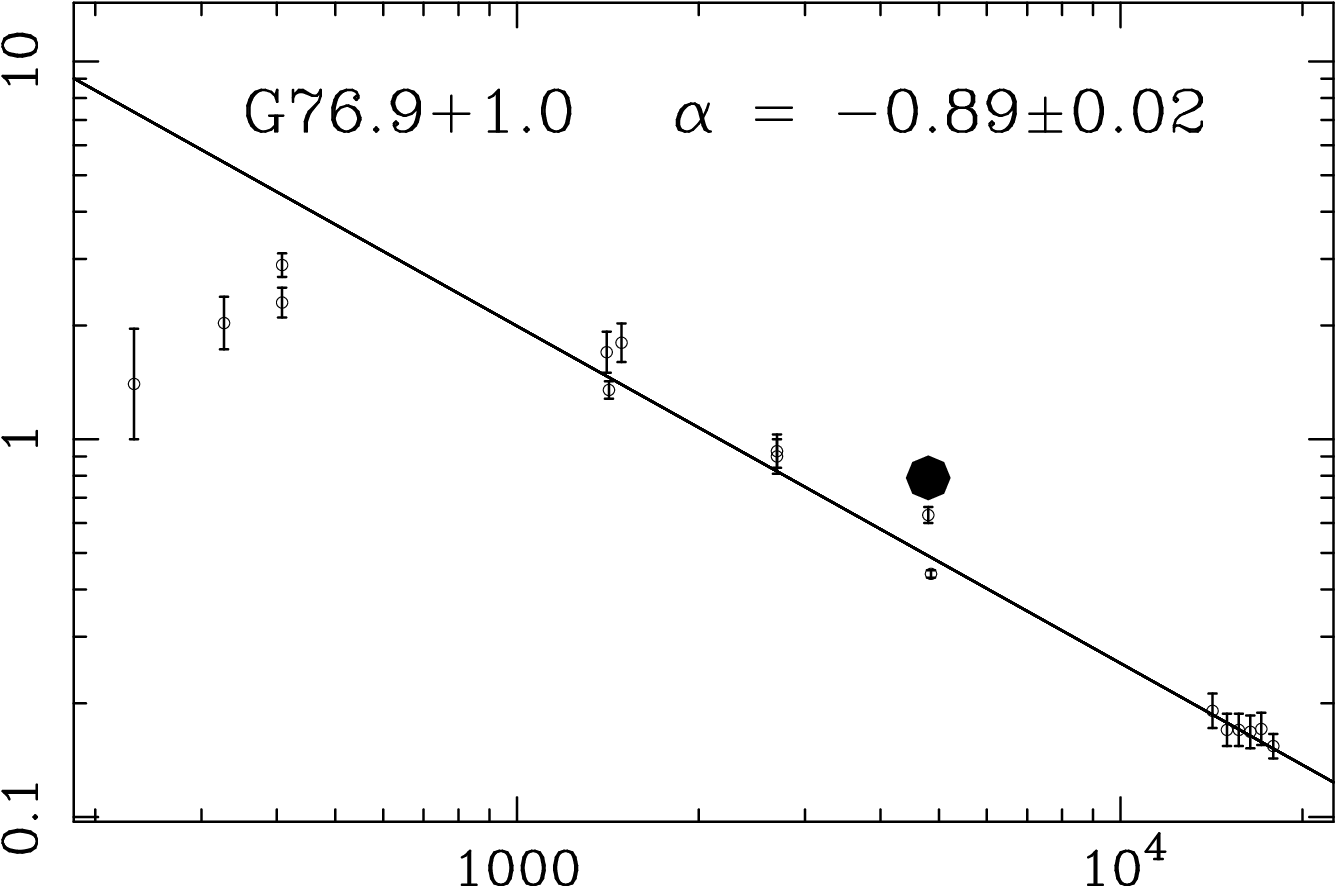}}
\\[1mm]
\resizebox{0.325\textwidth}{!}{\includegraphics[angle=-90]{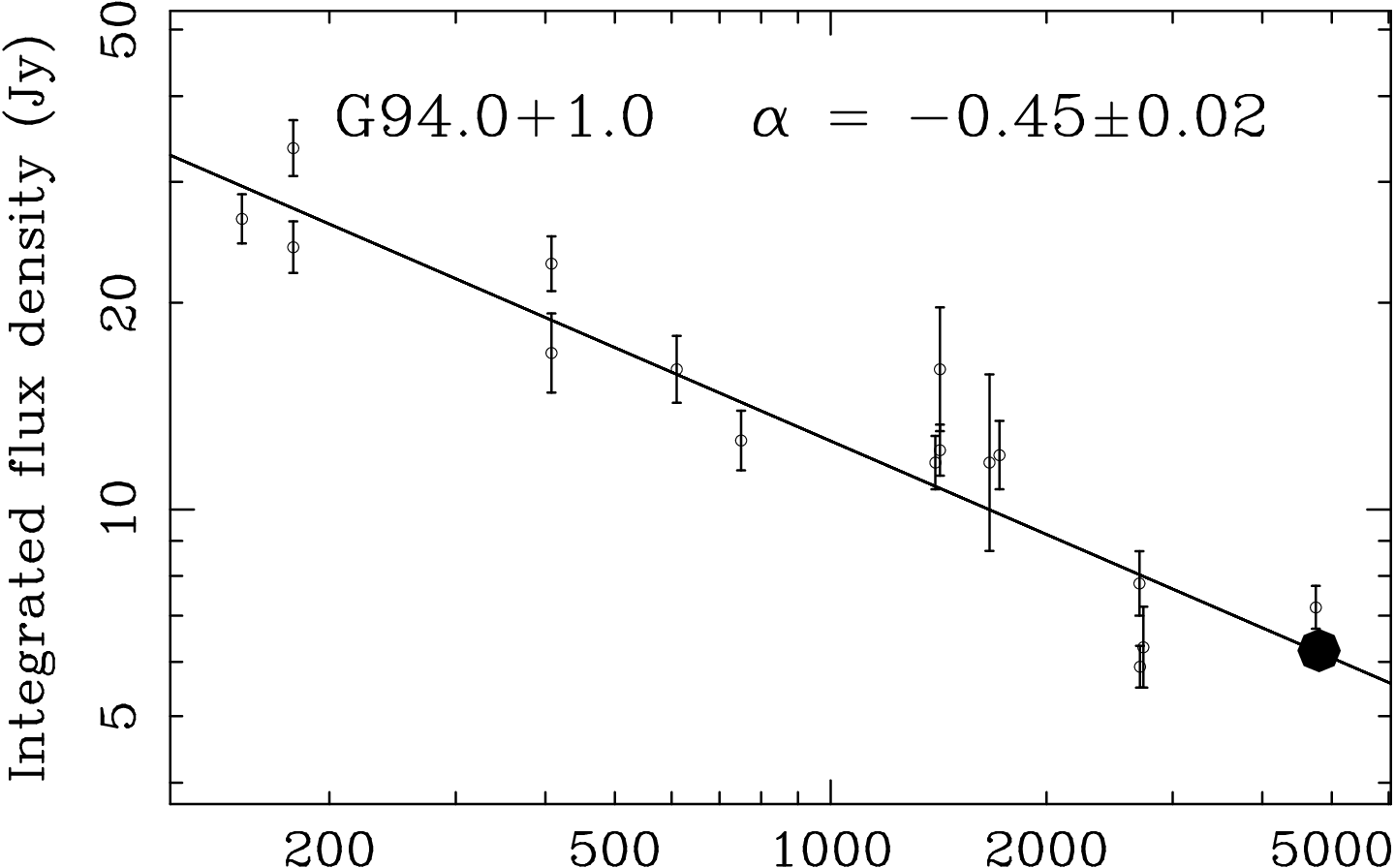}}
\resizebox{0.31\textwidth}{!}{\includegraphics[angle=-90]{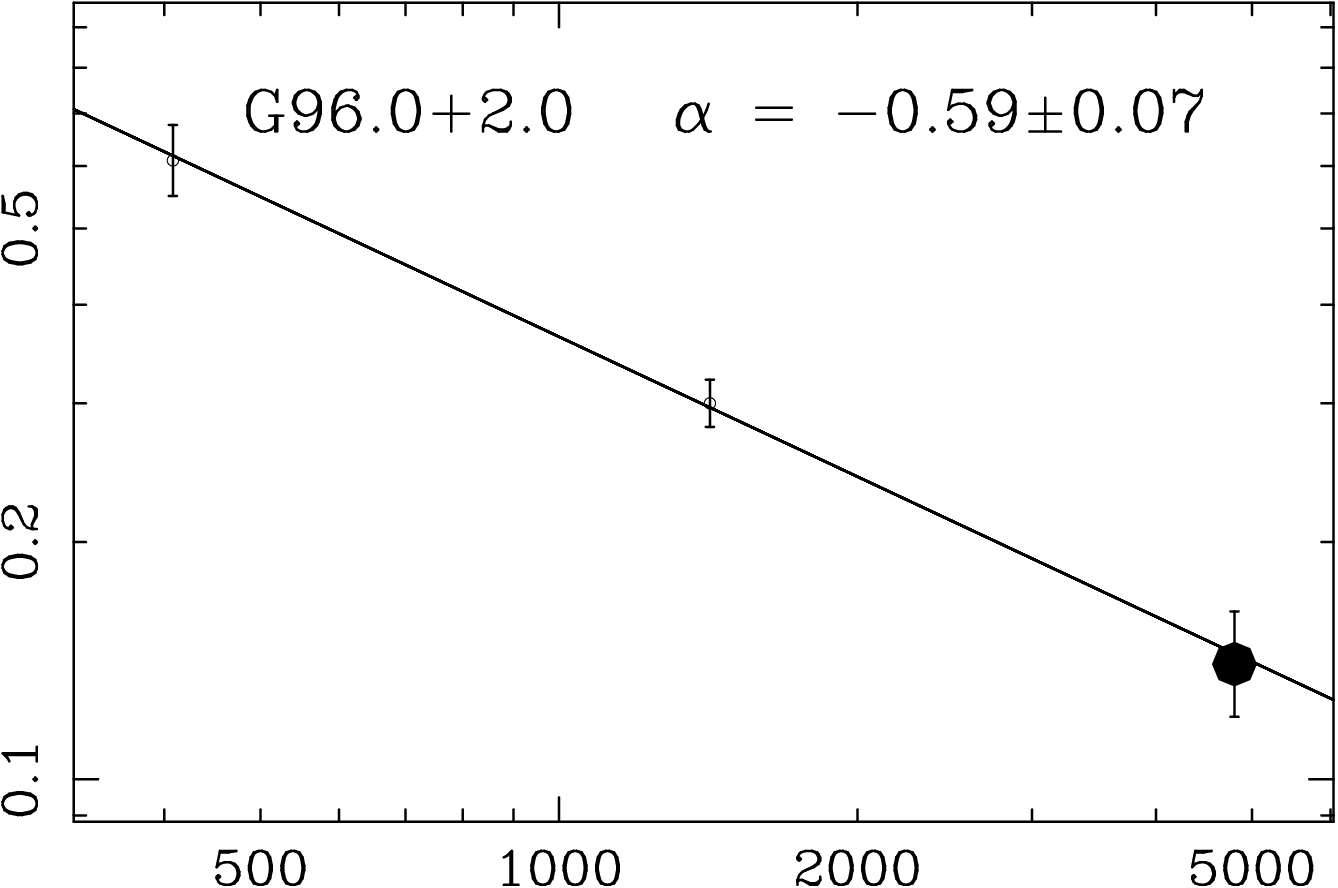}}
\resizebox{0.31\textwidth}{!}{\includegraphics[angle=-90]{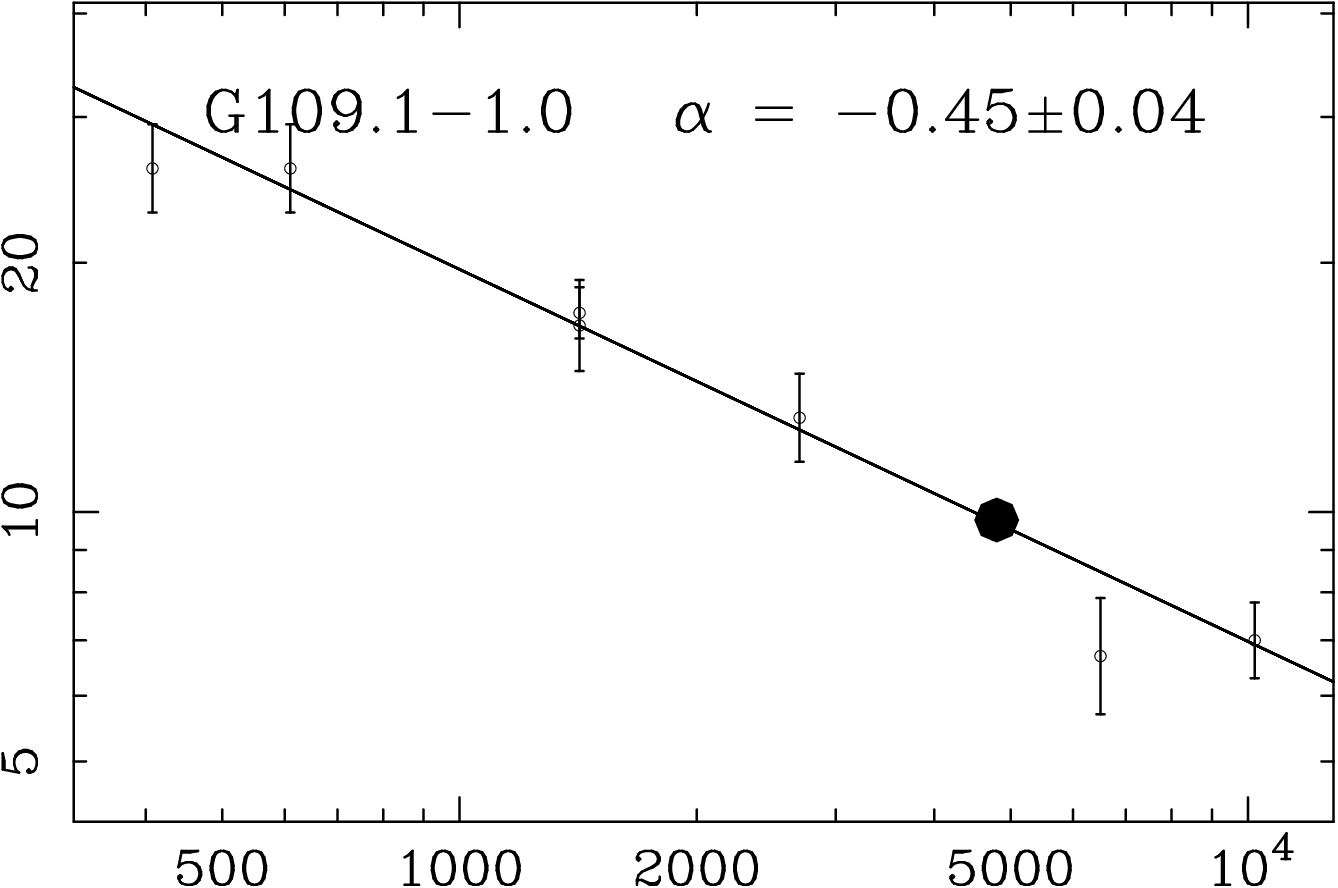}}
\\[1mm]
\resizebox{0.325\textwidth}{!}{\includegraphics[angle=-90]{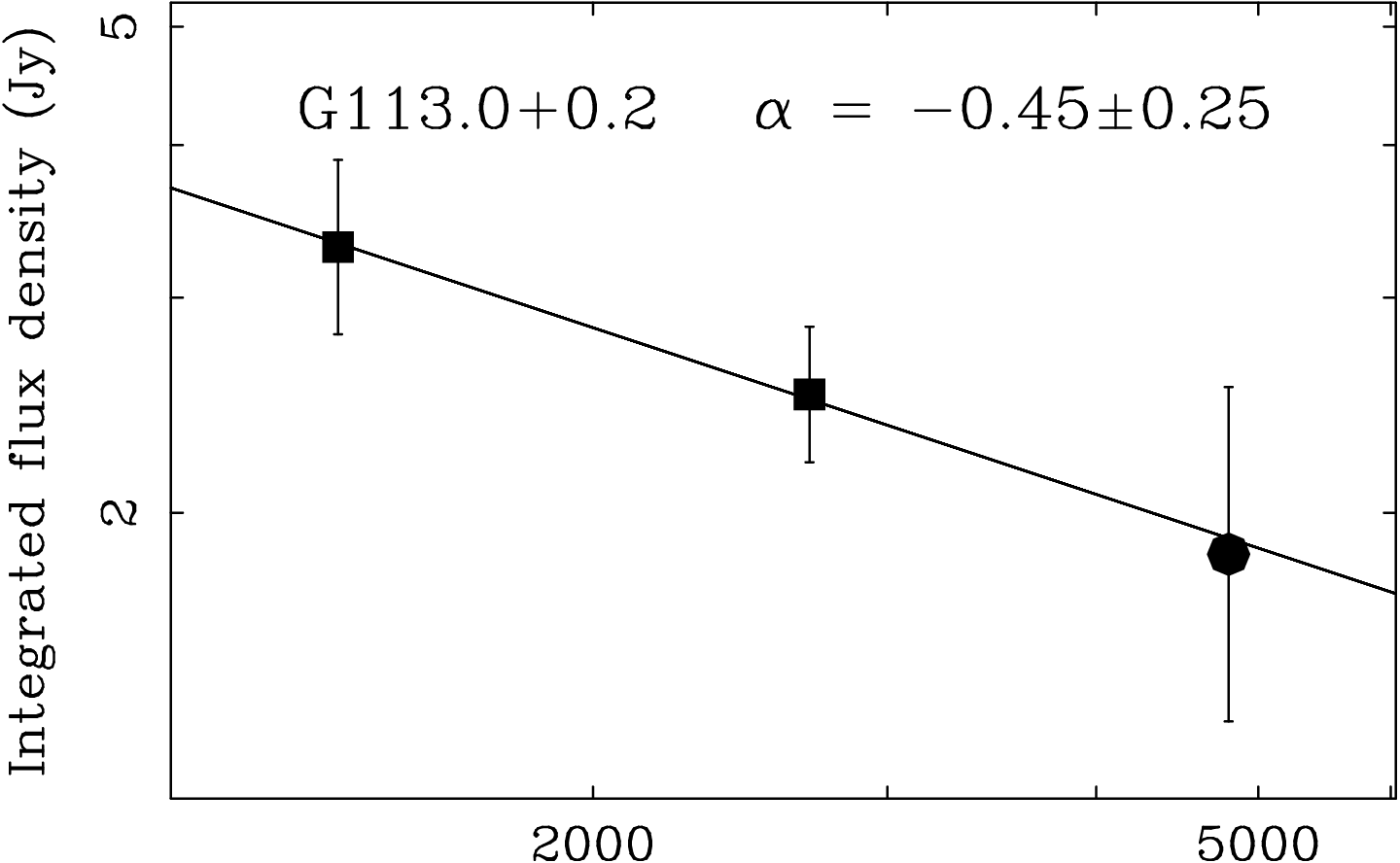}}
\resizebox{0.31\textwidth}{!}{\includegraphics[angle=-90]{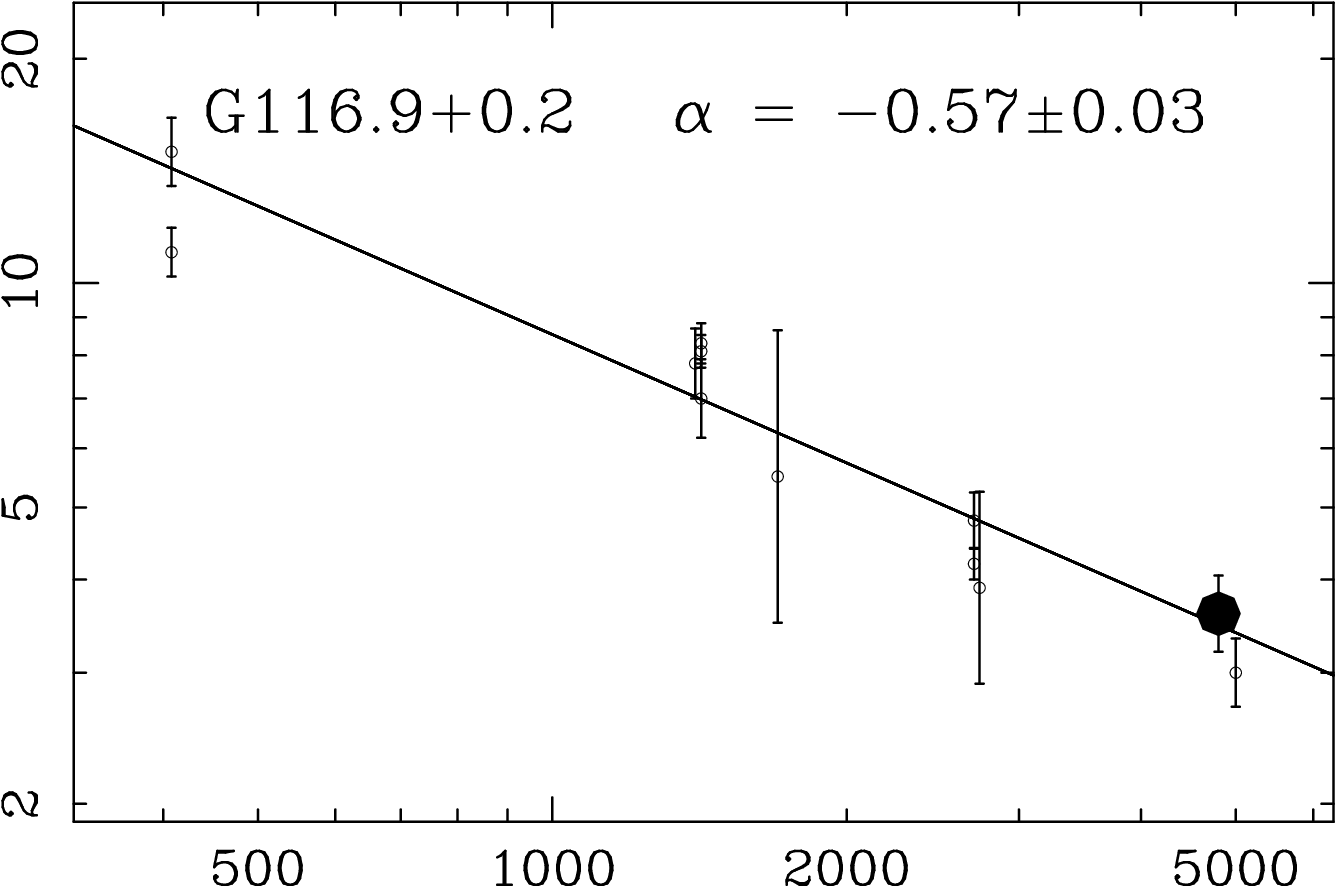}}
\resizebox{0.31\textwidth}{!}{\includegraphics[angle=-90]{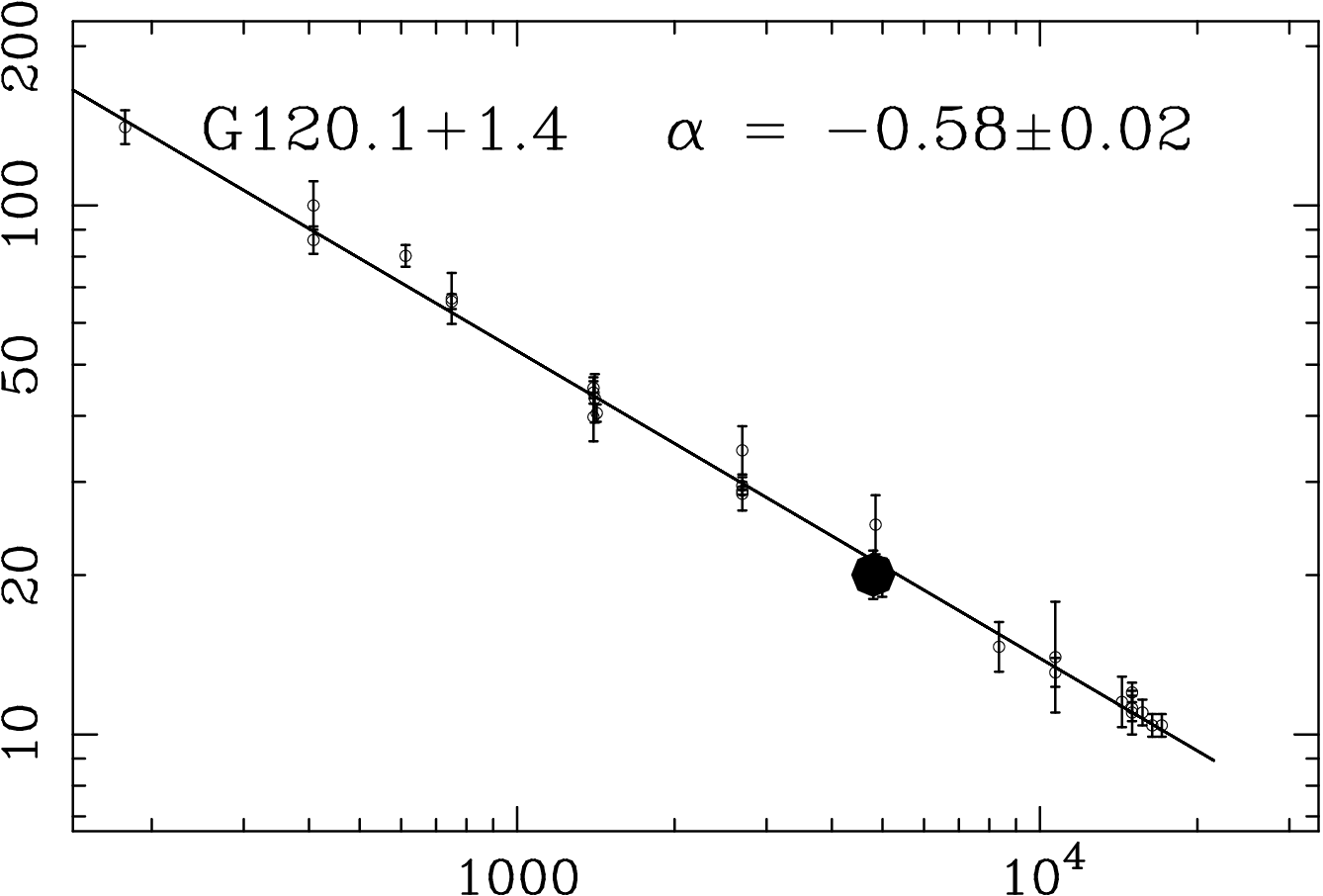}}
\\[1mm]
\resizebox{0.325\textwidth}{!}{\includegraphics[angle=-90]{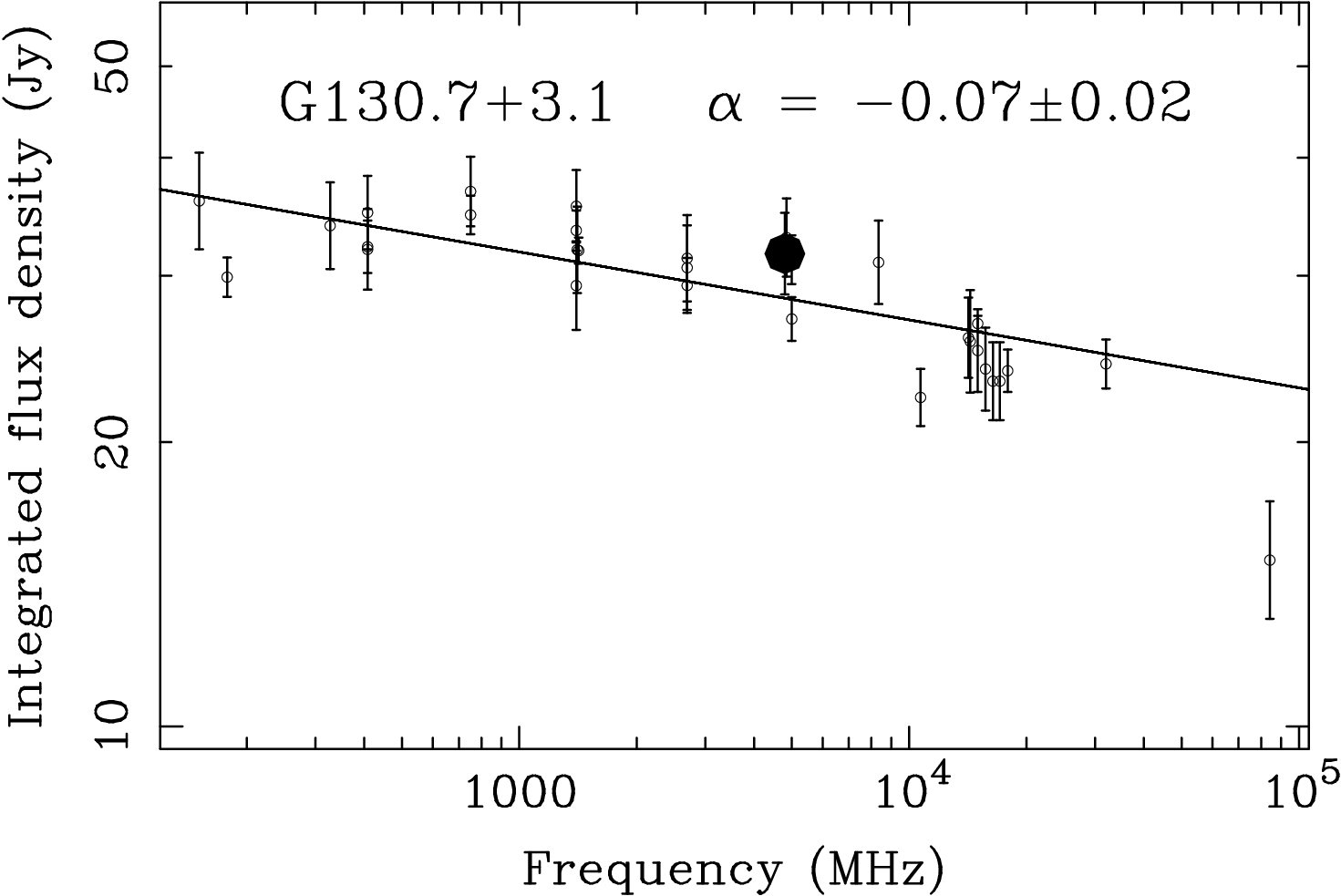}}
\resizebox{0.31\textwidth}{!}{\includegraphics[angle=-90]{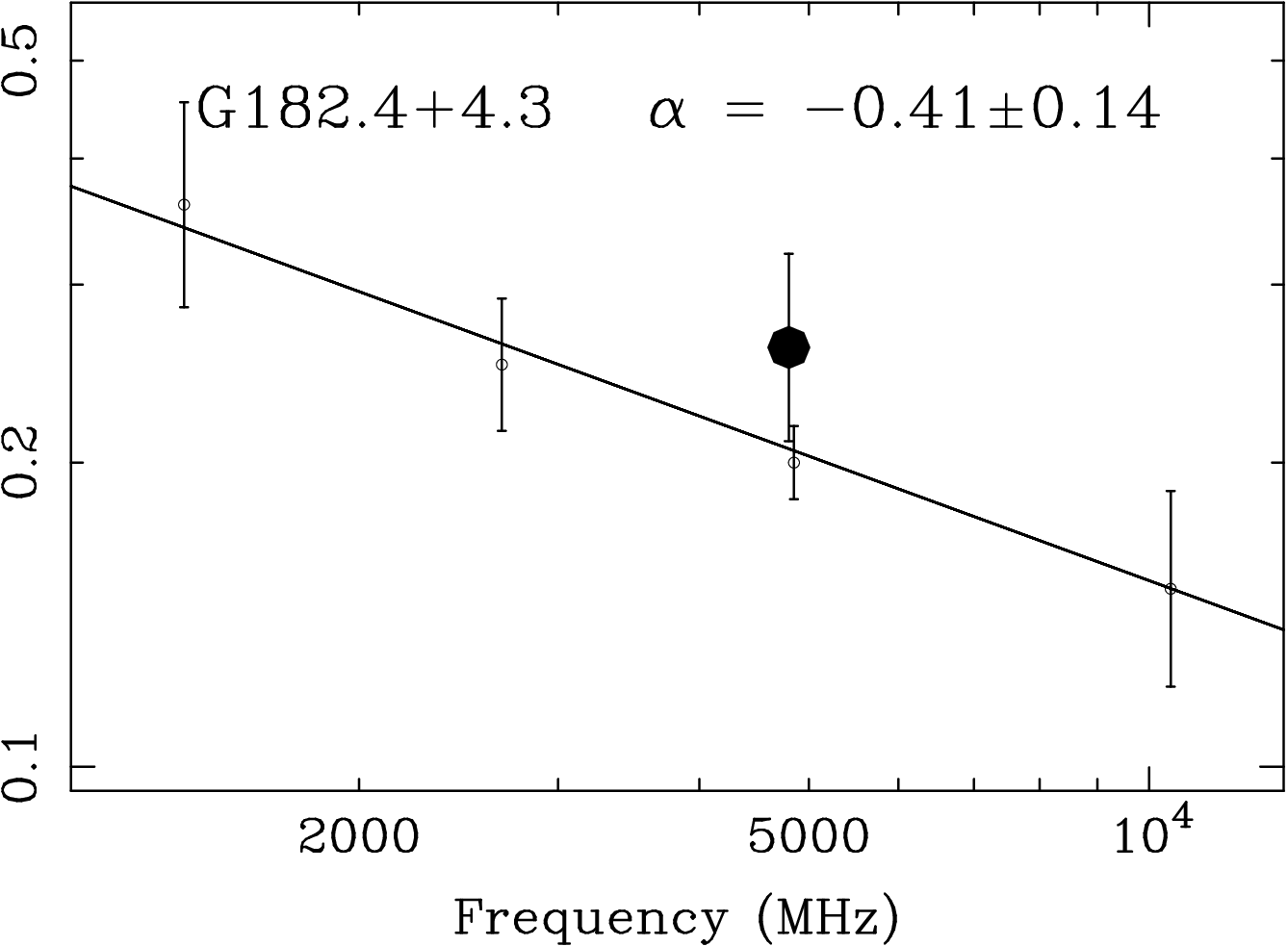}}
\caption{--continued.}
\end{figure*}

For nearly half of the SNRs in Fig.~\ref{spectra}, the new $\lambda$6~cm measurements 
are the only or amongst the highest frequency data, and are therefore 
important to constrain the high-frequency spectra. These SNRs are: G11.1$-$1.0, 
G15.1$-$1.6, G15.4+0.1, G15.9+0.2, G16.2$-$2.7, G16.4$-$0.5, G16.7+0.1, 
G17.4$-$2.3, G17.8$-$2.6, G20.0$-$0.2, G20.4+0.1, G36.6+2.6, 
G43.9+1.6, G53.6$-$2.2, G55.7+3.4, G59.8+1.2, G68.6$-$1.2, G69.7+1.0, 
G94.0+1.0, G96.0+2.0, G113.0+0.2, and G116.9+0.2, 

For 13 SNRs, improved spectra have been determined by combining the flux 
densities at $\lambda$6~cm, $\lambda$11~cm, and $\lambda$21~cm, which are 
further proved by the TT-plot \citep{tpkp62} results (see for example Fig.~\ref{tt17.8}). 
The relation $\alpha=\beta+2$ is used to convert the spectral index $\beta$ 
from TT-plot into $\alpha$.
These SNRs are: G15.1$-$1.6, 
G16.2$-$2.7, G16.4$-$0.5, G17.4$-$2.3, G17.8$-$2.6, G20.4+0.1, G36.6+2.6, 
G43.9+1.6, G53.6$-$2.2, G55.7+3.4, G59.8+1.2, G68.6$-$1.2, and G113.0+0.2. 
For some of these SNRs the early measurements were not used for 
spectral fitting, because they, as outliers, largely deviate from the spectra 
based on the new data at $\lambda$6~cm, $\lambda$11~cm, and $\lambda$21~cm, e.g. 
the measurements of G16.2$-$2.7 by \citet{tru99}, of G16.4$-$0.5 and 
G20.4+0.1 by \citet{bgg+06}, of G55.7+3.4 by \citet{gssw77}, and 
of G68.6$-$1.2 by \citet{kffu06}. The reason for the inconsistency is unclear. 
The spectra obtained by us are more reliable as they agree with the results 
from TT-plots.

For three SNRs, G21.5$-$0.9, G31.9+0.0 (3C 391), and G74.9+1.2 (CTB 87), the 
spectral turnover at high frequencies can be confirmed. The spectral break 
above 32~GHz for G21.5$-$0.9 was suggested by \citet{srh+89}. Below 32~GHz, the 
spectrum is very flat with a spectral index of $\alpha=-0.06\pm0.03$, 
consistent with that given by \citet{mr87a}. Above 32~GHz, the spectral index 
is $\alpha=-0.41\pm0.09$. The spectral break for 3C 391 was noted by 
\citet{mr94a}. Above 1~GHz, the spectral index of $\alpha=-0.54\pm0.02$ is 
consistent with that by \citet{mr94a}. Below 1~GHz the spectral index is 
$\alpha=-0.02\pm0.04$. \citet{blkd05} ascribed the spectral turnover to 
absorption and obtained an opacity of 1.1 at 74~MHz, which needs confirmation 
at even lower frequencies. The spectral break above 11~GHz for CTB 87 was reported 
by \citet{mr87a} according to their 32~GHz measurement. Below 11~GHz, the 
spectral index obtained by \citet{mr87a} is $\alpha=-0.26$, which is consistent 
with our result. Adding new flux densities at 10.35~GHz \citep{lmd+00} and 
16~GHz \citep{hsg+09} confirms the spectral break with a spectral index of 
$\alpha=-0.71\pm0.18$ above 11~GHz. The frequency turnover for G27.8$+$0.6 and 
G130.7+3.1 is less certain and needs more high-frequency observations for  
confirmation. 

Some SNRs, such as G15.1$-$1.6, G20.4+0.1, and G59.8+1.2 are probably new 
plerions as they have very flat spectra of $\alpha=-0.01\pm0.09$, 
$\alpha=-0.08\pm0.09$, and $\alpha=-0.03\pm0.05$. G16.8$-$1.1 seems misidentified 
and is likely an \ion{H}{II} region as discussed below. 
The spectral index $\beta$ from the TT-plots is the weighted 
average of the spectral indices from two pairs: $\lambda$6~cm and 
$\lambda$11~cm, and $\lambda6$~cm and $\lambda$21~cm. 
We comment below on 
those SNRs for which the newly determined spectral indices deviate by more than 
$3\times\sigma$ from earlier results or when the previous spectrum was very 
uncertain as indicated by the question mark in Table~\ref{snr_s}. 

\begin{itemize}
\item{G11.1$-$1.0.} \citet{bgg+06} obtained a spectral index of $\alpha=-0.5$ 
between $\lambda$90~cm and $\lambda$11~cm data and $\alpha=-0.6$ between 
$\lambda$90~cm and $\lambda$20~cm measurements. A fit of the three data points 
by \citet{bgg+06} yields a spectral index of $\alpha=-0.48\pm0.05$, consistent 
with our result of $\alpha=-0.41\pm0.02$.

\item{G15.1$-$1.6.} The spectral index of $\alpha=-0.8$ obtained by 
\citet{rfrj88} is very uncertain as it was based on data only at two 
frequencies. We obtained a spectral index of 
$\alpha=-0.01\pm0.09$, which is that of a thermal source. However, 
\citet{bac+08} made optical observations, which convincingly showed that 
G15.1$-$1.6 is a SNR.

\item{G17.4$-$2.3.} The spectral index $\alpha=-0.46\pm0.12$ was obtained by 
fitting the new flux densities at $\lambda$6~cm, $\lambda$11~cm, and $\lambda$21~cm. 
This result is consistent with the average value $\alpha=-0.52\pm0.03$ 
using the TT-plot method.

\item{G17.8$-$2.6.} With the new flux densities at $\lambda$6~cm, 
$\lambda$11~cm, and $\lambda$21~cm we obtained a spectral index  
$\alpha=-0.50\pm0.07$, which is in good agreement with the value 
$\alpha=-0.52\pm0.13$ derived from TT-plots (Fig.~\ref{tt17.8}).
\begin{figure}[!htbp]
\resizebox{0.45\textwidth}{!}{\includegraphics[angle=-90]{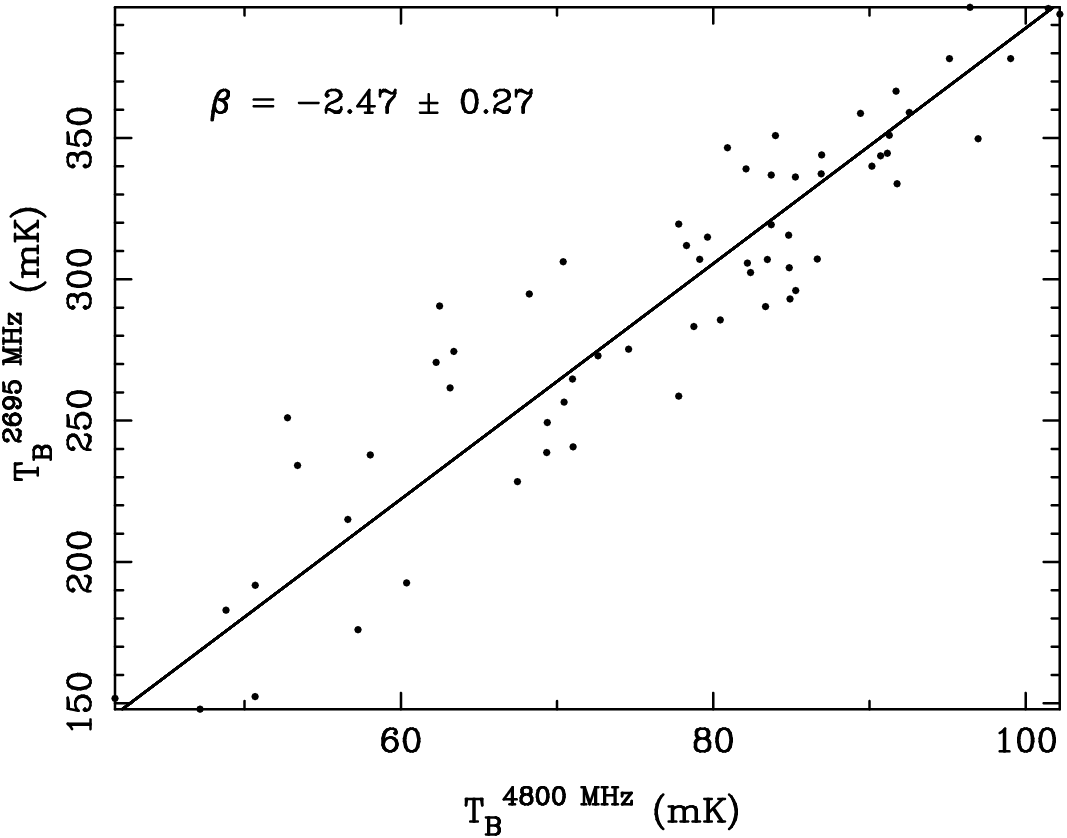}}
\resizebox{0.45\textwidth}{!}{\includegraphics[angle=-90]{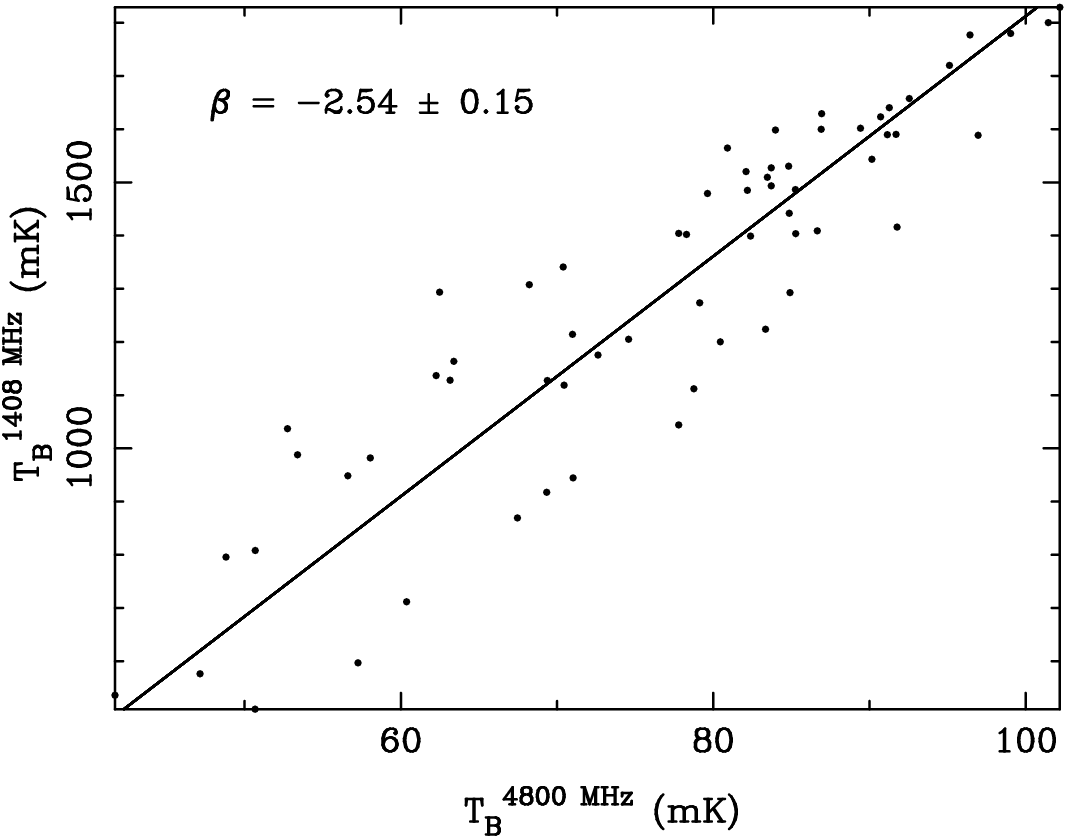}}
\caption{TT-plots of G17.8$-$2.6 between $\lambda$6~cm (4800~MHz) and $\lambda$11~cm 
(2695 MHz), and between $\lambda$6~cm and $\lambda$21~cm (1408~MHz).}
\label{tt17.8}
\end{figure} 

\item{G20.4+0.1.} The VLA flux densities presented by \citet{bgg+06} 
(included in Fig.~\ref{spectra}) are significantly lower than the flux 
densities we derived from single-dish observations. The spectrum 
presented here, which is characterized by $\alpha=-0.08\pm0.09$ and based on 
new flux densities at $\lambda$6~cm, 
$\lambda$11~cm, and $\lambda$21~cm, is significantly flatter than 
the value $\alpha=-0.4$ reported  
by \citet{bgg+06}. The average spectral index derived from TT-plots among the three bands is $\alpha=-0.09\pm0.04$, which agrees with the integrated flux density spectrum displayed in Fig.~\ref{snr_s}. 

\item{G36.6+2.6.} The spectral index $\alpha=-0.67\pm0.12$ was 
obtained from the 
new flux density values at $\lambda$6~cm, $\lambda$11~cm, and $\lambda$21~cm 
only.

\item{G43.9+1.6.} The spectral index reported by \citet{rfrj88} is indicated as very 
uncertain. We derived a value $\alpha=-0.47\pm0.06$ from the integrated flux 
densities. The average spectral index $\alpha=0.59\pm0.10$ from TT-plots between the Urumqi $\lambda$6~cm and 
the Effelsberg $\lambda$11~cm and $\lambda$21~cm data is consistent with the 
spectrum shown in Fig.~\ref{spectra}.

\item{G53.6$-$2.2 (3C 400.2).} The published spectral index $\alpha=-0.76$ 
listed in Table~\ref{snr_s} results from the low VLA flux density at 1465~MHz 
by \citet{dggw94}. The derived spectral index in the current work is $\alpha=-0.50\pm0.02$ after including the higher single-dish 
$\lambda$21~cm flux density from the Effelsberg survey. 

\item{G57.2+0.8 (4C 21.53).} The derived value $\alpha=-0.62\pm0.01$ is slightly 
larger than $\alpha=-0.67$ by \citet{hsg+09}, which could be ascribed to a 
higher $\lambda$6~cm flux density.

\item{G59.8+1.2.} This SNR was found to have a flat spectrum rather than a 
steep spectrum as it was reported earlier by \citet{rfrj88}. Optical observations by 
\citet{bmx+05} showed a large \ion{S}{II}/H$\alpha$ ratio, which confirms 
this source as non-thermal. It may be classified as a pulsar wind nebula (PWN) 
instead of a classical shell-type SNR. This object consists of an elliptically 
shaped source and a tail. Based on TT-plots we obtained spectral indices of 
$\alpha = -0.13\pm0.06$ between $\lambda$6~cm and $\lambda$11~cm 
and $\alpha = -0.03\pm0.10$ between $\lambda$6~cm and $\lambda$21~cm for the 
source component. Their weighted average spectral index is 
$\alpha = -0.09\pm0.05$, almost consistent with $\alpha = -0.03\pm0.05$ from 
integrated flux densities as shown in Fig.~\ref{spectra}. The spectral index 
does not change when including the tail. Unfortunately no polarized emission is 
visible in the $\lambda$6~cm maps. There is no ROSAT X-ray emission associated, 
and no pulsar reported so far is in the direction of G59.8+1.2. More detailed 
investigations are required to firmly establish its classification.

\item{G68.6$-$1.2.} This SNR was discovered by \citet{rfrj88}. Based on the new 
flux densities at $\lambda$6~cm, $\lambda$11~cm, and $\lambda$21~cm we derived 
a spectral index $\alpha=-0.22\pm0.09$. The $\lambda$21~cm flux density
quoted by \citet{kffu06} was not included in the fit, as it largely 
deviates from the spectrum.

\item{G76.9+1.0.} This object strongly resembles the PWN DA 495 \citep{lhw93}.
Available flux densities indicate a spectral turnover at about 1~GHz. Fitting 
the measurements above 1~GHz and including the new data at 16~GHz by 
\citet{hsg+09} we obtained a spectral index $\alpha=-0.89\pm0.02$. This 
steepening probably stems from synchrotron aging similar to what is seen in 
DA 495 \citep{klr+08}. 

\item{G113.0+0.2.} This SNR with strong polarized emission was discovered by 
\citet{kur05}. However, so far no flux density could be quoted as it is a 
very weak and extended SNR. Besides the flux density at $\lambda$6~cm, we also 
measured its flux density at $\lambda$11~cm and $\lambda$21~cm from the 
Effelsberg surveys. The spectral index is $\alpha=-0.45\pm0.25$ supported by 
the TT-plot result of $\alpha=-0.43\pm0.12$.

\end{itemize}

\subsection{Polarization}

The ``background filtering" method \citep{sr79} cannot be used to remove large scale 
polarization $U$ and $Q$ data. To show the intrinsic polarized emission from 
the SNRs, we subtracted a hyper-plane defined by values at the four corners of 
the $U$ and $Q$ images of the SNRs extracted from the survey. We then 
re-calculated PI and PA. 

\begin{figure*}[!htbp]
\centering
\resizebox{0.32\textwidth}{!}{\includegraphics[angle=-90]{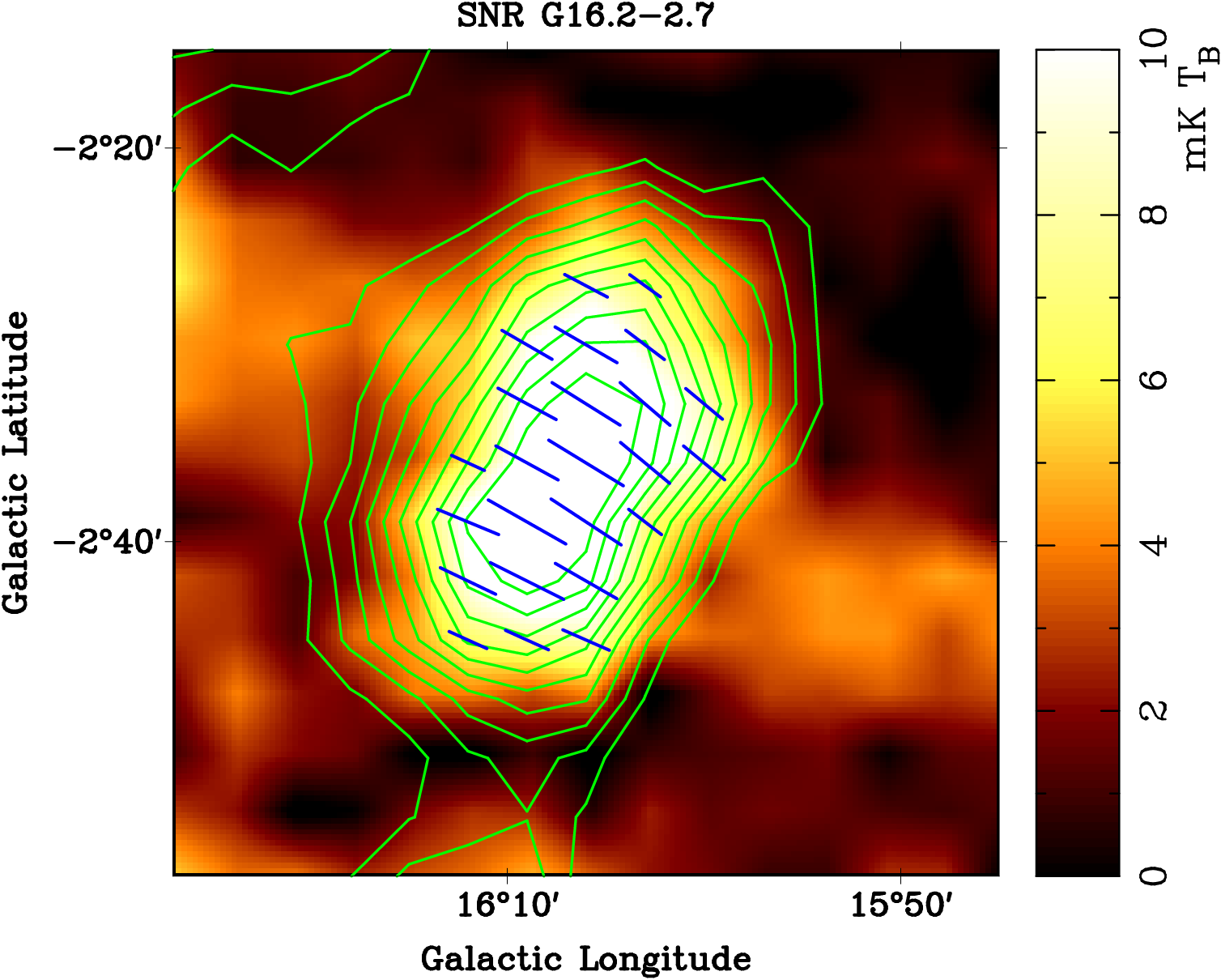}}
\resizebox{0.32\textwidth}{!}{\includegraphics[angle=-90]{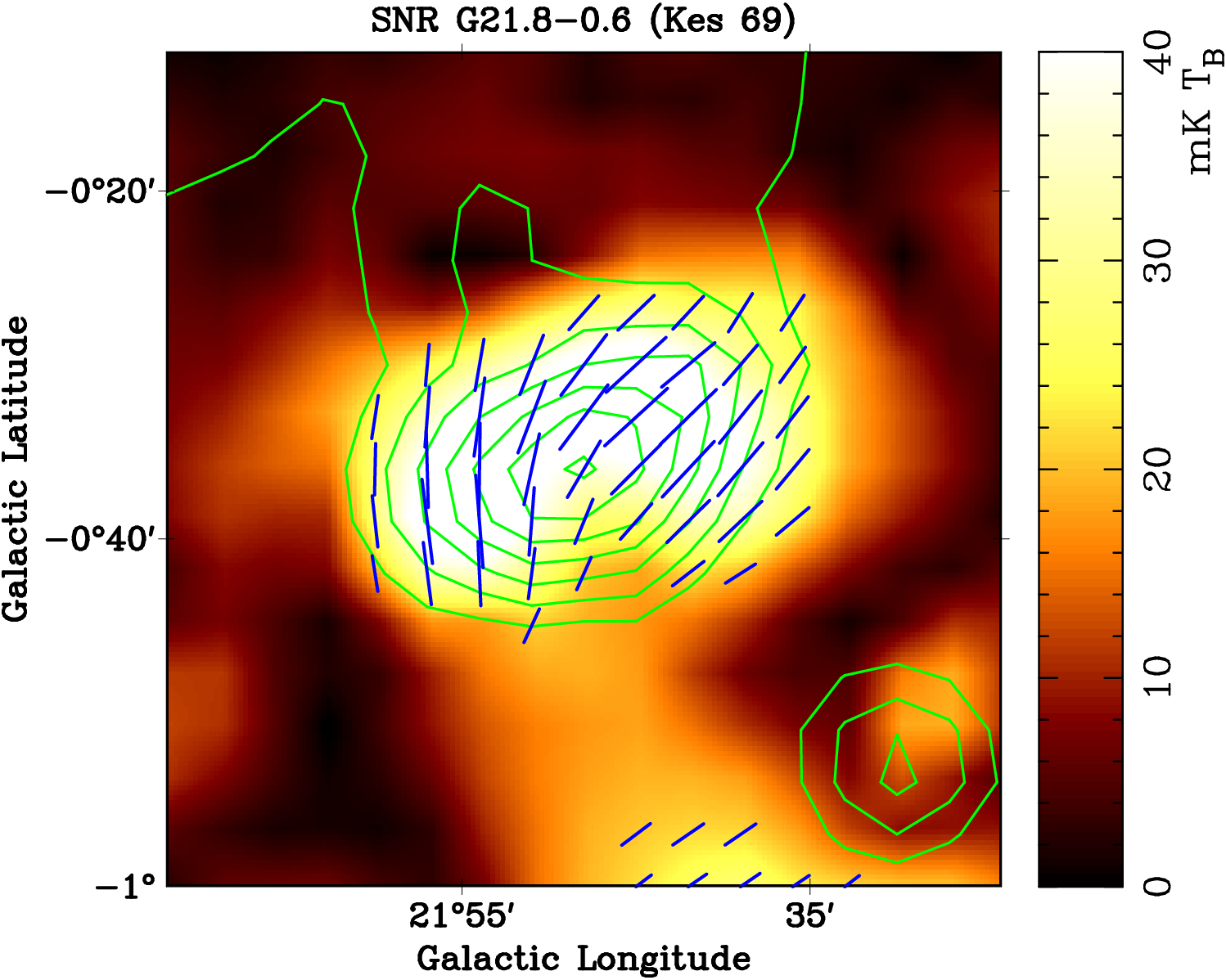}}
\resizebox{0.32\textwidth}{!}{\includegraphics[angle=-90]{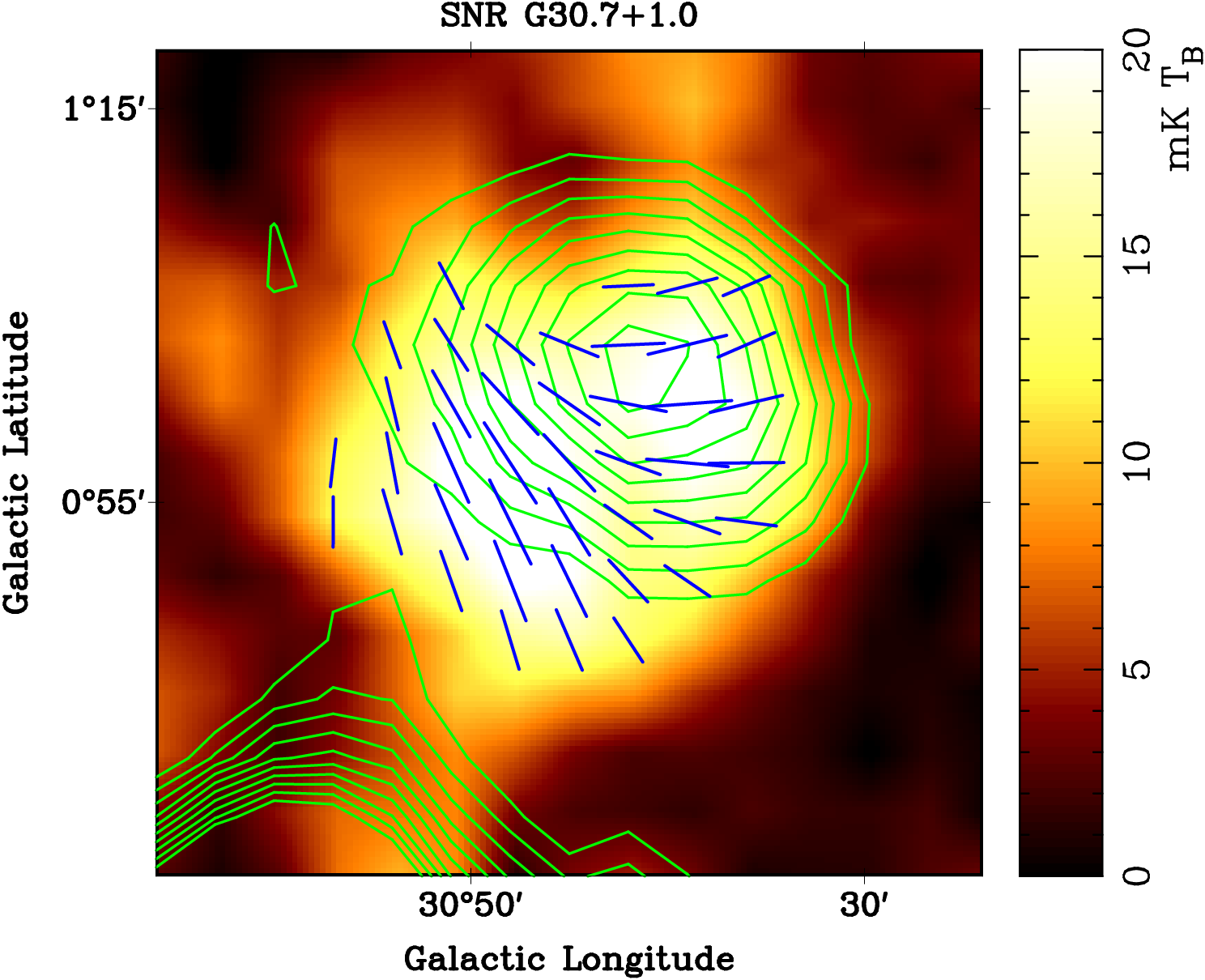}}\\[1mm]
\resizebox{0.32\textwidth}{!}{\includegraphics[angle=-90]{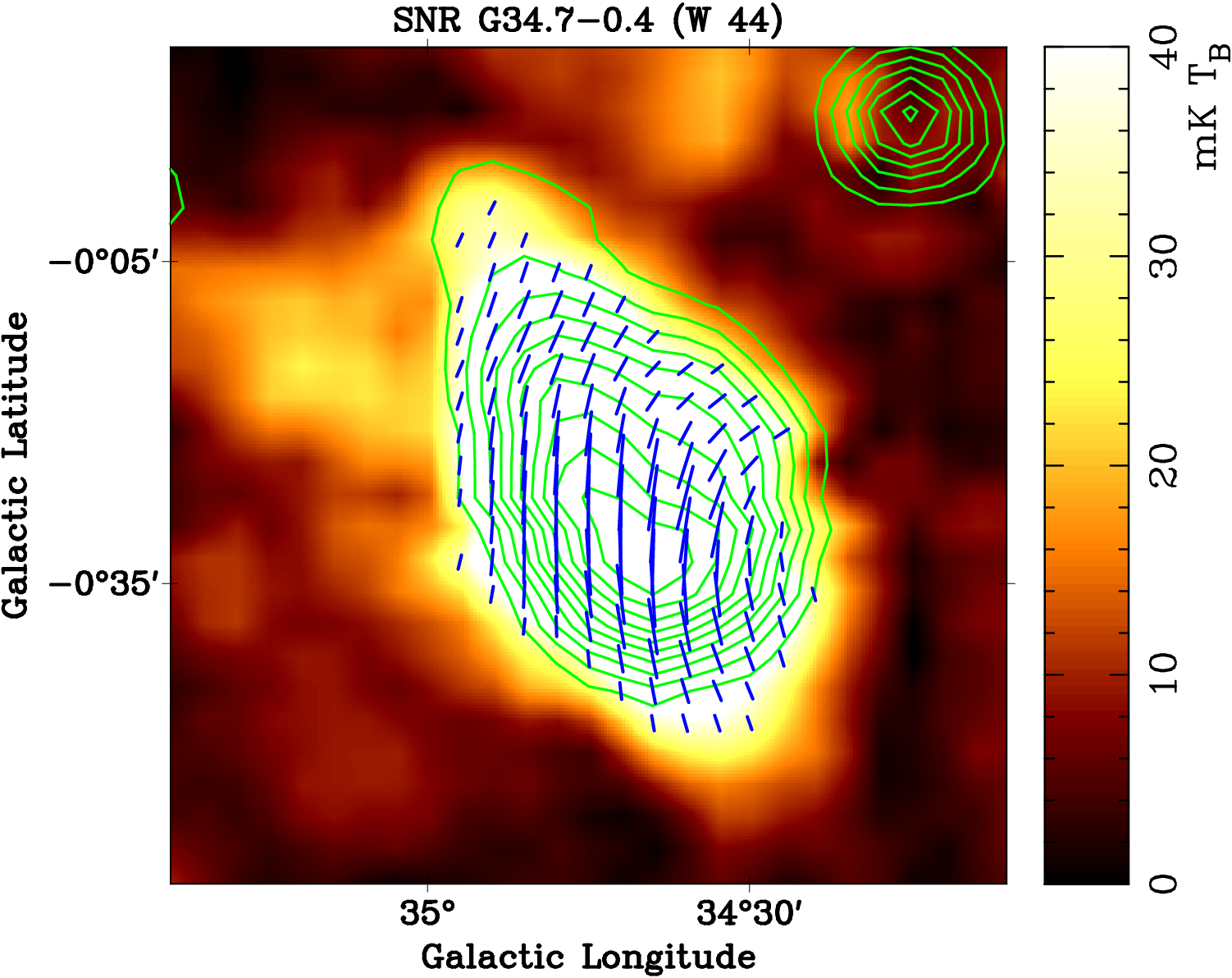}}
\resizebox{0.32\textwidth}{!}{\includegraphics[angle=-90]{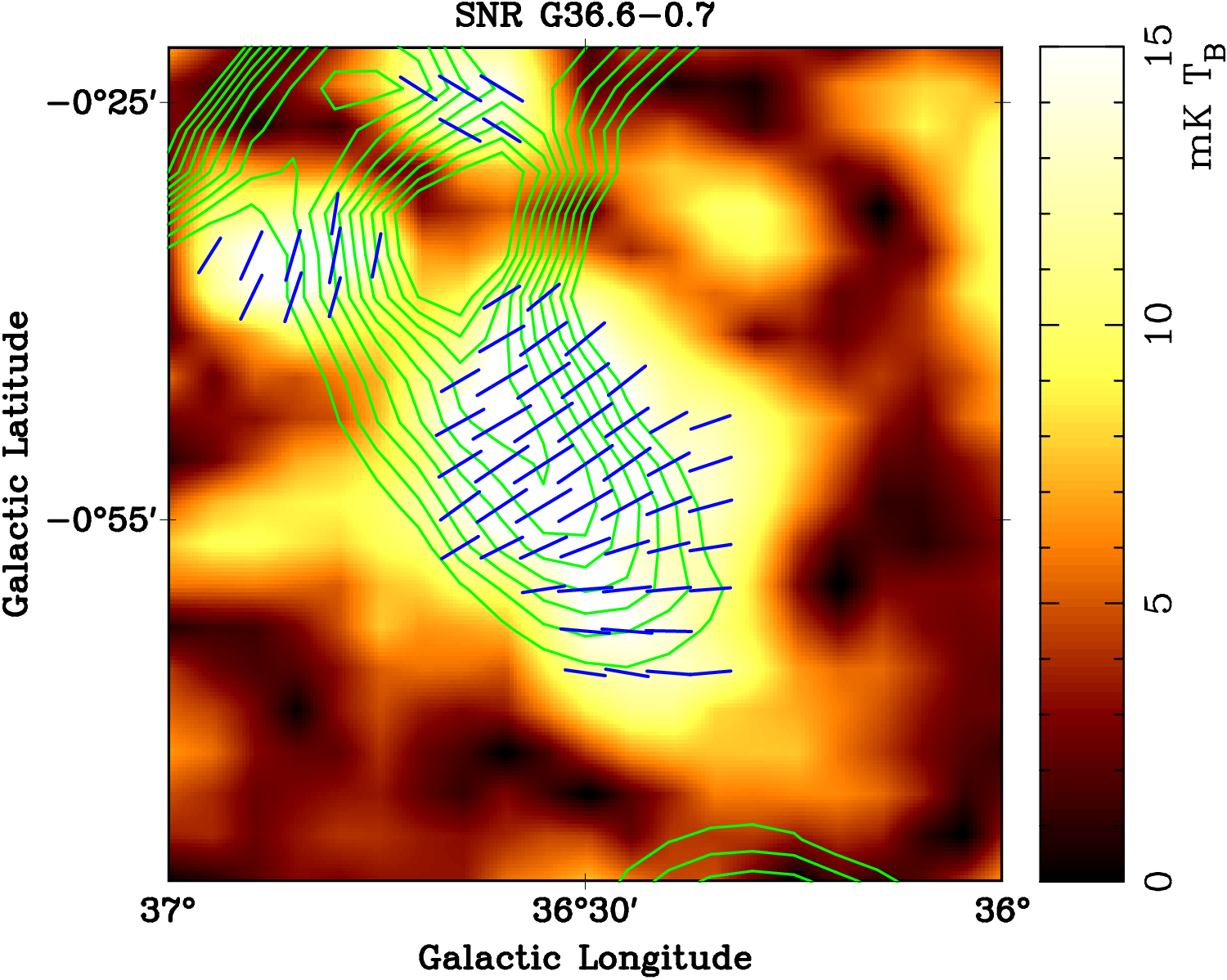}}
\resizebox{0.32\textwidth}{!}{\includegraphics[angle=-90]{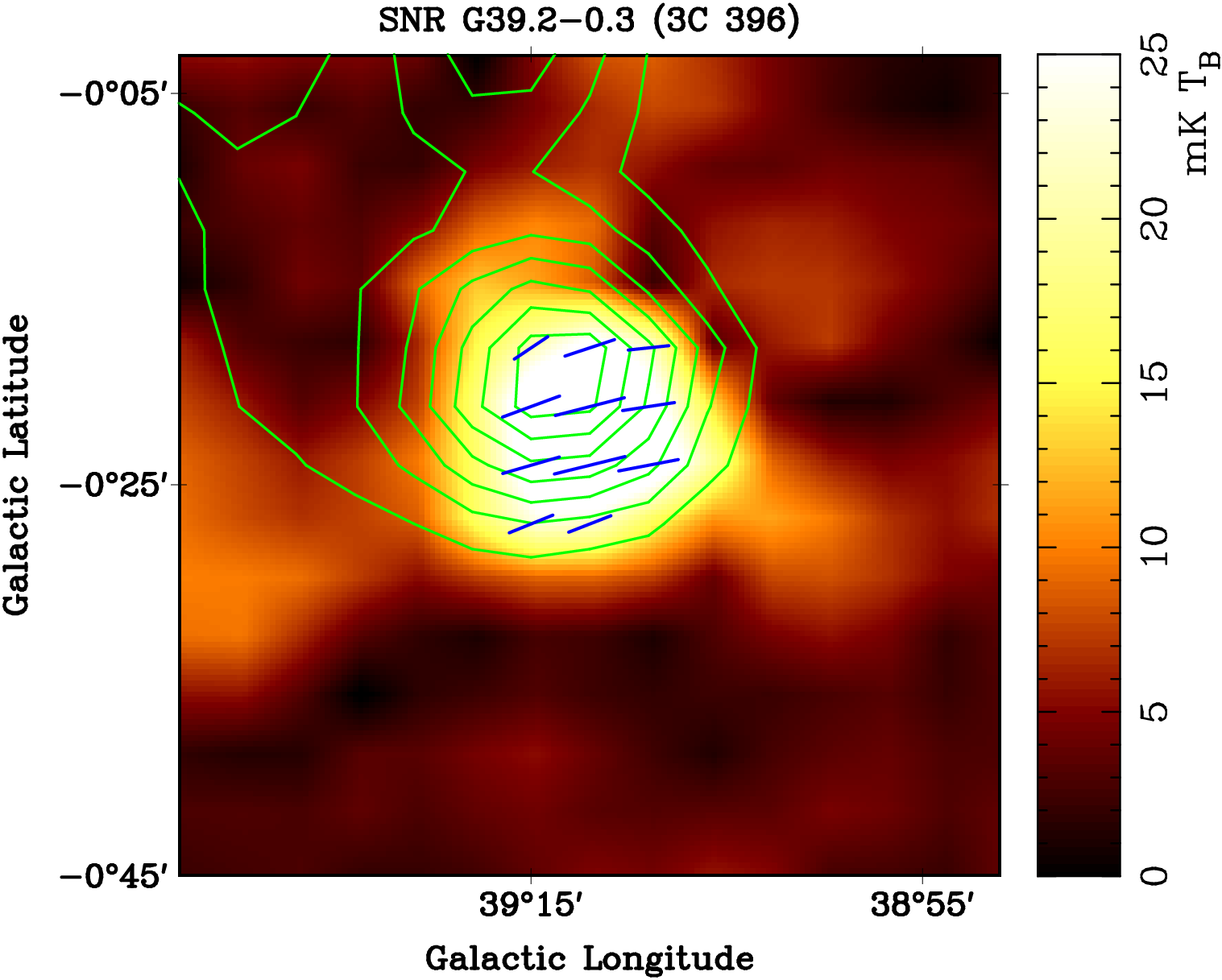}}\\[1mm]
\resizebox{0.32\textwidth}{!}{\includegraphics[angle=-90]{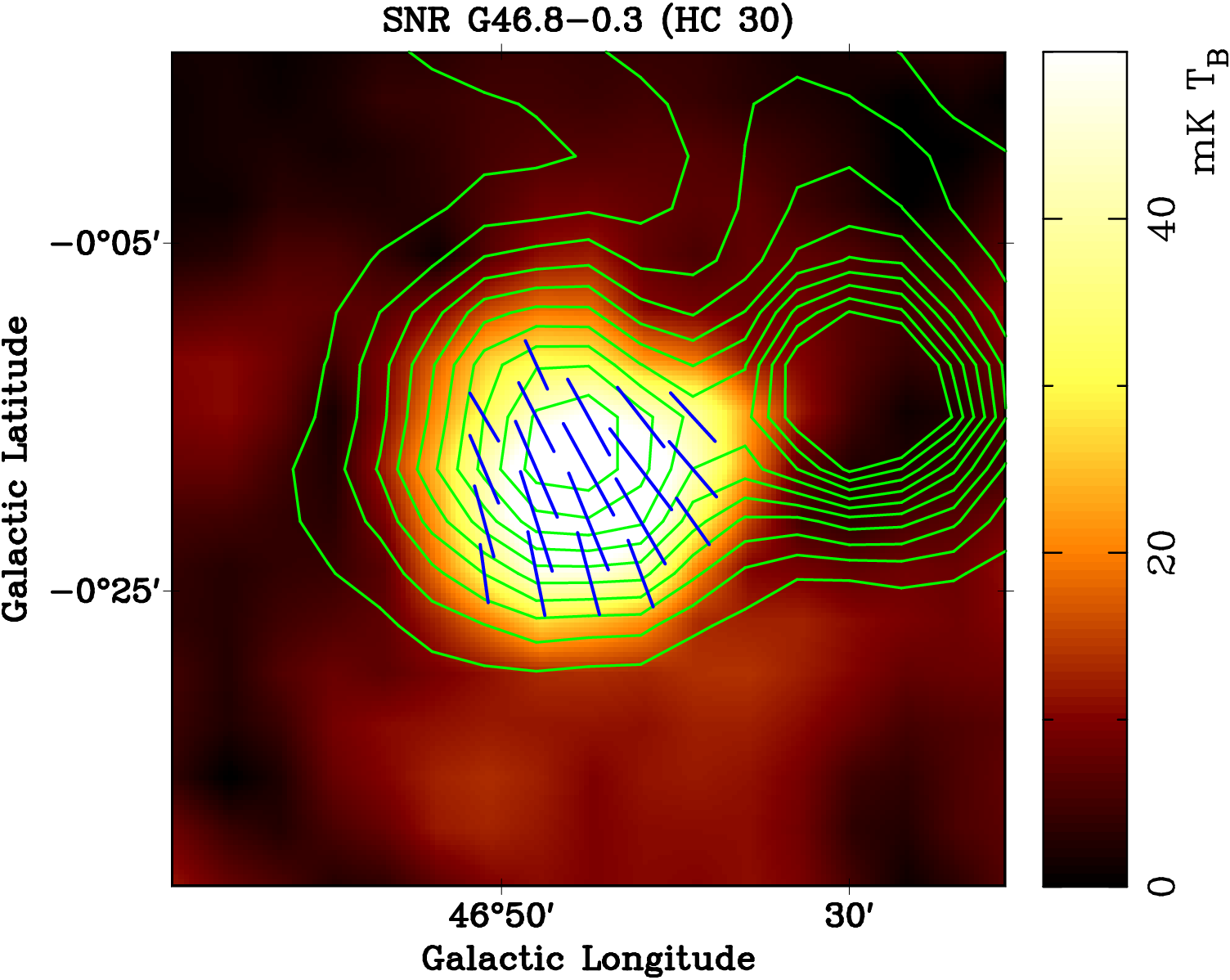}}
\resizebox{0.32\textwidth}{!}{\includegraphics[angle=-90]{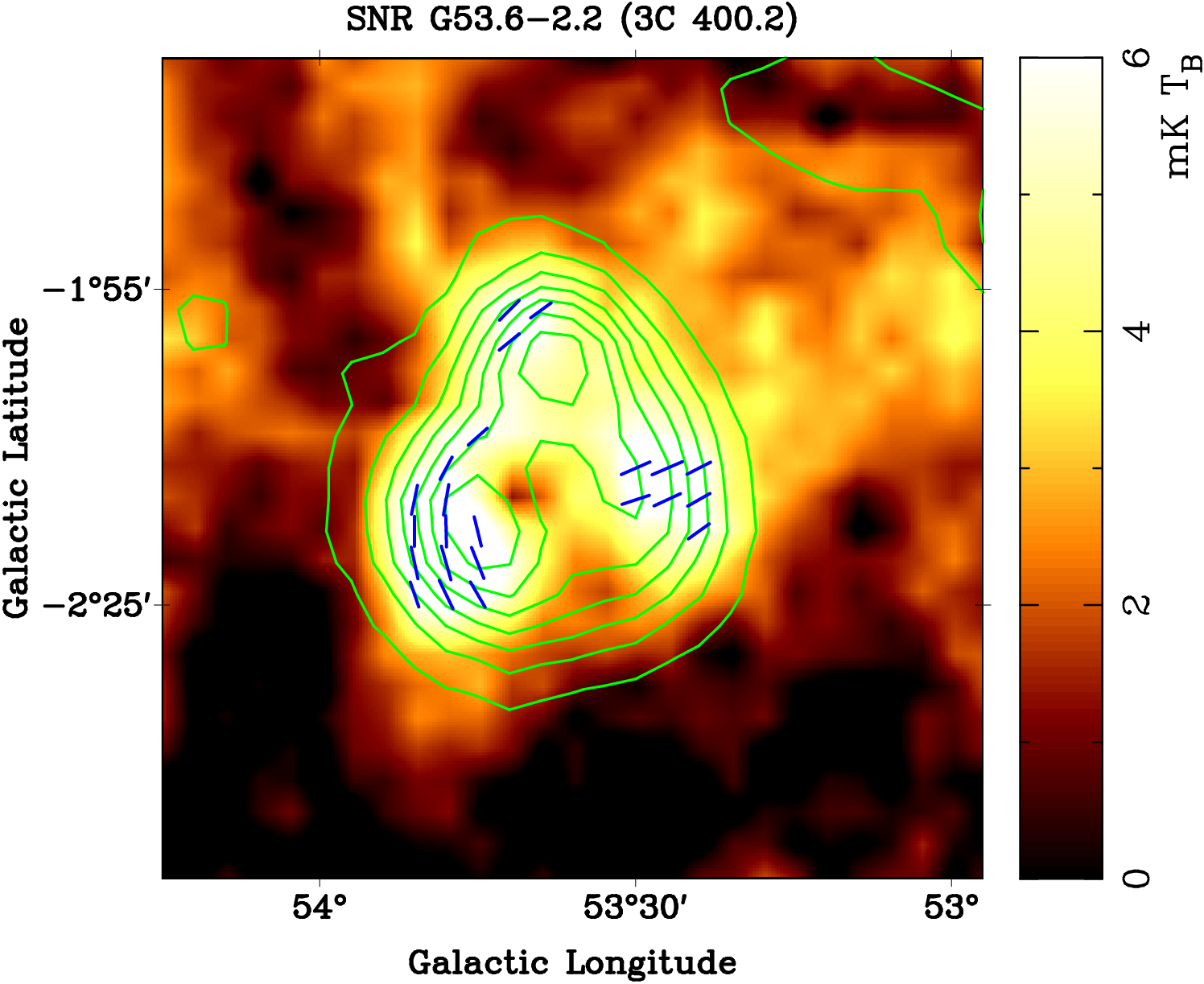}}
\resizebox{0.32\textwidth}{!}{\includegraphics[angle=-90]{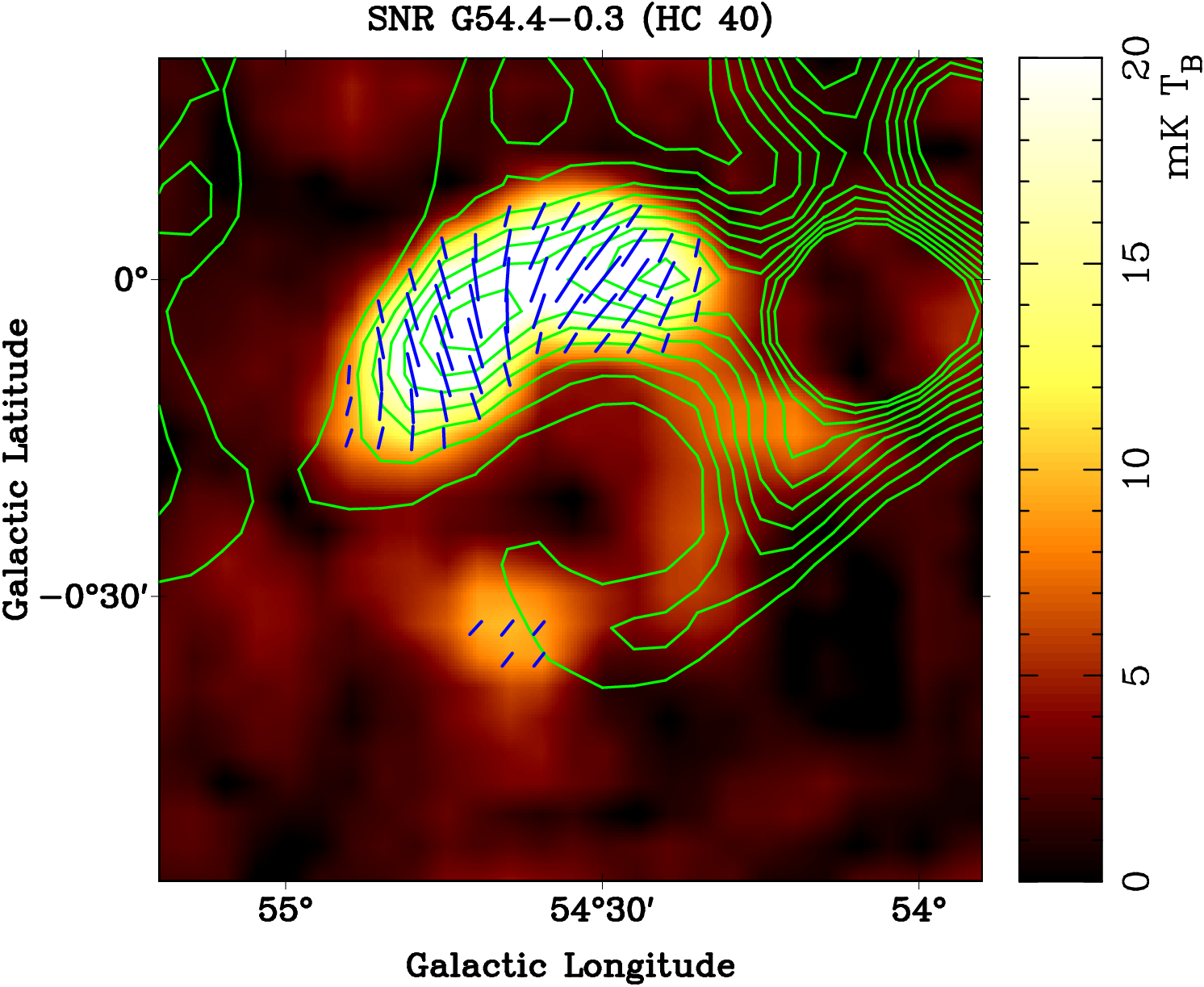}}\\[1mm]
\resizebox{0.32\textwidth}{!}{\includegraphics[angle=-90]{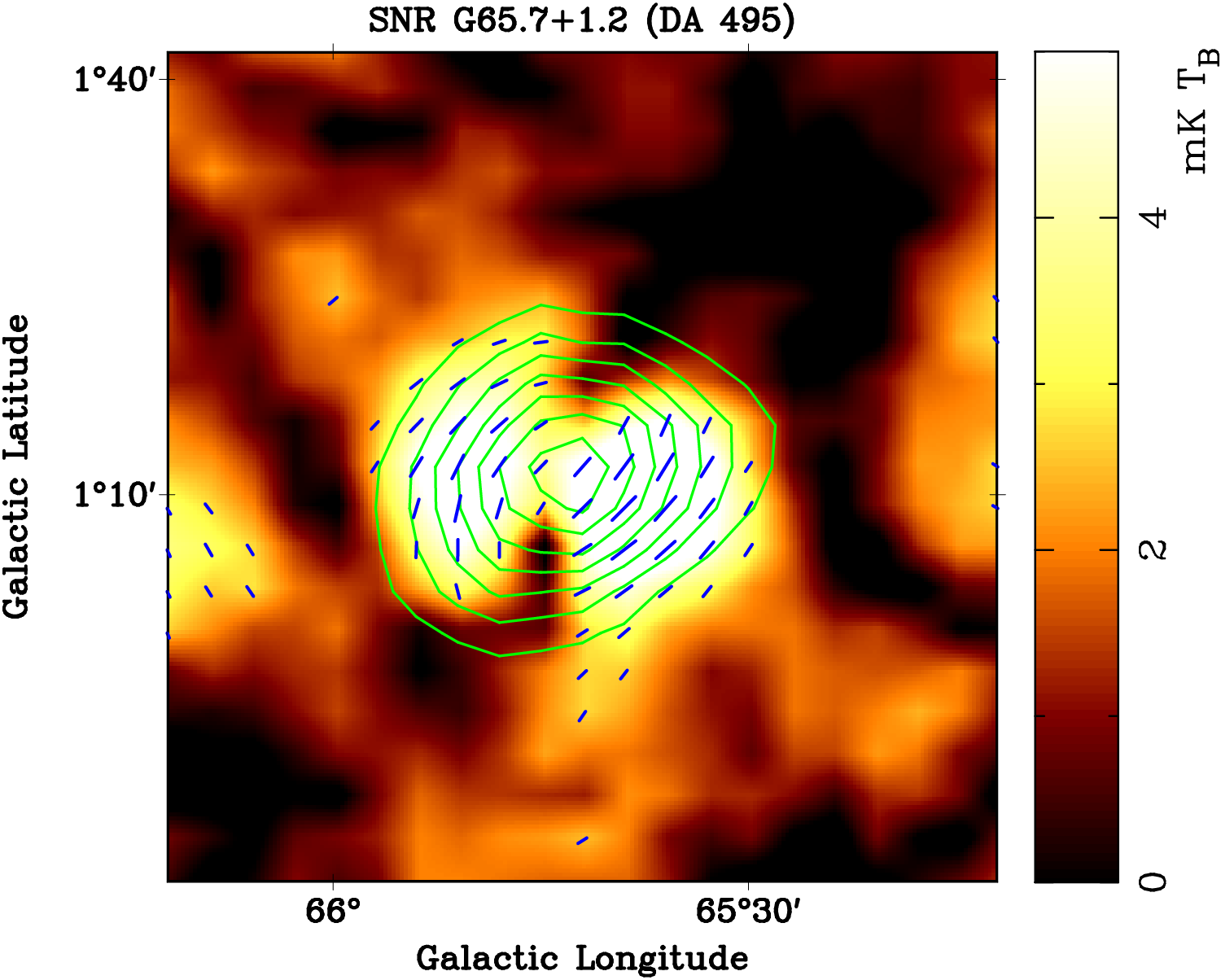}}
\resizebox{0.32\textwidth}{!}{\includegraphics[angle=-90]{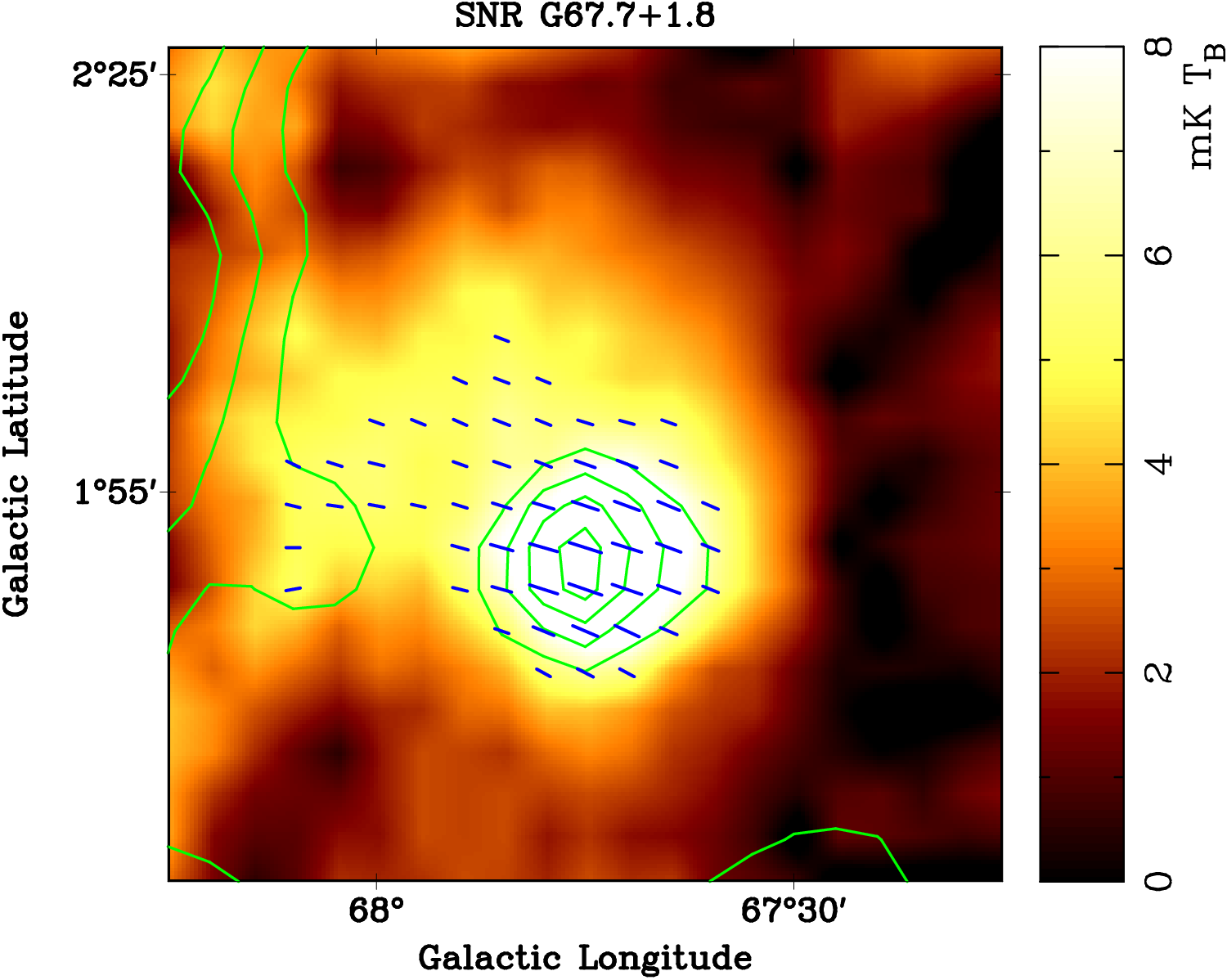}}
\resizebox{0.32\textwidth}{!}{\includegraphics[angle=-90]{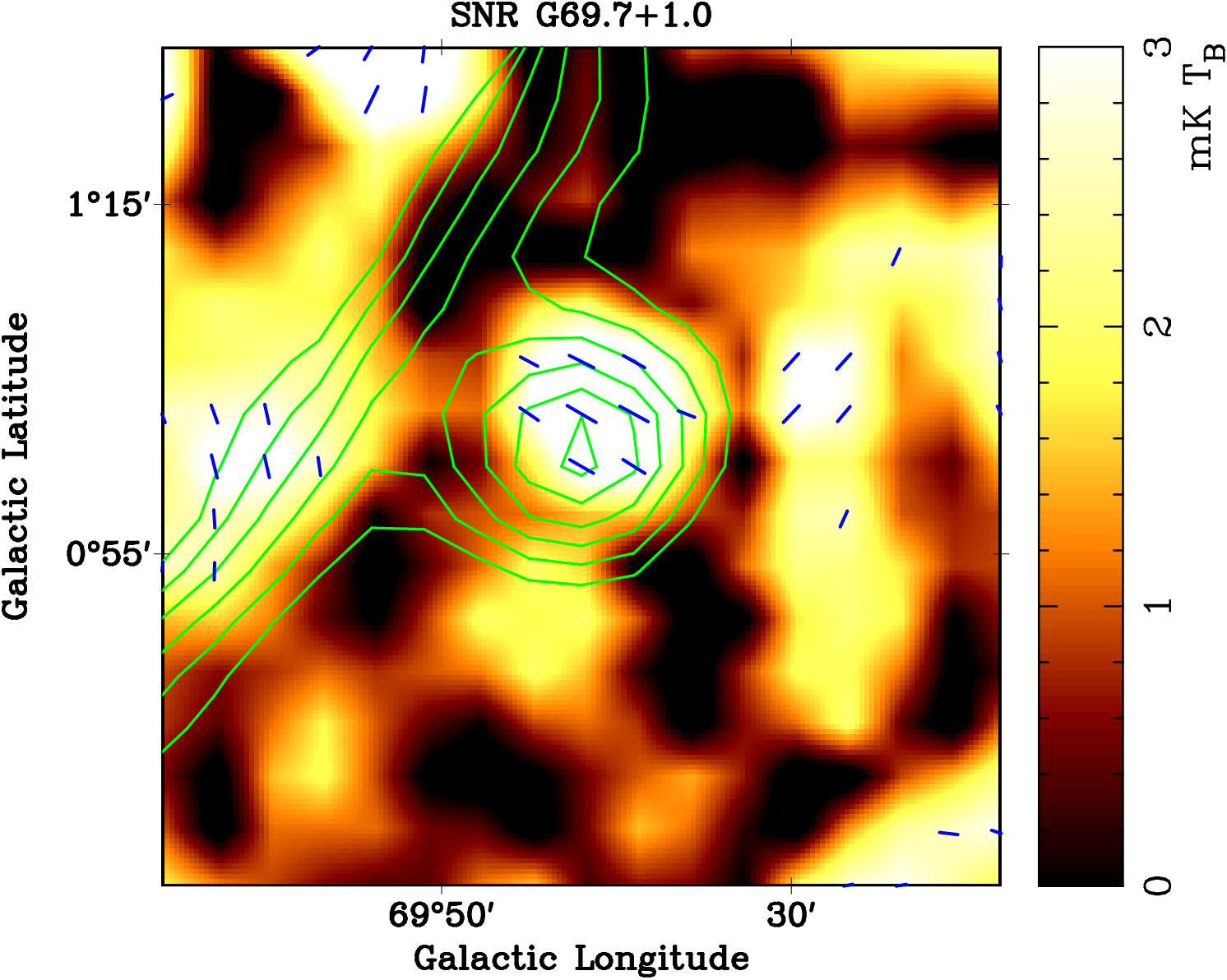}}
\caption{$\lambda$6~cm images of SNRs. Polarized intensity is 
encoded in images, while contours show total intensities. Bars indicate 
B-vectors (observed E-vectors + $90\degr$). The starting levels and the contour 
step intervals (both in mK~$T_{\rm B}$) are for G16.2$-$2.7: 20 and 6, for 
G21.8$-$0.6: 800 and 250, for G30.7$+$1.0: 10 and 15, for G34.7$-$0.4: 500 and 
300, for G36.6$-$0.7: 20 and 10, for G39.2$-$0.3: 250 and 150, for G46.8$-$0.3: 
50 and 50, for G53.6$-$2.2: 6 and 15, for G54.4$-$0.3: 50 and 20, for 
G65.7+1.2: 30 and 15, for G67.7+1.8: 5 and 8, and for G69.7+1.0: 30 and 8.
}
\label{p_snr}
\end{figure*}\addtocounter{figure}{-1}
\begin{figure*}
\resizebox{0.32\textwidth}{!}{\includegraphics[angle=-90]{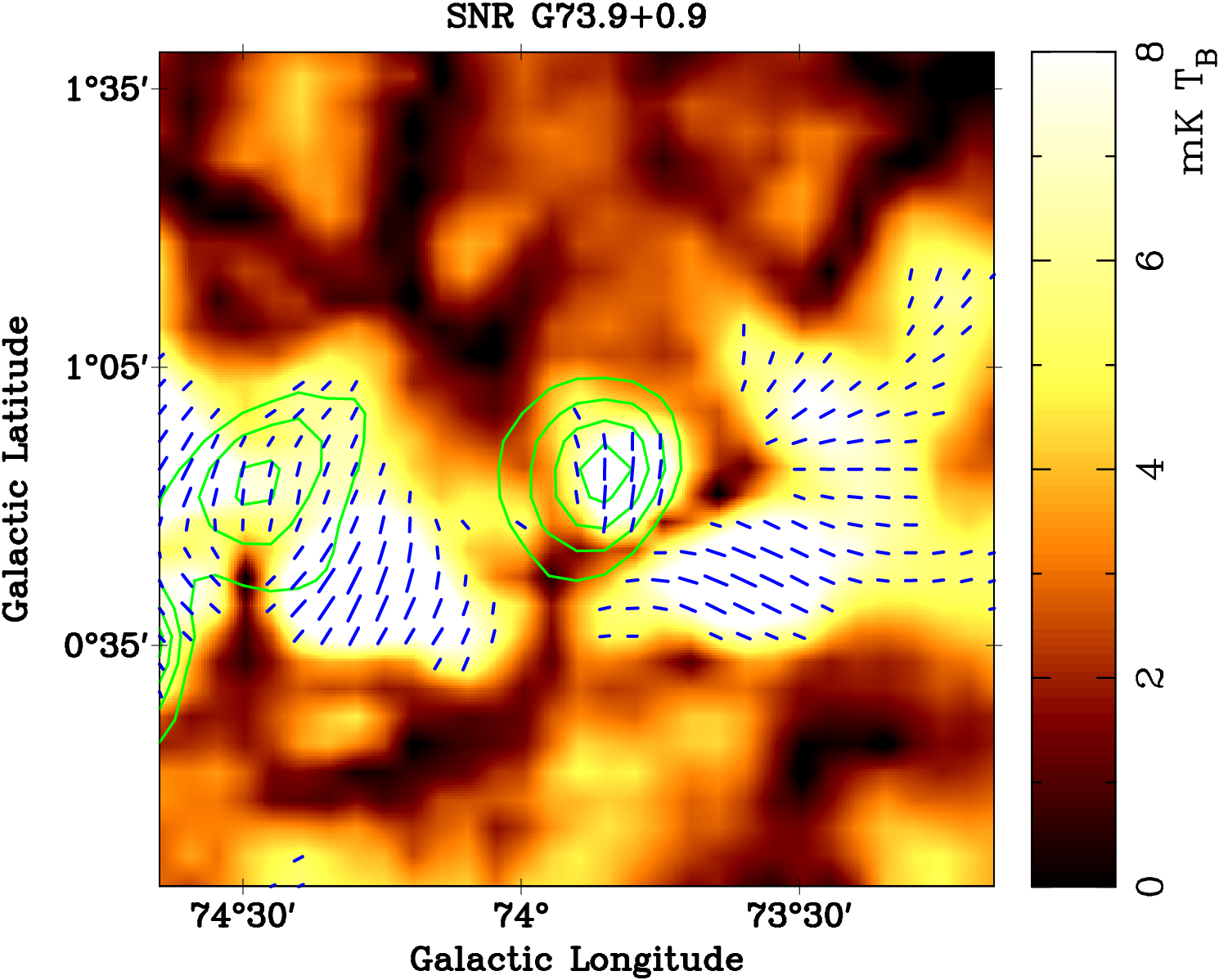}}
\resizebox{0.32\textwidth}{!}{\includegraphics[angle=-90]{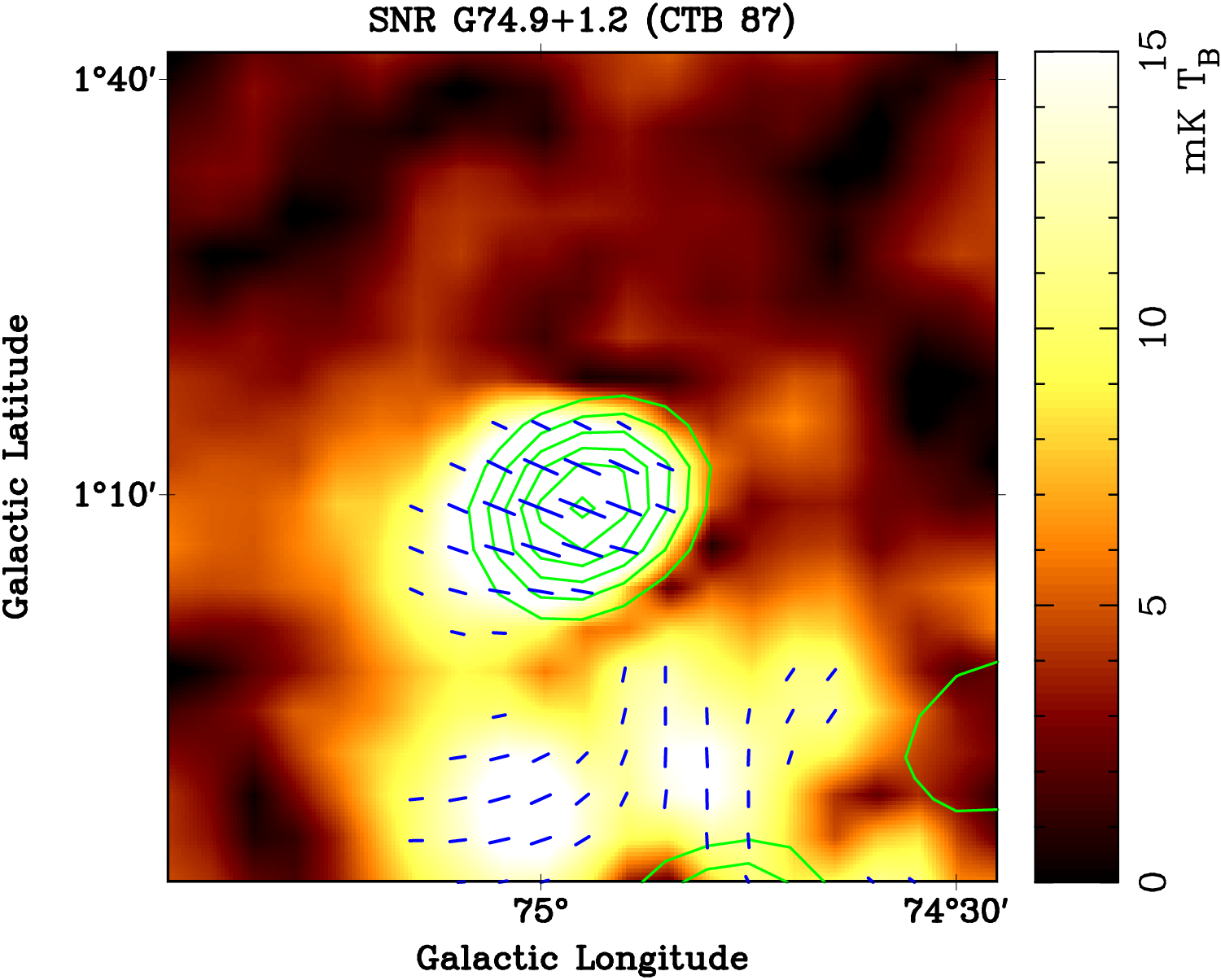}}
\resizebox{0.32\textwidth}{!}{\includegraphics[angle=-90]{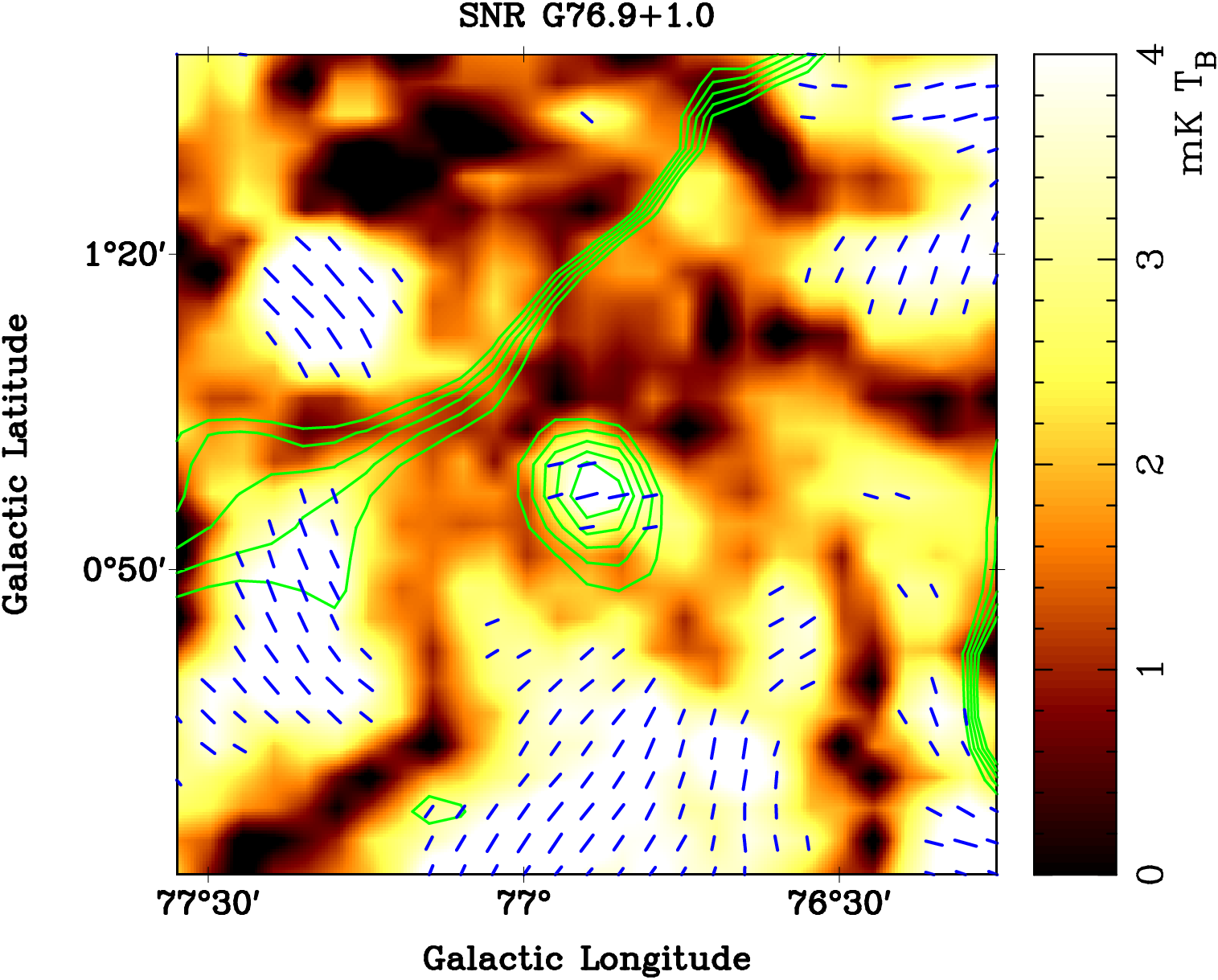}}\\[1mm]
\resizebox{0.32\textwidth}{!}{\includegraphics[angle=-90]{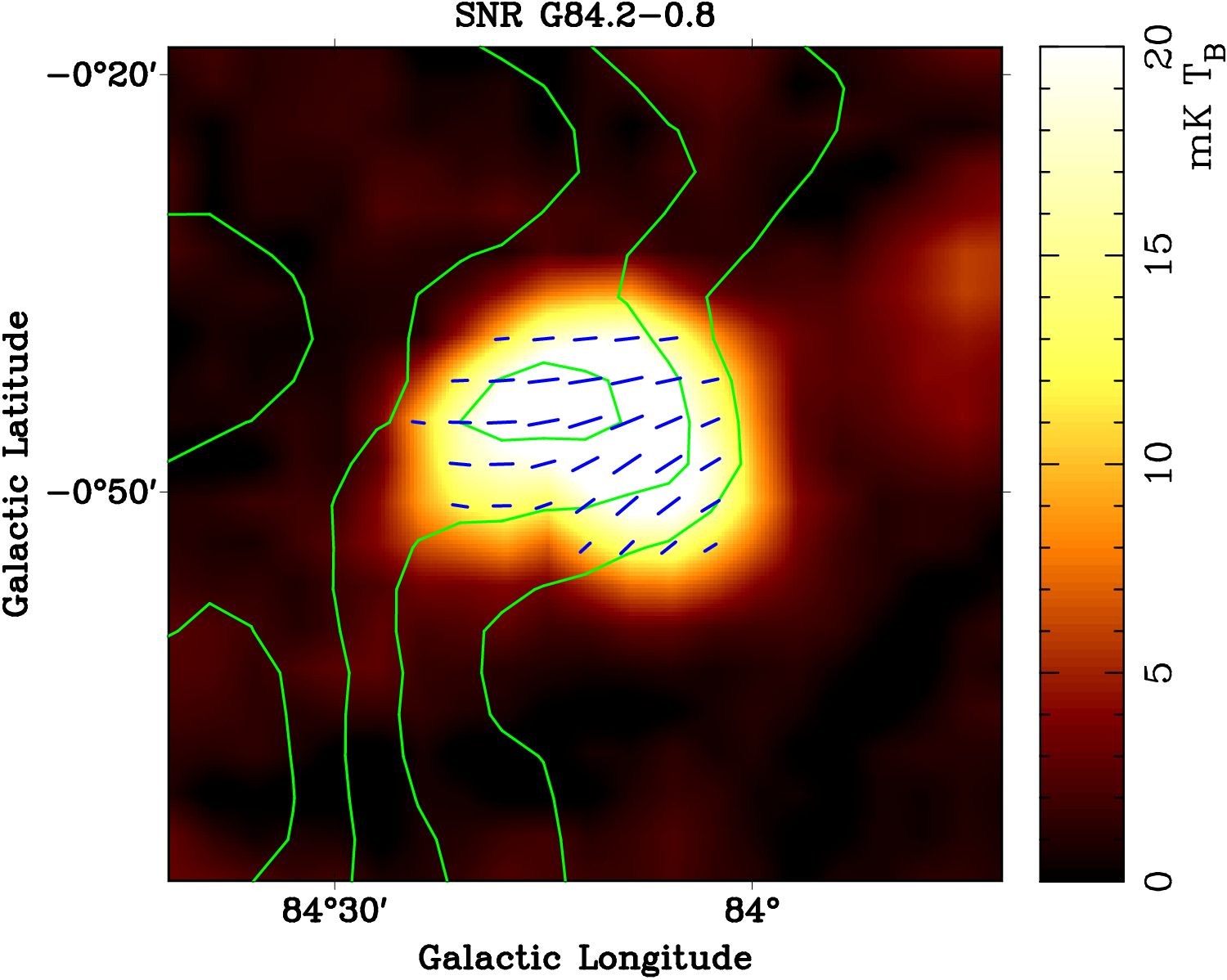}}
\resizebox{0.32\textwidth}{!}{\includegraphics[angle=-90]{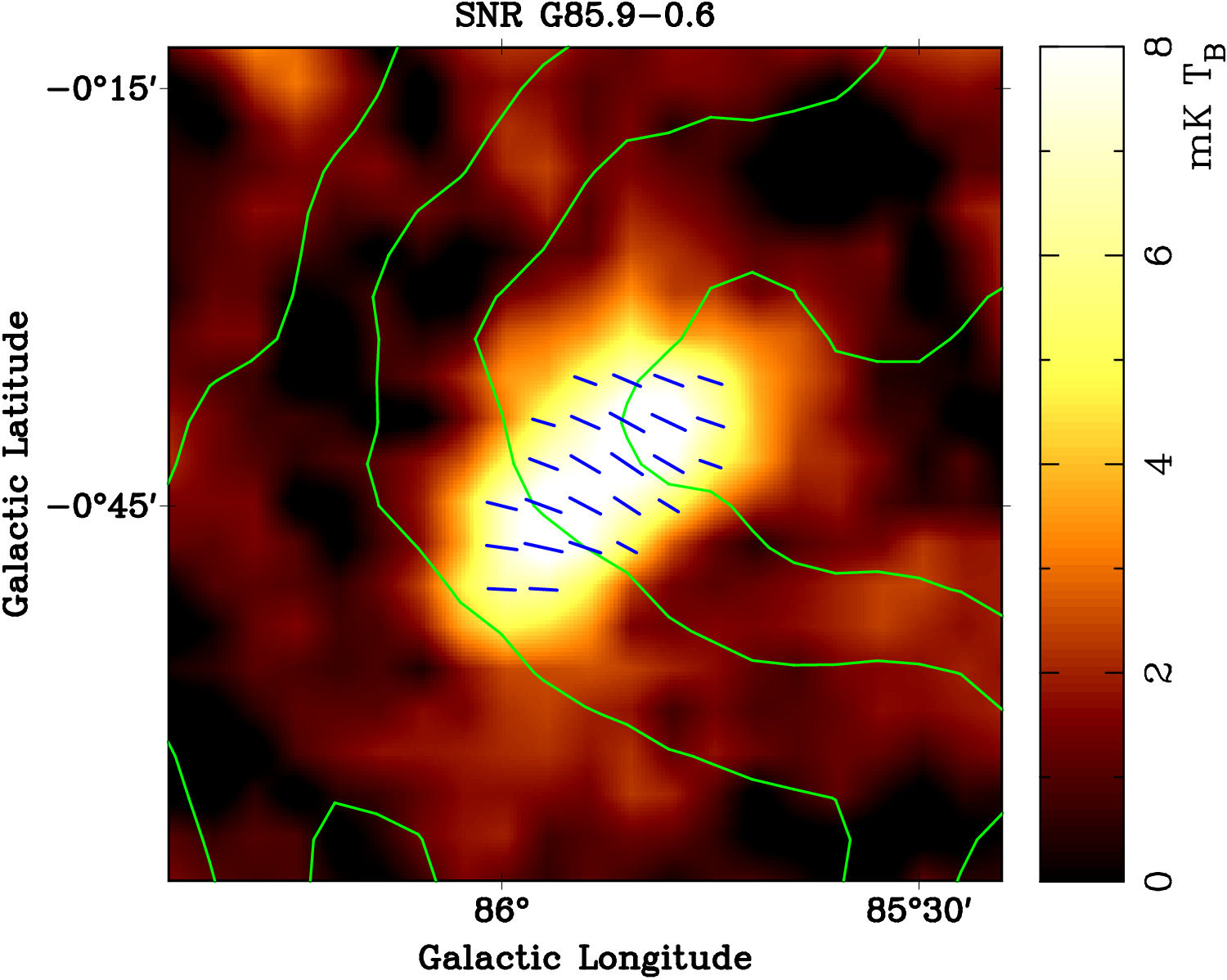}}
\resizebox{0.32\textwidth}{!}{\includegraphics[angle=-90]{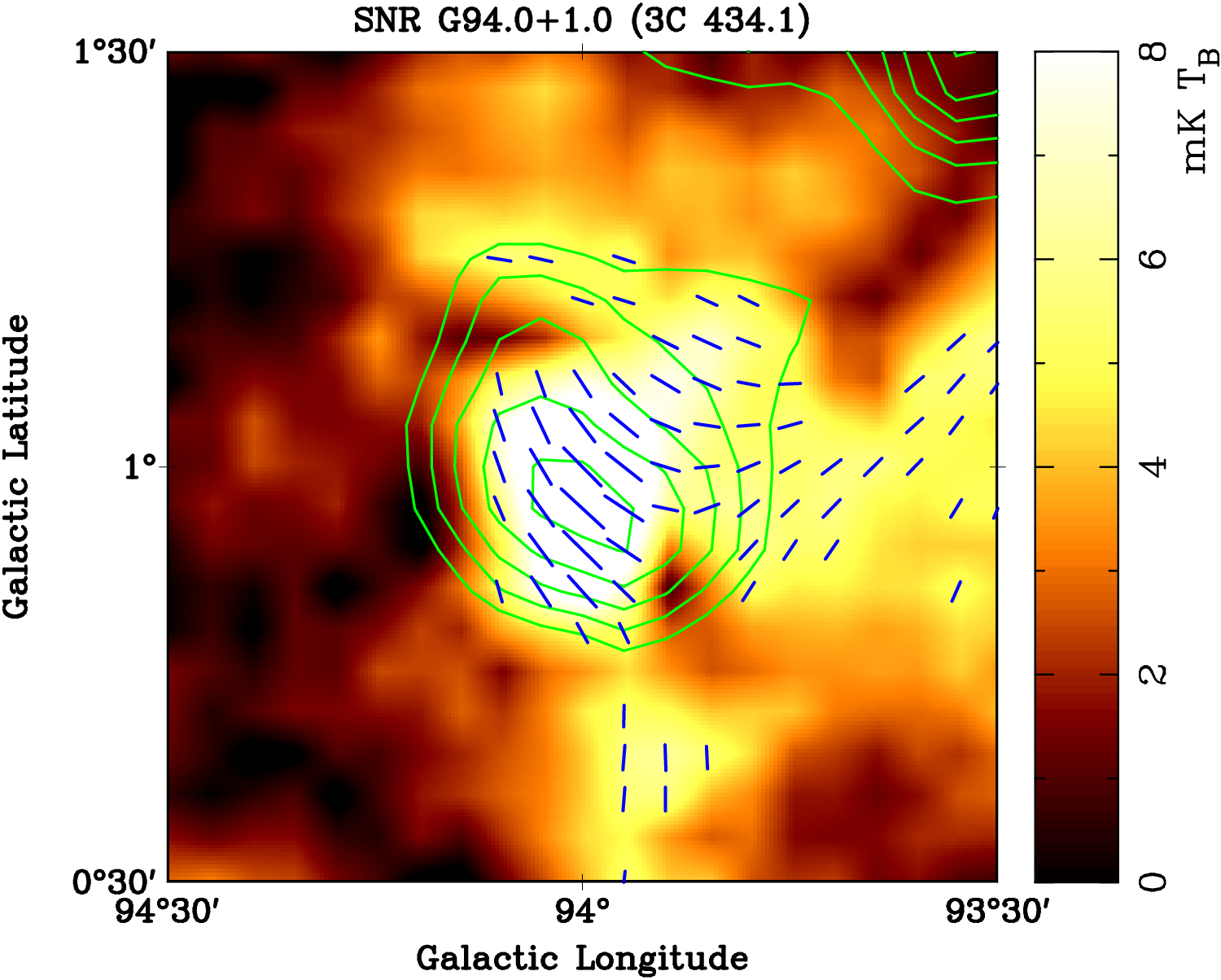}}\\[1mm]
\resizebox{0.32\textwidth}{!}{\includegraphics[angle=-90]{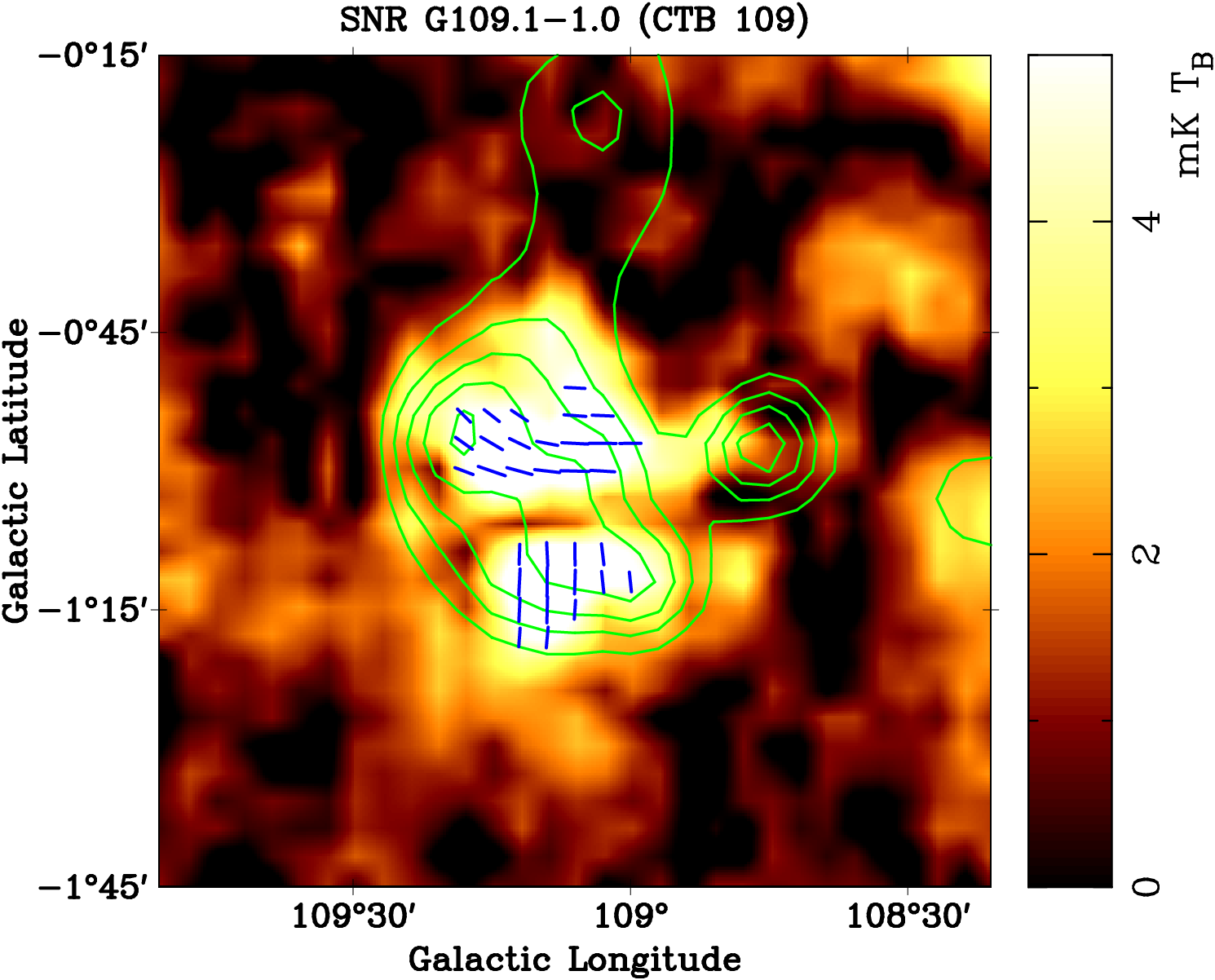}}
\resizebox{0.32\textwidth}{!}{\includegraphics[angle=-90]{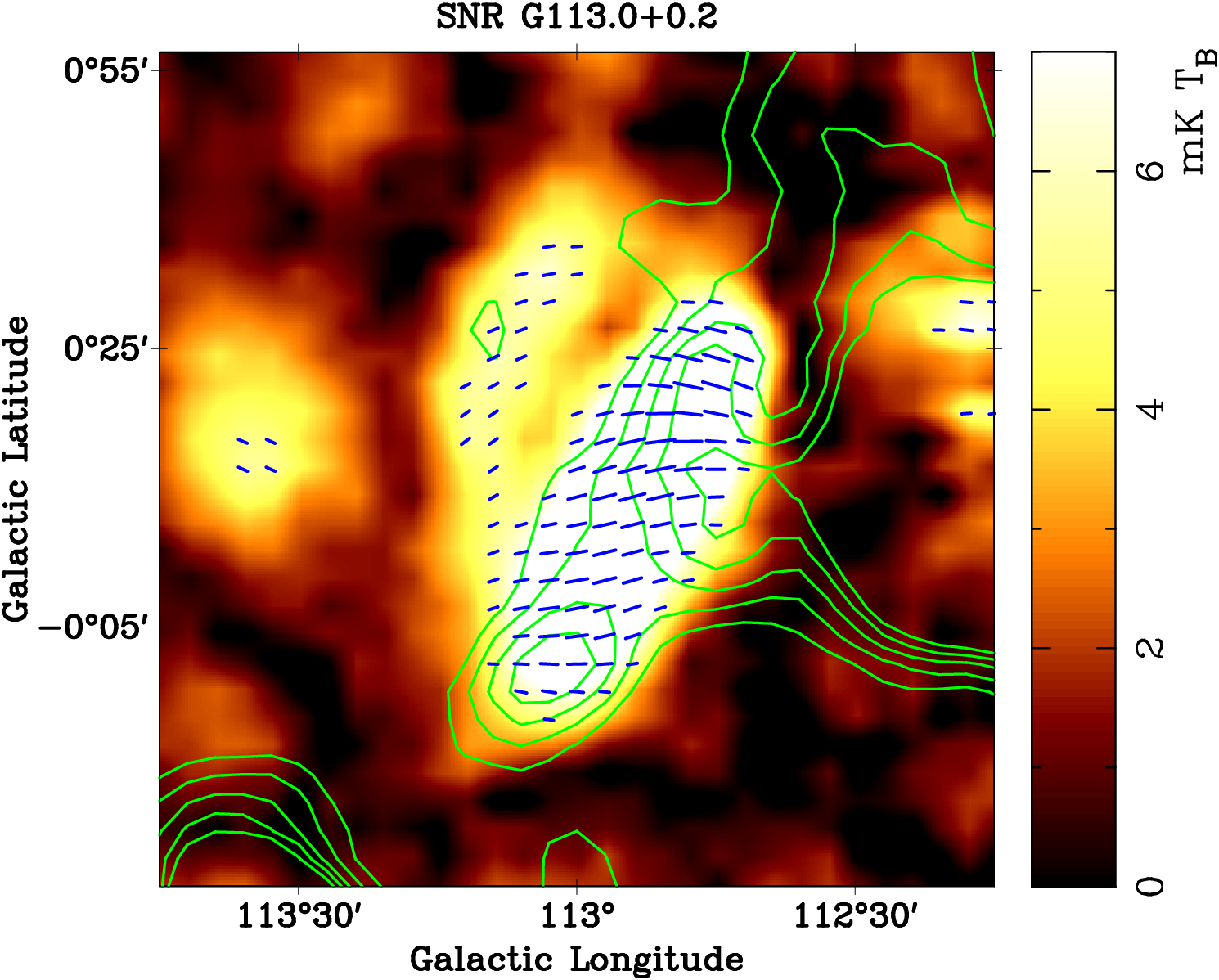}}
\resizebox{0.32\textwidth}{!}{\includegraphics[angle=-90]{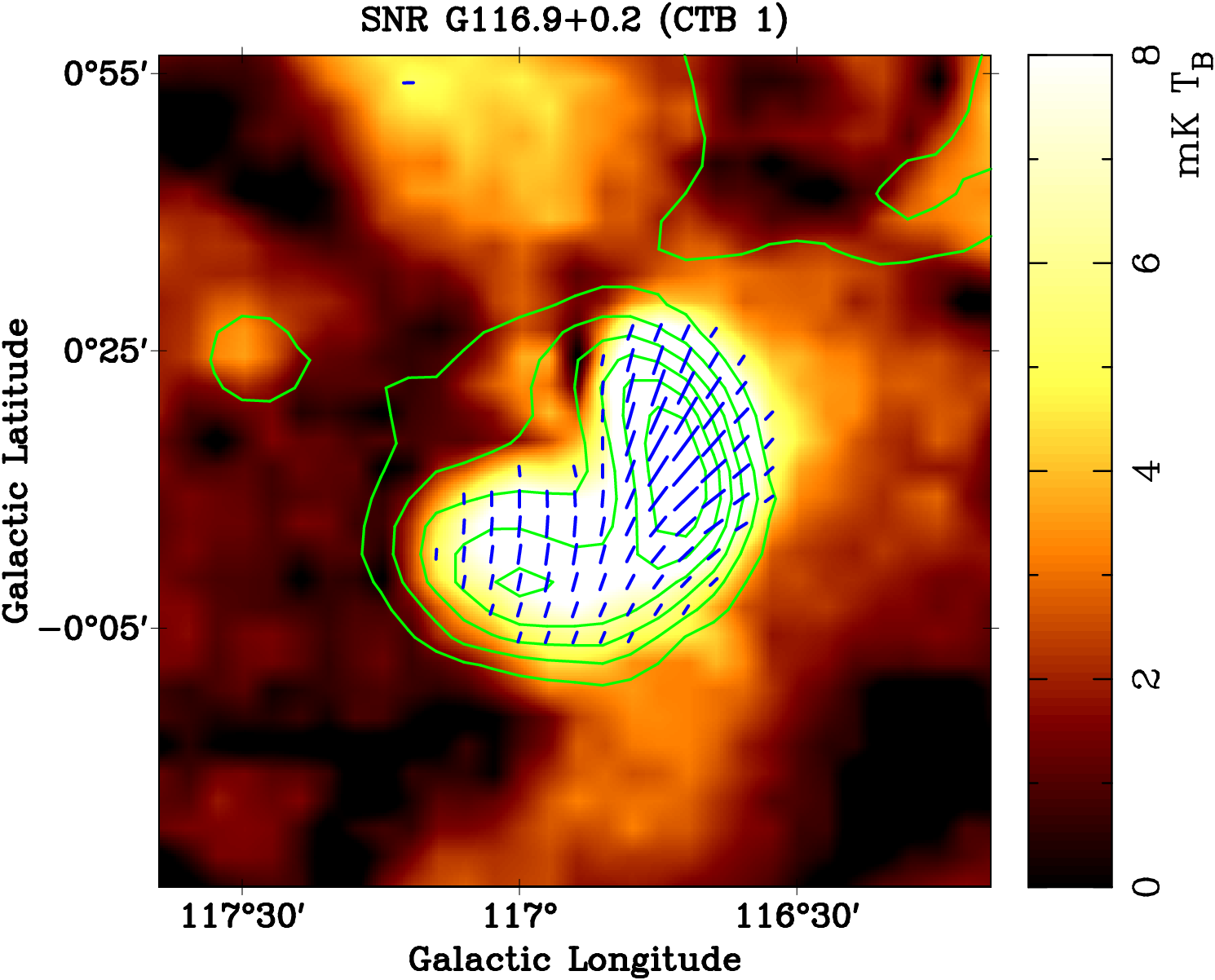}}\\[1mm]
\resizebox{0.32\textwidth}{!}{\includegraphics[angle=-90]{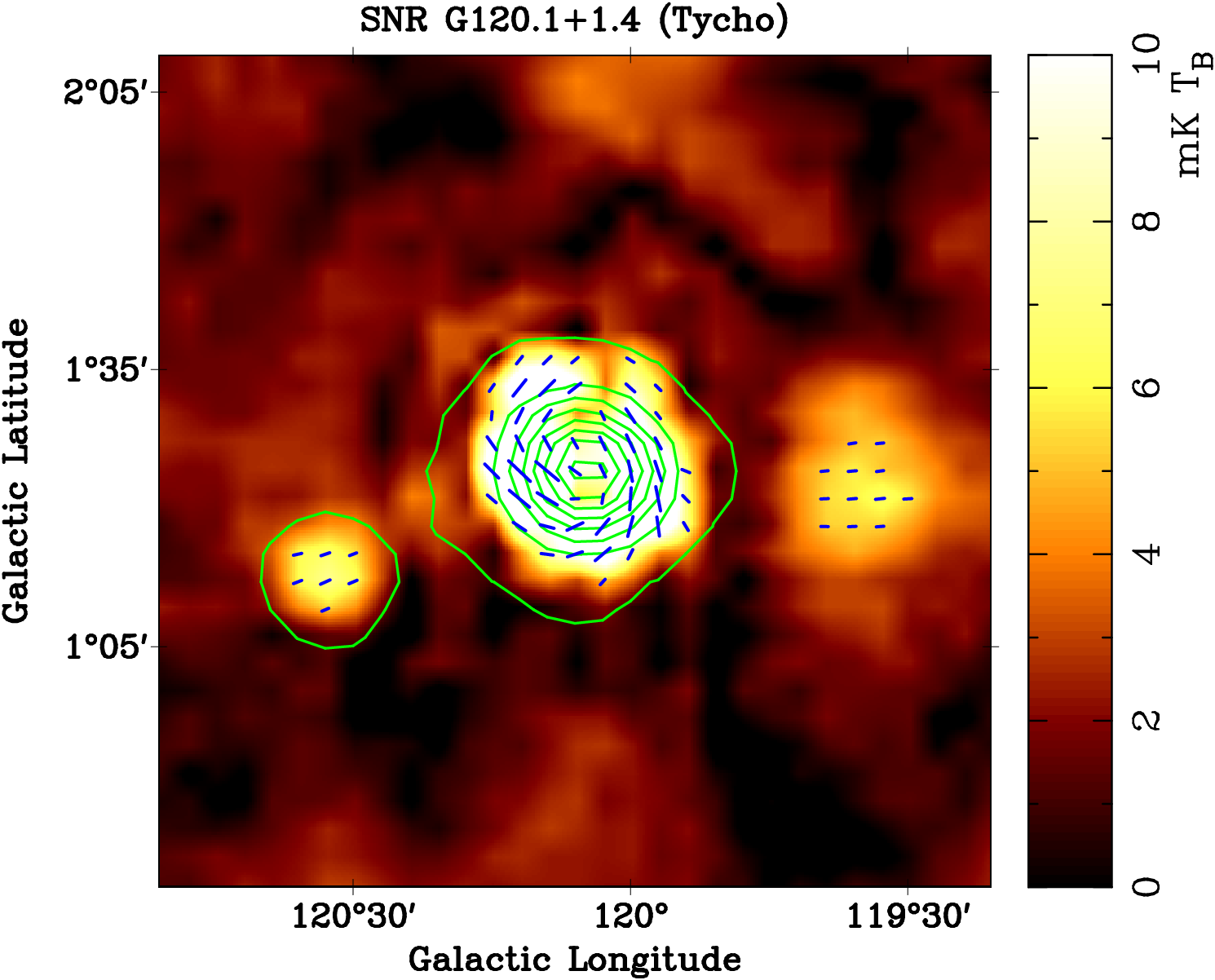}}
\resizebox{0.32\textwidth}{!}{\includegraphics[angle=-90]{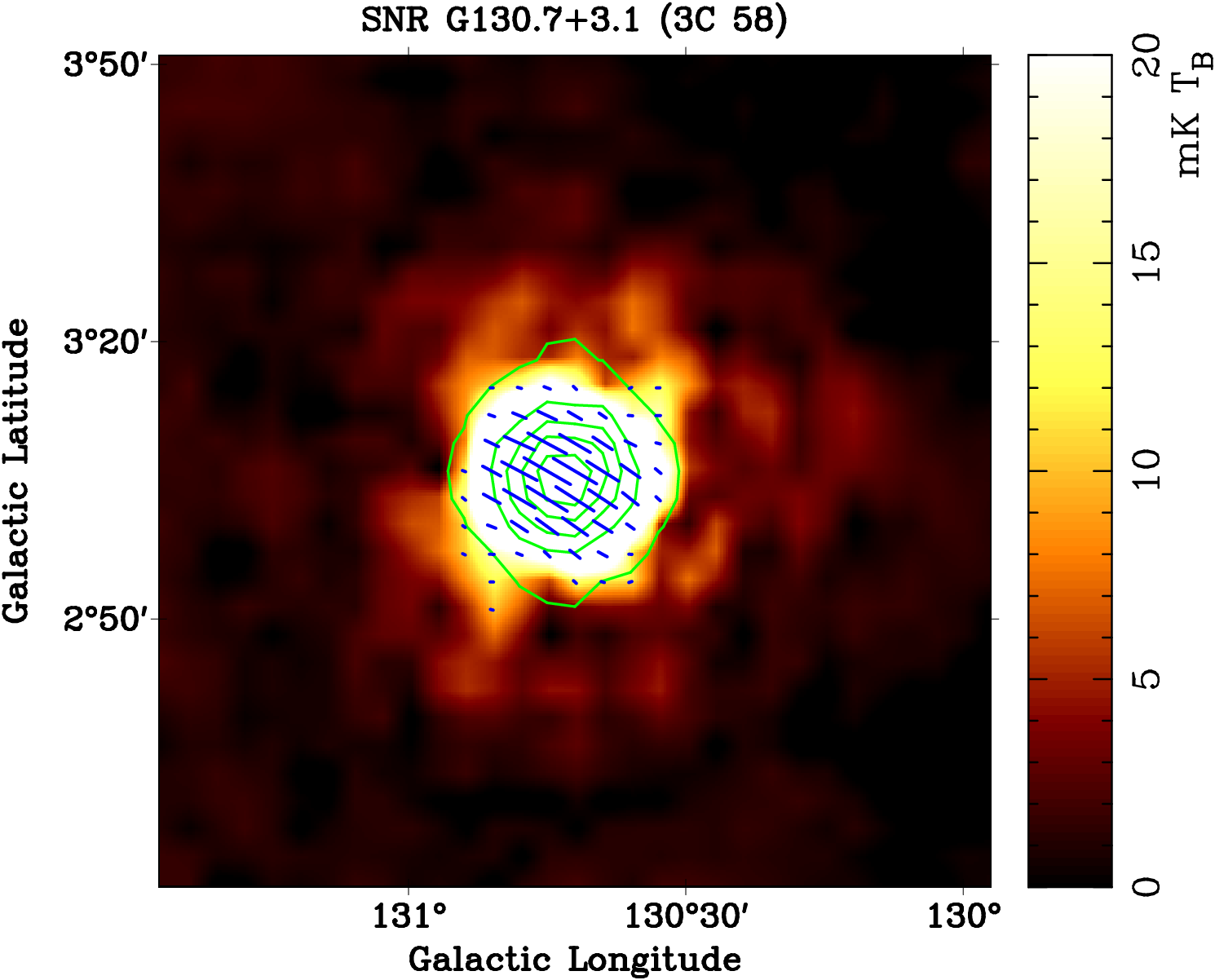}}
\resizebox{0.32\textwidth}{!}{\includegraphics[angle=-90]{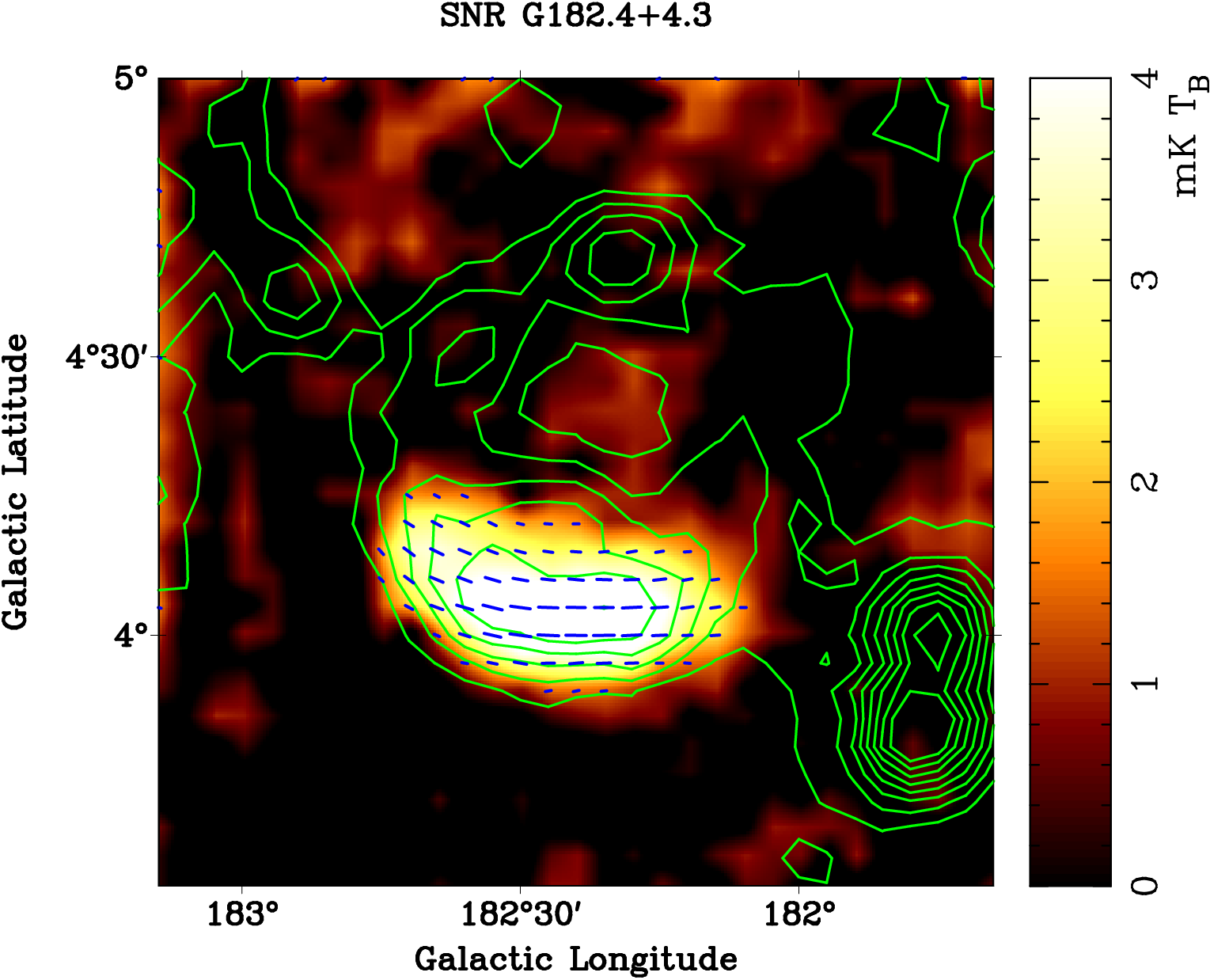}}
\caption{--continued. The starting levels and the contour step intervals (both 
in mK~$T_{\rm B}$) are for G73.9+0.9: 250 and 50, for G74.9+1.2: 250 and 
100, for G76.9+1.0: 65 and 10, for G84.2-0.8: 400 and 100, for G85.9-0.6: 200 
and 100, for G94.0+1.0: 120 and 30, for G109.1$-$1.0: 100 and 50, for  
G113.0+0.2: 30 and 8, for G116.9+0.2: 35 and 15, for G120.1+1.4: 50 and 300,
for G130.7+3.1: 100 and 800, and for G182.4+4.3: 2 and 2.
}
\end{figure*}

Polarized emission at $\lambda$6~cm was detected from 25 small SNRs. Their 
images are shown in Fig.~\ref{p_snr} except for Cas~A. 
The integrated polarization flux density and the average polarization percentage 
were also estimated. For SNRs G67.7+1.8 and G76.9+1.0, intrinsic polarized 
emission could not be separated from the surroundings. The results for the 
remaining SNRs are listed in Table~\ref{ssnr} except for Cas~A. Note that the 
$9\farcm5$ beam mostly covers a significant fraction of the SNRs, which may 
cause beam depolarization. Therefore, the measured values should be 
considered as lower limits. The 
polarization percentages that we derived for some objects are definitely small 
compared to the results obtained with a smaller beam, for example 13\% against 32\% 
for G30.7+0.1 \citep{rfr+86}, 5\% compared to 24\% for DA~495 \citep{klr+08}, 
and 1\% versus 6\% for G73.9+0.9 \citep{rfr+86}. This can be ascribed to beam 
depolarization.

\addtocounter{table}{+1}
\begin{table}
\centering
\caption{Integrated polarized flux density and average percentage polarization 
for 22 SNRs at $\lambda$6~cm observed with a $9\farcm5$ beam.}
\label{ssnr}
\begin{tabular}{cr@{$\pm$}lc}
\hline\hline
SNR & \multicolumn{2}{c}{$S_{PI,\,{\rm 6\,cm}}$ (mJy)} & $S_{PI}/S_I$ (\%)\\
\hline
G16.2$-$2.7  &  150 &  14  &  12 \\
G21.8$-$0.6  &  790 &  50  &   3 \\
G30.7+1.0    &  370 &  22  &  13 \\
G34.7$-$0.4  & 5000 & 255  &   4 \\
G36.6$-$0.7  &  365 &  28  &   - \\
G39.2$-$0.3  &  240 &  15  &   3 \\
G46.8$-$0.3  &  595 &  32  &   8 \\
G53.6$-$2.2  &  220 &  15  &   6 \\
G54.4$-$0.3  &  795 &  41  &   -\\
G65.7+1.2    &   87 &  20  &   5\\
G69.7+1.0    &   27 &   5  &   4\\
G73.9+0.9    &   70 &   7  &   1\\
G74.9+1.2    &  334 &  30  &   5\\
G84.2$-$0.8  &  464 &  45  &   -\\
G85.9$-$0.6  &  130 &  10  &   -\\
G94.0+1.0    &  170 &  17  &   3\\
G109.1$-$1.0 &  212 &  20  &   2\\
G113.0+0.2   &  491 &  50  &  27\\
G116.9+0.2   &  503 &  50  &  14\\
G120.1+1.4   &  275 &  30  &   1\\
G130.7+3.1   & 1924 & 200  &   6\\
G180.2+4.3   &  139 &  14  &  53\\
\hline
\end{tabular}
\tablefoot{In cases where the total integrated intensity could not be measured
because of confusion no average percentage polarization could be given.}
\end{table}

It is for the first time that polarized emission was detected from SNRs 
G16.2$-$2.7, G69.7+1.0, G84.2$-$0.8, and G85.9$-$0.6. SNRs 
G84.2$-$0.8 and G85.9$-$0.6 are probably located behind the \ion{H}{II} complex 
W~80 \citep{klfl01,ulgk03}. It is therefore difficult to observe their 
polarization at lower frequencies such as 1.4~GHz. The detection of 
polarization from these objects finally confirms them as SNRs.

In case there were early polarization measurements with single dishes at 
$\lambda$6~cm for the SNRs, such as G21.8$-$0.6 by \citet{kvh74}, G30.7+1.0 and 
G73.9+0.9 by \citet{rfr+86}, G36.6$-$0.7 by \citet{frr+87}, G65.7+1.2 (DA 495) 
by \citet{klr+08}, and G182.4+4.3 by \citet{kfr98}, their polarization 
morphologies are quite similar compared to what we obtained, for example the 
bipolar magnetic field structure of DA 495. 

RMs for some SNRs were obtained in the case that polarization observations at other 
wavelengths were available. To illustrate this we take 
the sources G34.7$-$0.4 (W44) and G116.9+0.2 (CTB 1). 
W~44 shows a complex pattern in its polarization angle 
distribution. We retrieved polarization data from the Effelsberg $\lambda$11~cm 
polarization survey \citep{jfr87,drrf99} and compared them with the present 
$\lambda$6~cm map. The RM is estimated to be about $-55$~rad~m$^{-2}$ 
and $-105$~rad~m$^{-2}$ towards the southern ($b<-0\fdg5$) and northern 
($b>-0\fdg5$) parts of W~44, respectively.
 The pulsar PSR J1856$+$0113 ($l=34\fdg56$, $b=-0\fdg5$) associated 
with W~44 has a RM of $-140\pm30$ rad~m$^{-2}$ \citep{hml+06}, which is roughly 
consistent with our estimate. CTB 1 is an evolved SNR. The B-vectors 
follow the shell at 10.6~GHz \citep{rei02} and deviate by about 40$\degr$ 
from the shell at $\lambda$6~cm. According to these data we calculated a RM of 
about $-$180~rad m$^{-2}$  in the western shell.

RMs can also be estimated for some shell-type SNRs based on the morphology by 
assuming that the intrinsic magnetic field is tangential \citep[e.g.][]{rei02}, 
such as G54.4$-$0.3 (HC 40). This SNR shows B-vectors that deviate 
significantly from the shell direction, which indicates an RM of about 
250$\pm$100~rad~m$^{-2}$.

\subsection{G16.8$-$1.1 is an HII region}
A flat spectrum with $\alpha\sim0.16$ was earlier derived by \citet{rfr+86}, which in principle could indicate 
either a thermal source or a crab-like SNR. \citet{rfr+86} classified 
it as a SNR, because strong polarized emission was observed with the Effelsberg 
100-m telescope at $\lambda$6~cm resulting in a percentage polarization of 
about 15\%. The slightly inverted spectrum was thought to be influenced by the 
compact \ion{H}{II} region Sh~2-50 coinciding with the SNR. It should be in 
front of G16.8$-$1.1, because it causes depolarization. The pulsar 
PSR B1822$-$14 
\citep{cl86} is seen within the area of G16.8$-$1.1 showing a very high RM of 
$-$899~rad~m$^{-2}$. Its relation to G16.8$-$1.1 is still unclear. 

\begin{figure}[!htbp]
\centering
\resizebox{0.48\textwidth}{!}{\includegraphics[angle=-90]{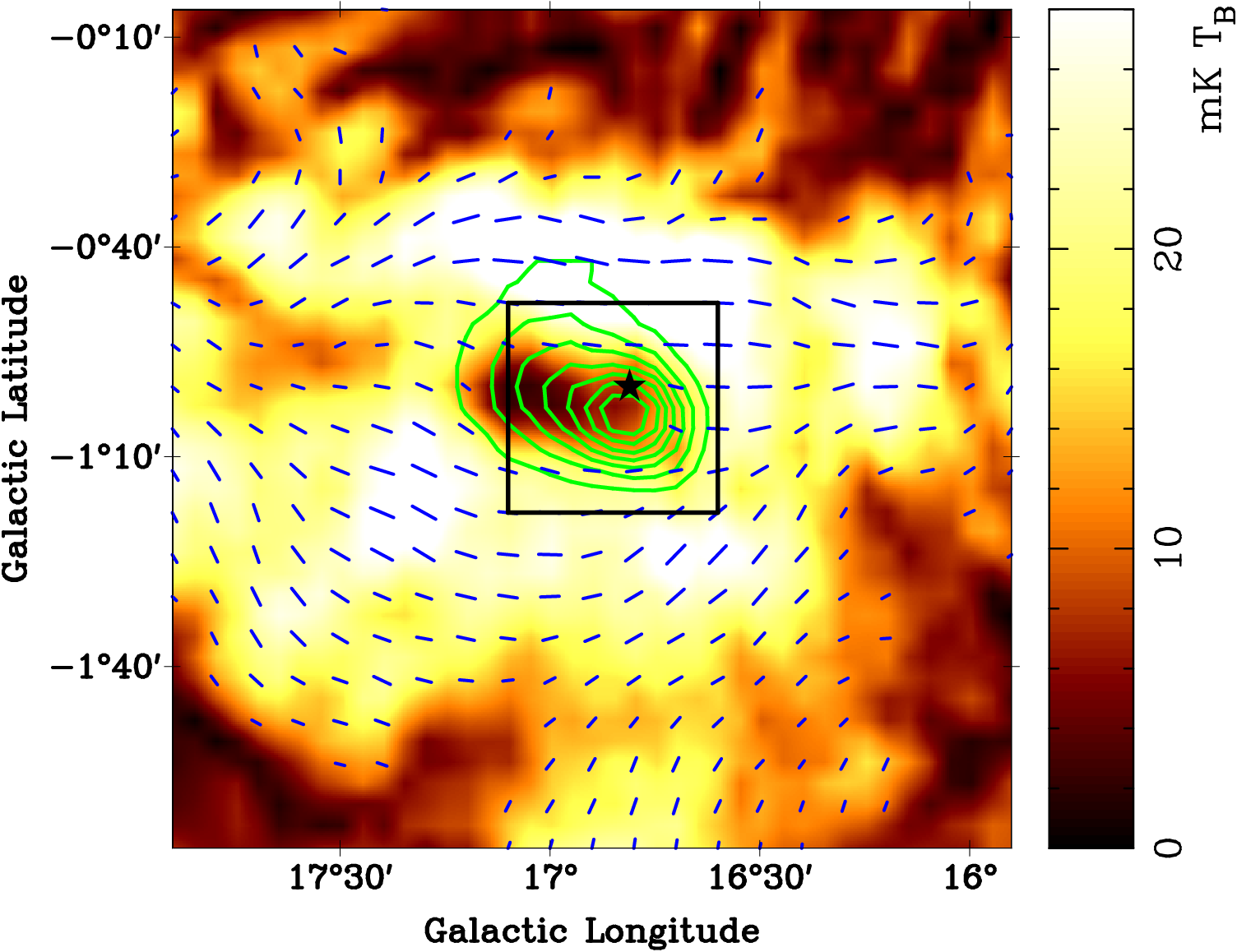}}
\resizebox{0.48\textwidth}{!}{\includegraphics[angle=-90]{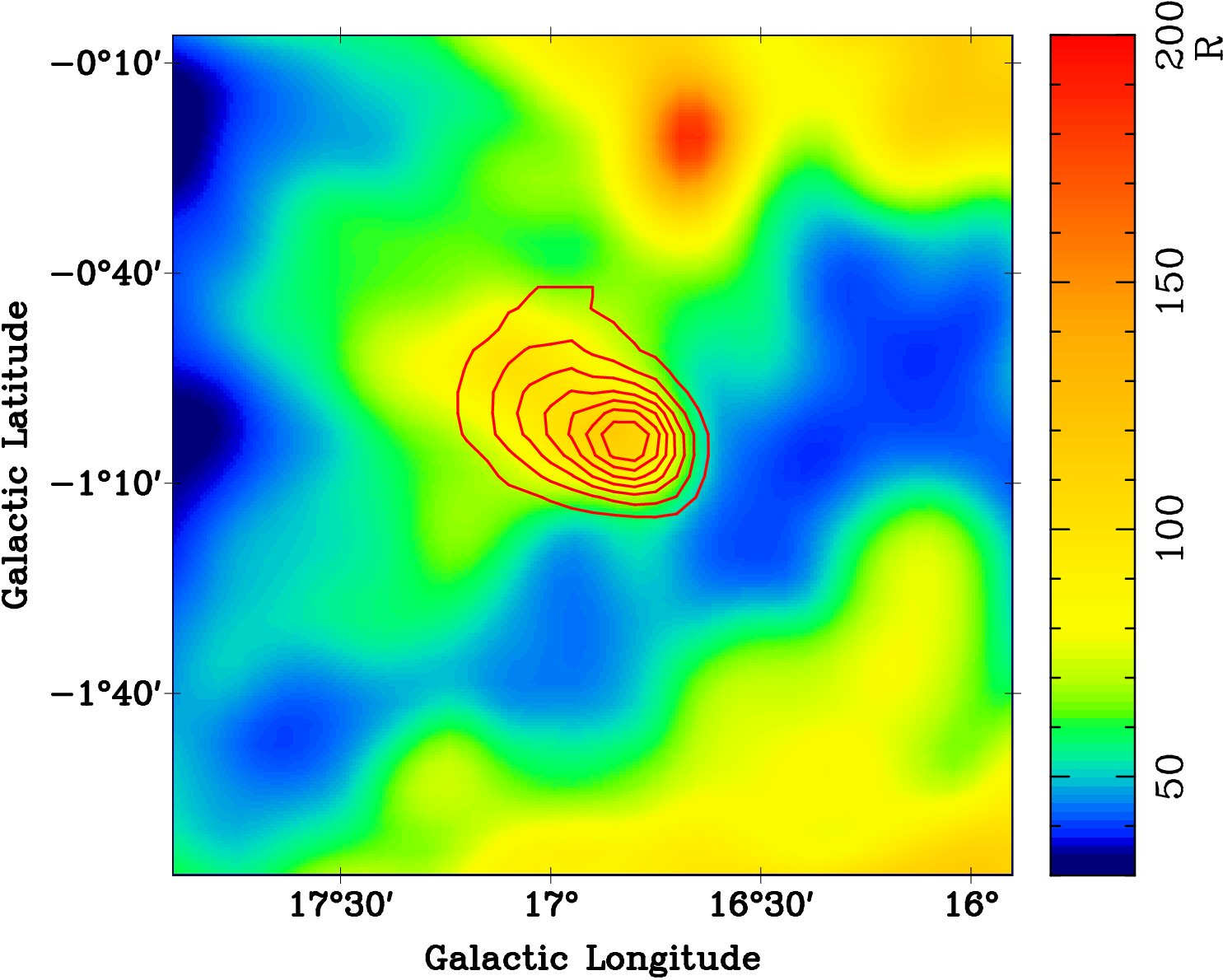}}
\resizebox{0.48\textwidth}{!}{\includegraphics[angle=-90]{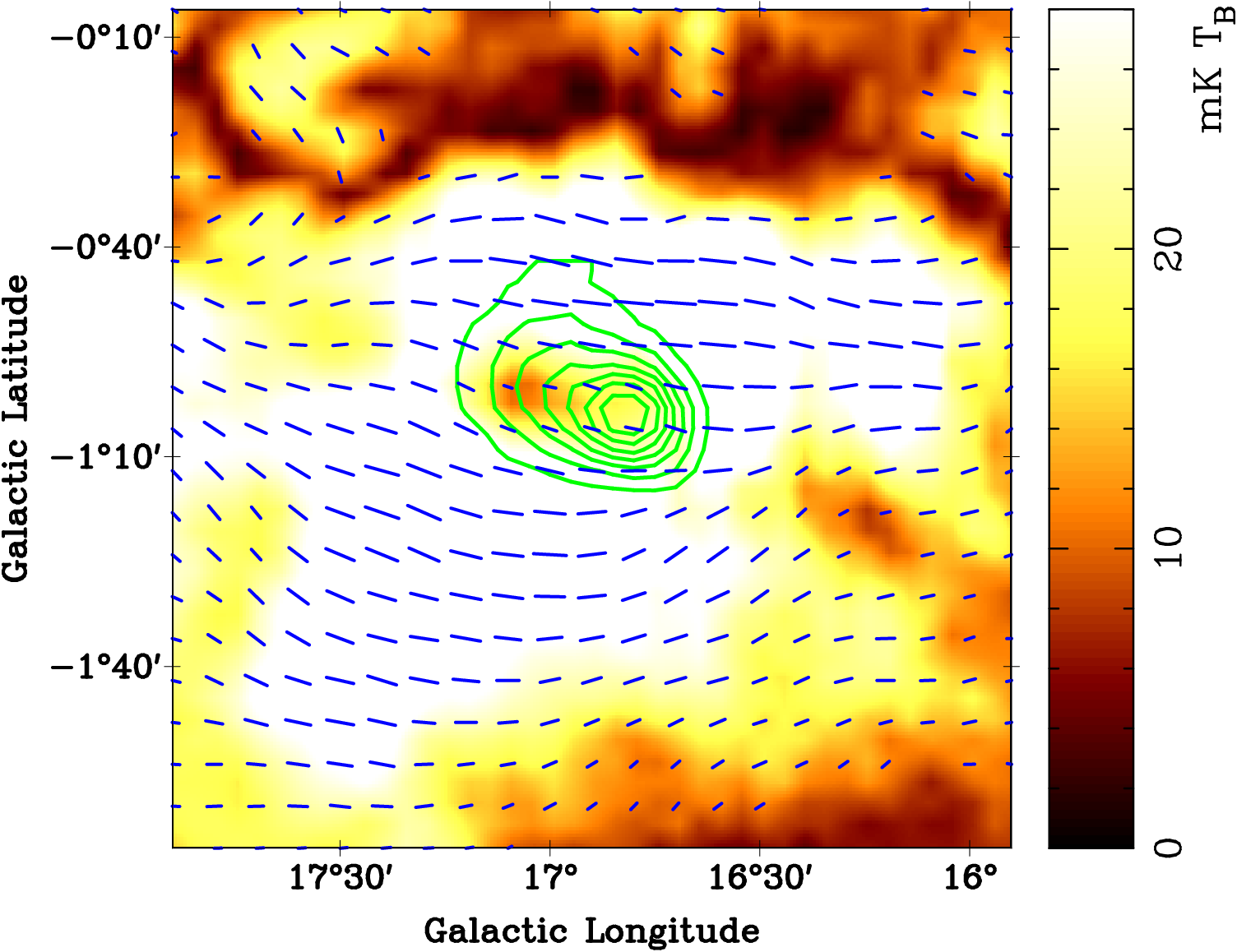}}
\caption{Images of G16.8$-$1.1. The contours display the total intensity at 
$\lambda$6~cm, starting at 50~mK~$T_{\rm B}$ and running in steps of 
50~mK~$T_{\rm B}$. The image shows the observed $\lambda$6~cm polarization 
intensity in the {\it top} panel, the H$\alpha$ intensity in the {\it middle} 
panel, and the absolutely calibrated $\lambda$6~cm polarization intensity 
in the {\it bottom} panel. The bars indicate B-vectors with their lengths 
proportional to polarized intensity with a low intensity cutoff of 
1~mK~$T_{\rm B}$. The box indicates the mapped region by \citet{rfr+86}. The 
star indicates the position of the pulsar PSR B1822$-$14.}
\label{g16.8}
\end{figure}

The $0\fdg5\times0\fdg5$ Effelsberg map presented by \citet{rfr+86} covers G16.8$-$1.1, 
but does not show its surroundings. On a larger $2\degr\times2\degr$ map 
(Fig.~\ref{g16.8}, top panel) extracted from the new $\lambda$6~cm survey, 
strong polarization intensity in the surrounding area of G16.8$-$1.1 is also  
visible. The central area of the source, however, seems to be nearly 
unpolarized. The edge areas of the smaller Effelsberg $\lambda$6~cm map 
\citep{rfr+86} almost coincide with strong polarized emission 
(Fig.~\ref{g16.8}) as visible in the Urumqi $\lambda$6~cm survey. Standard 
baseline subtraction assumes zero emission at the edges of a map. If this 
assumption does not hold, the polarized emission level in a map is not correct. 
Thus the main argument for a SNR identification of G16.8$-$1.1 becomes 
questionable and the entire G16.8$-$1.1 complex might be considered as thermal. 
Actually the $\lambda$6~cm total intensity of G16.8$-$1.1 fairly well matches 
the H$\alpha$ emission \citep{fin03} from the \ion{H}{II} region Sh 2-50 
(Fig.~\ref{g16.8}, middle panel), indicating that they are probably the same 
object. 

To check whether G16.8$-$1.1 acts as a Faraday screen 
\citep[e.g.][]{shr+07,grh+10}, we show the $\lambda$6~cm polarization intensity 
with large-scale emission restored according to WMAP data \citep{srh+11} in the 
bottom panel in Fig.~\ref{g16.8}. There is indication of insignificant rotation of polarization angles but some depolarization towards G16.8$-$1.1, which 
suggests that G16.8$-$1.1 is either within the polarized emission region or at a smaller distance than the polarized emission. More data are needed to settle this relation. 
Sh~2-50 is probably associated with the Scutum super shell at a distance of 
about 3.3~kpc \citep{csb+00}, which is close to the polarization horizon at 
$\lambda$6~cm as discussed by \citet{srh+11}. 

In case that G16.8$-$1.1 is thermal, the distance of 
5.1 kpc to the pulsar PSR B1822-14, 
based on the NE2001 model, will be reduced. Its large negative RM might be 
affected by G16.8$-$1.1 in a way already described by \citet{mwkj03} for 
pulsars shining through \ion{H}{II} regions.

\subsection{Cas~A}

\object{Cas A} is the strongest radio source seen beyond the solar system and included
in the $\lambda$6~cm survey section presented by \cite{xhr+11}. The Cas~A area 
was observed several times between 2004 and 2008. The peak brightness 
temperature at $\lambda$6~cm is around 95~K. For this large intensity, we have 
checked that the linearity of the receiving system holds. 

Cas~A is a young SNR whose intensity shows a secular decrease with the rate 
determined by \citet{bgpw77} as
\begin{equation}\label{barr}
\frac{{\rm d}S}{S{\rm d}t}=
\left[(0.97\pm0.04)-(0.30\pm0.04)\lg\left(\frac{\nu}{\rm GHz}\right)\right]\%\,\,
{\rm yr^{-1}}.
\end{equation}

At 4.8~GHz the decreasing rate is 0.77\% per year. The absolute spectrum for 
Cas~A was also determined by \citet{bgpw77}. In epoch 1965.0, the spectral 
index is $\alpha=-0.792$ between 0.3~GHz and 31~GHz and the flux density at 
4.8~GHz is 921.4~Jy. The expected flux density at $\lambda$6~cm in epoch 2004 
is 684$\pm$18~Jy and in epoch 2008 is 663$\pm$19~Jy. The measured flux density 
at $\lambda$6~cm is 688$\pm$35~Jy, which agrees with the expectations based on
\citet{bgpw77} within the errors. We note that \citet{hdd+08} proposed 
a lower decrease rate for Cas~A based on new observations as
\begin{equation}\label{hdd}
\frac{{\rm d}S}{S{\rm d}t}=
\left[(0.68\pm0.04)-(0.15\pm0.04)\lg\left(\frac{\nu}{\rm GHz}\right)\right]\%\,\,
{\rm yr^{-1}}.
\end{equation}

Applying Eq.~(\ref{hdd}) we calculated a decrease rate of 0.58\% at 
$\lambda$6~cm, resulting in a flux density of 736$\pm$20~Jy in epoch 2004 and 
718$\pm$21~Jy in epoch 2008. These values are somewhat larger than what we 
measured. 

Cas~A has a size of about $5\arcmin$ and cannot be resolved in the 
$\lambda$6~cm survey. As a very young shell-type SNR the magnetic field is 
expected to be radial. This can be seen from Effelsberg 32~GHz 
observations, 
which resolve Cas~A \citep{rei02}. We measured a polarization flux density at 
$\lambda$6~cm of 38$\pm$4~Jy for this SNR, corresponding to an average 
polarization percentage of about 6\%. 

\section{Summary}

We studied small SNRs with angular sizes less than $1\degr$ in the Sino-German 
$\lambda$6~cm polarization survey of the Galactic plane. Integrated flux 
densities of 51 SNRs were obtained. Fitting these measurements together with 
previous observations at other wavelengths, we obtained spectra of 50 SNRs. 
For half of the SNRs, the $\lambda$6~cm measurements provide by far the 
highest frequency data, and therefore play an important role on constraining the 
spectra of these SNRs. We have determined spectra for SNRs G15.1$-$1.6, 
G16.2$-$2.7, G16.4$-$0.5, G17.4$-$2.3, G17.8$-$2.6, G20.4+0.1, G36.6+2.6, 
G43.9+1.6, G53.6$-$2.2, G55.7+3.4, G59.8+1.2, G68.6$-$1.2, and G113.0+0.2, by 
mainly using the flux densities from the $\lambda$6~cm survey, 
along with the 
$\lambda$11~cm and $\lambda$21~cm Effelsberg surveys. Note that spectra of 
these SNRs were poorly determined up to present. G16.8$-$1.1 is most likely an 
\ion{H}{II} region and not a SNR. 

We were also able to extract polarization images of 25 SNRs. For SNRs 
G16.2$-$2.7, G69.7+1.0, G84.2$-$0.8, and G85.9$-$0.6, the 
polarized emission is detected for the first time. For some SNRs, RMs could be 
estimated.

We conclude that it is important to observe SNRs at high frequencies to 
accurately determine their spectra and study their intrinsic polarization 
properties. 

\begin{acknowledgements}
We like to thank the staff of the Urumqi Observatory for qualified 
assistance during the installation of the receiving system and the survey 
observations. In particular we are grateful to Otmar Lochner for the 
construction of the $\lambda$6~cm receiver, installation and commissioning. 
Maozheng Chen and Jun Ma helped with the installation of the $\lambda$6~cm 
receiving system and maintained it since 2004. We thank Prof. Ernst F\"urst for 
his support of the survey project and critical reading of the manuscript. 
The MPG and the NAOC supported the 
construction of the Urumqi $\lambda$6~cm receiving system by special funds. The 
Chinese survey team is supported by the National Natural Science foundation of 
China (10773016, 10833003, 10821061) and the National Key Basic Research 
Science Foundation of China (2007CB815403). XHS acknowledges financial support 
by the MPG and by Prof. Michael Kramer during his stay at MPIfR Bonn.
Some data in this paper are based on observations with the 100-m telescope of the MPIfR at Effelsberg.
XYG is supported by the Young Researcher Grant of NAOC. 
We thank the anonymous referee for the very helpful comments which has 
significantly improved the paper.

\end{acknowledgements}

\bibliographystyle{aa}
\bibliography{/home/xhsun/MPIfR/bibtex}

\longtab{1}{
\begin{longtable}{lr@{$\pm$}llr@{$\pm$}lr@{$\pm$}llr@{$\pm$}ll}
\caption{\label{snr_s} SNRs with flux densities measured from the $\lambda$6~cm survey.}\\
\hline\hline
Name & \multicolumn{2}{c}{Prev. $S_{\rm 6\,cm}$ (Jy)} & Ref. & 
\multicolumn{2}{c}{New $S_{\rm 6\,cm}$ (Jy)} & 
\multicolumn{2}{c}{Prev. $\alpha$} & Ref. & 
\multicolumn{2}{c}{New $\alpha$} & Ref.
\\\hline 
\endfirsthead
\caption{continued.}\\
\hline\hline
Name & \multicolumn{2}{c}{Prev. $S_{\rm 6\,cm}$ (Jy)} & Ref. & 
\multicolumn{2}{c}{New $S_{\rm 6\,cm}$ (Jy)} & 
\multicolumn{2}{c}{Prev. $\alpha$} & Ref. & 
\multicolumn{2}{c}{New $\alpha$} & Ref.
\\\hline 
\endhead
\hline
\endfoot
\object{G11.1$-$1.0} & 5.6 & 0.6 &3 & 3.40 & 0.25 & \multicolumn{2}{c}{$-$0.6}& 2 & $-$0.41&0.02&1--5\\
\object{G11.2$-$0.3}  & 9.6 & 0.5  & 11 & 8.95 & 0.48 & $-$0.50&0.02& 11& $-$0.48&0.01&1, 4, 5, 6 -- 19\\
\object{G15.1$-$1.6}  & \multicolumn{2}{c}{$\ldots$}& $\ldots$  &  4.81 & 0.27& \multicolumn{2}{c}{$-$0.8?}&142&$-$0.01&0.09&1, 4\\
\object{G15.4+0.1}    & \multicolumn{2}{c}{$\ldots$}& $\ldots$ &  2.07 & 0.18 & \multicolumn{2}{c}{$-$0.6}& 2 & $-$0.62&0.03&1, 2\\
\object{G15.9+0.2}    & 1.9 & 0.2 &21 &  1.95 & 0.29 & \multicolumn{2}{c}{$-$0.63}&22& $-$0.63&0.03&1, 4, 5, 8, 14, 16, 20--22\\ 
\object{G16.2$-$2.7}  & \multicolumn{2}{c}{$\ldots$} &$\ldots$&  1.28 & 0.10 & $-$0.51&0.10&23 & $-$0.42&0.10&23\\
\object{G16.4$-$0.5}  & \multicolumn{2}{c}{$\ldots$} &$\ldots$&  2.97 & 0.26 & \multicolumn{2}{c}{$-$0.7}&2& $-$0.26&0.15&2 \\
\object{G16.7+0.1}    & 0.95 &   0.10 & 24 & 1.23 & 0.11 & \multicolumn{2}{c}{$-$0.6}& 24& $-$0.56&0.05&1, 5, 14, 24 \\
\object{G16.8$-$1.1}  & 7.4 & 1.2 & 25 & 7.39 & 0.39 & \multicolumn{2}{c}{0.16}& 25 & 0.25&0.07&1, 5, 25\\
\object{G17.4$-$2.3}  & \multicolumn{2}{c}{$\ldots$}&$\ldots$&  2.33 & 0.23 & \multicolumn{2}{c}{$-$0.8?}&142&$-$0.46&0.12&4\\
\object{G17.8$-$2.6}  & \multicolumn{2}{c}{$\ldots$}&$\ldots$&  2.28 & 0.13 & \multicolumn{2}{c}{$-$0.3?}&142& $-$0.50&0.07&\\
\object{G18.8+0.3} & 15.7 & 3 & 31  & 15.29 & 0.89 & \multicolumn{2}{c}{$-$0.42}& 22 & $-$0.46&0.02&1, 4, 8, 10, 15, 22, 26--32\\
\object{G18.9$-$1.1}  & 23.8 & 2.4 & 34 & 19.57 & 1.01 & \multicolumn{2}{c}{$-$0.4}& 34 & $-$0.39&0.03&1, 4, 33--35\\
\object{G20.0$-$0.2}  & 10.4 & 1 & 36 & 9.23 & 0.54 & $-$0.04&0.06& 143& $-$0.06&0.04&1, 4, 5, 14, 36\\
\object{G20.4+0.1}    & \multicolumn{2}{c}{$\ldots$}&$\ldots$& 7.50 & 0.50 & \multicolumn{2}{c}{$-$0.4}&2& $-$0.08&0.09&2, 4\\
\object{G21.5$-$0.9}  & 7.2 & 1 & 45 & 6.54 & 0.37 & \multicolumn{2}{c}{0.0}& 42& $-$0.06&0.03&1, 4--6, 10, 13, 15, 28, 37--45\\
   \multicolumn{9}{c}{\hfill}                &$-$0.41&0.09&\\
\object{G21.8$-$0.6}  & 29 & 3 & 9& 24.03 & 1.29 & \multicolumn{2}{c}{$-$0.5}&10&$-$0.56&0.03&1, 4, 8--10, 13, 15, 17, 30, 32, 41, 46, 47\\
\object{G27.8+0.6}    & 18 &    2& 48 & 20.98 & 1.08 & $-$0.3&0.1&48&$-$0.25&0.07&1, 48\\
\object{G29.7$-$0.3} & 3.4 &0.7 & 31 & 3.62 & 0.58 & \multicolumn{2}{c}{$-$0.66}&12&$-$0.63&0.02&1, 6--8, 10, 12, 16--18, 27, 31, 40, 41, 49, 50\\ 
\object{G30.7+1.0}    & 3.4 &    0.4 &  25 & 2.93 & 0.19 & \multicolumn{2}{c}{$-$0.34}& 25 & $-$0.38&0.05&10, 25\\
\object{G31.9+0.0} & 10 & 1 & 31 &  8.94 & 0.56 & \multicolumn{2}{c}{$-$0.5}&63&$-$0.02&0.04&1, 5--8, 10, 14--16, 27, 31, 49--64\\
  \multicolumn{9}{c}{\hfill}                &$-$0.54&0.02&\\
\object{G33.6+0.1}    & 11.4 &    1.1 & 20 & 9.44 & 0.54 & \multicolumn{2}{c}{$-$0.4}&10&$-$0.51&0.02&1, 5--8, 10, 14, 16, 20, 52, 65--69\\
\object{G34.7$-$0.4} & 127 & 13 & 1 & 117.55 & 6.00 & $-$0.37 & 0.02 & 70 & $-$0.37&0.02&1, 8, 10, 26, 27, 31, 32, 41, 47, 51, 52, \\
\multicolumn{11}{c}{\hfill} & 60, 62, 64, 66, 70--84 \\
\object{G36.6+2.6}    & \multicolumn{2}{c}{$\ldots$} &$\ldots$&  0.39 & 0.04 & \multicolumn{2}{c}{$-$0.5?} & 142 & $-$0.67&0.12&\\
\object{G39.2$-$0.3} & 8.7 & 0.9 & 86&  8.84 & 0.53 & 0.42 & 0.02 &91&$-$0.34&0.01&1, 5--8, 10, 13--18, 30, 49, 51, 62, \\
  \multicolumn{11}{c}{\hfill}&64, 66, 68, 85-92\\
\object{G40.5$-$0.5}  & \multicolumn{2}{c}{$\ldots$}&$\ldots$&  6.39 & 0.34 & 0.41&0.05&93&$-$0.41&0.08&1, 28, 64, 93, 94\\
\object{G41.1$-$0.3} & 15 & 1.5 & 1 & 18.52 & 1.07 & \multicolumn{2}{c}{$-$0.48}&12 & $-$0.50&0.01&1, 5, 7, 8, 10, 12, 14--16, 32, 47, 56, 60, 62, \\
      \multicolumn{11}{c}{\hfill} & 64--66, 89, 94\\
\object{G43.3$-$0.2} & 17 & 2 &9& 19.10 & 0.98 & \multicolumn{2}{c}{$-$0.48}&12 & $-$0.46&0.01&1, 7--9, 12, 14, 16, 17, 94--98\\
\object{G43.9+1.6}    & \multicolumn{2}{c}{$\ldots$} &$\ldots$&  4.55 & 0.24 & \multicolumn{2}{c}{0.0?}&142& $-$0.47&0.06&\\
\object{G46.8$-$0.3} & 7.1 & 0.7 & 65& 7.02 & 0.18 & \multicolumn{2}{c}{$-$0.53}&22&$-$0.54&0.02&1, 8, 20, 22, 28, 32, 60, 65, 89, 98 \\
\object{G53.6$-$2.2} & \multicolumn{2}{c}{$\ldots$}&$\ldots$&  4.00 & 0.22 & $-$0.76&0.02&99&$-$0.50&0.02&26, 32, 47, 60, 62, 64, 66, 99, 100\\ 
\object{G55.7+3.4}    & \multicolumn{2}{c}{$\ldots$} &$\ldots$&  0.52 & 0.03 & $-$0.6&0.1&101&$-$0.34&0.11&101\\ 
\object{G57.2+0.8} & 0.62 & 0.06 &102&  0.74 & 0.04 & \multicolumn{2}{c}{$-$0.67}&104&$-$0.62&0.01&5, 14, 98, 102--107\\ 
\object{G59.8+1.2}    & \multicolumn{2}{c}{$\ldots$}&$\ldots$&  1.43 & 0.08 & \multicolumn{2}{c}{$-$0.5}&142& $-$0.03&0.05&\\
\object{G63.7+1.1}    & 1.16 &    0.10 &108&  1.12 & 0.06 & \multicolumn{2}{c}{$-$0.30}&104&$-$0.24&0.02&5, 14, 98, 102, 104, 106, 108\\    
\object{G65.7+1.2} & 1.79 & 0.1 &111&  1.95 & 0.10 & \multicolumn{2}{c}{$-$0.59}&111& $-$0.57&0.01&1, 5, 32, 66, 98, 107, 109--112\\
\object{G67.7+1.8}    & 0.42 &  0.05 & 98&  0.30 & 0.03 & $-$0.49&0.05&110&$-$0.61&0.01&5, 98, 104, 110\\ 
\object{G68.6$-$1.2}  & \multicolumn{2}{c}{$\ldots$}&$\ldots$& 0.80 & 0.04 & \multicolumn{2}{c}{0.0?}&142&$-$0.22&0.09&110\\
\object{G69.7+1.0}    & 0.58 &  0.05 &113&  0.68 & 0.07 & $-$0.70&0.06&110&$-$0.71&0.05&5, 110, 113, 114\\
\object{G73.9+0.9}    & 6.7 &   0.5 & 25 & 6.17 & 0.34 & $-$0.23&0.03&110&$-$0.23&0.03&5, 25, 110, 115\\
\object{G74.9+1.2} & 7.5 & 0.7 &119& 6.35 & 0.35 & $-$0.29&0.02&110&$-$0.26&0.02&5, 15, 28, 42, 49, 66, 98, 104, 110, 115--121 \\
   \multicolumn{9}{c}{\hfill}                &$-$0.71&0.18&\\
\object{G76.9+1.0}    & 0.63 &    0.03 &122&  0.79 & 0.07 & $-$0.60&0.02&110&$-$0.89&0.02&5, 40, 98, 104, 110, 122--124 \\    
\object{G94.0+1.0} & 7.2 &  0.5 &126&  6.23 & 0.35 & $-$0.48&0.02&110&$-$0.45&0.02&32, 47, 51, 62, 64, 110, 125--129\\
\object{G96.0+2.0}    & \multicolumn{2}{c}{$\ldots$}&$\ldots$ &  0.14 & 0.02 & $-$0.45&0.13&110&$-$0.59&0.07&110\\
\object{G109.1$-$1.0} &\multicolumn{2}{c}{$\ldots$}&$\ldots$&  9.78 & 0.52 & $-$0.50&0.04&110&$-$0.45&0.04&110, 130--132\\
\object{G113.0+0.2}   &\multicolumn{2}{c}{$\ldots$}&$\ldots$& 1.85 & 0.50 & \multicolumn{2}{c}{$\ldots$}&$\ldots$&$-$0.45&0.25&\\
\object{G116.9+0.2} &3.0 & 0.3 &20&  3.60 & 0.40 & $-$0.61&0.03&110&$-$0.57&0.03&20, 32, 47, 110, 133--135\\
\object{G120.1+1.4} &21.1 & 0.4 &62& 20.03 & 2.00 & $-$0.65&0.01&110&$-$0.58&0.02& 28, 40, 49, 51, 56, 61, 62, 64, 78,\\
      \multicolumn{11}{c}{\hfill}& 94, 102, 104, 107, 110, 136--138\\
\object{G130.7+3.1} &31.2 &  1.8 & 136& 31.66 & 3.00 & $-$0.07&0.01&110&$-$0.07&0.02&8, 28, 40, 42, 44, 49, 61, 62, 64,\\ 
     \multicolumn{11}{c}{\hfill}& 78, 94, 102, 104, 107, 110, 136, 138--140\\
\object{G182.4+4.3}   &0.20 &  0.02 & 141& 0.26 & 0.05 & $-$0.44&0.10&141&$-$0.41&0.14&141\\
\end{longtable}
\tablebib{
 (1) \citet{adg+70},    (2) \citet{bgg+06},     (3) \citet{ch87},
 (4) \citet{gd70},      (5) \citet{rrf90},      (6) \citet{bk75},      
 (7) \citet{ds75},      (8) \citet{gre74},      (9) \citet{gs70},     
(10) \citet{kas92},    (11) \citet{kr01},      (12) \citet{mr87b}, 
(13) \citet{mwgm69},   (14) \citet{rfh+84},    (15) \citet{rwb+70},
(16) \citet{sle77},    (17) \citet{sg70},      (18) \citet{sw76},
(19) \citet{trk02},    (20) \citet{abk77},     (21) \citet{ccg75},    
(22) \citet{dgg+96},   (23) \citet{tru99},     (24) \citet{hvbl89},  
(25) \citet{rfr+86},   (26) \citet{cgc75},     (27) \citet{kes68},
(28) \citet{lmd+00},   (29) \citet{mck+89},    (30) \citet{md75},
(31) \cite{mil69},     (32) \citet{wil73},     (33) \citet{bt88}, 
(34) \citet{frr+85},   (35) \citet{pvv88},     (36) \citet{bh85}, 
(37) \citet{bk76},     (38) \citet{bwd01},     (39) \citet{cc74}, 
(40) \citet{frrr90},   (41) \citet{md74},      (42) \citet{mr87a},   
(43) \citet{sest89},   (44) \citet{srh+89},    (45) \citet{ww76},
(46) \citet{kvh74},    (47) \citet{vk74},      (48) \citet{rfs84},
(49) \citet{fff+74},   (50) \citet{gsw67},     (51) \citet{ben62},
(52) \citet{bk69},     (53) \citet{bk71},      (54) \citet{blkd05},
(55) \citet{cdg+71},   (56) \citet{cdl65},     (57) \citet{cha74},
(58) \citet{dmka73},   (59) \citet{gssw79},    (60) \citet{hc69},
(61) \citet{kpt68},    (62) \citet{kpw69},     (63) \citet{mr94a},
(64) \citet{pwh66},    (65) \citet{ccc75},     (66) \citet{dd75},
(67) \citet{gmw69},    (68) \citet{gwm75},     (69) \citet{gsd+09},
(70) \citet{cdbk07},   (71) \citet{dgw65},     (72) \citet{dwbw80},
(73) \citet{gdk+97},   (74) \citet{har62},     (75) \citet{klns60},
(76) \citet{kuz62},    (77) \citet{kv69},      (78) \citet{kv72},    
(79) \citet{lmh61},    (80) \citet{les60},     (81) \citet{mor65},
(82) \citet{sch63},    (83) \citet{wes58},     (84) \citet{wil63},
(85) \citet{adps79},   (86) \citet{bcbw81},    (87) \citet{bh87},   
(88) \citet{dps81},    (89) \citet{dwc70},     (90) \citet{hb69a}, 
(91) \citet{phs+90},   (92) \citet{sgb+07},    (93) \citet{dsp80},    
(94) \citet{dsp81},    (94) \citet{gbl75},     (95) \citet{hb69b}, 
(96) \citet{llk+01},   (97) \citet{mr94b},     (98) \citet{twg92},
(99) \citet{dggw94},  (100) \citet{gss75},    (101) \citet{gssw77},
(102) \citet{bwe91},  (103) \citet{clf85},    (104) \citet{hsg+09},
(105) \citet{ss84},   (106) \citet{tgc+96},   (107) \citet{wb92},    
(108) \citet{wtt97},  (109) \citet{dwy71},    (110) \citet{kffu06},
(111) \citet{klr+08}, (112) \citet{lc83},     (113) \citet{jun86},
(114) \citet{jfr88},  (115) \citet{pc90},     (116) \citet{dids75},
(117) \citet{gps80},  (118) \citet{mor82},    (119) \citet{whl91},
(120) \citet{wltp97}, (121) \citet{ws78},     (122) \citet{lhw93},
(123) \citet{lzzh97}, (124) \citet{mcg+11},   (125) \citet{fos05},
(126) \citet{gms+84}, (127) \citet{lhr85},    (128) \citet{mnst82},
(129) \citet{vg98},   (130) \citet{dow83},    (131) \citet{hhc+84},
(132) \citet{sth83},  (133) \citet{lrd82},    (134) \citet{rb81},
(135) \citet{tl06},   (136) \citet{hcd69},    (137) \citet{kehs79},
(138) \citet{rrf97},  (139) \citet{bkw01},    (140) \citet{gre86},
(141) \citet{kfr98},  (142) \citet{rfrj88},   (143) \citet{ode86}}

}

\end{document}